# Distributive Computability

Sebastiano Vigna

# Contents

# Chapter 1

# Introduction

This thesis presents a series of theoretical results and practical realisations about the theory of computation in distributive categories. Distributive categories have been proposed as a foundational tool for Computer Science in the last years, starting from the papers of R.F.C. Walters [Walters:1989a, Walters:1992a, Walters:1992b]. We shall focus on two major topics: *distributive computability*, i.e., a generalized theory of computability based on distributive categories, and the **IMP**$(G)$ language, which is a language based on the syntax of distributive categories. The link between the former and the latter is that the functions computed by **IMP**$(G)$ programs are exactly the distributively computable functions.

In order to put into a historical perspective the definitions we shall use, we would like to quote part of a letter of D.E. Knuth written in 1966 to the editor of the *Communications of the ACM* [Knuth:1966]:

> The preceding letter by Dr. Huber defines "algorithm" in terms of programming languages. I would like to take a slightly different point of view, in which algorithms are concepts which have existence apart from any programming language. [...] I will try to explain how the notion of "algorithm" can be mathematically formulated along these lines. Let us say that a computational method comprises a set $Q$ (finite or infinite) of "states", containing a subset $X$ of "inputs" and a subset $Y$ of "outputs"; and a function $F$ from $Q$ into itself. (These quantities are usually also restrictely to be finitely definable, in some sense that corresponds to what human beings can comprehend.) Such a computational method defines a computation for each $x$ in $X$ as follows: Let $q_0 = x$, and for $m \geq 0$ when $q_m$ has been defined let $q_{m+1} = F(q_m)$ if $q_m \notin Y$; if $q_m \in Y$, we say that the computational method terminates after $m$ steps, producing the output $q_m$. Clearly any program (i.e., a sentence in a programming language) is an example of a computational method in this sense. An algorithm is now defined to be a computational method that terminates in finitely many steps for each input $x$. Finally we can define the notion of a program representing a computational method: let the computational method $C$ be $(Q, X, Y, F)$ and let the program $P$ be $(Q', X', Y', F')$; $P$ represents $C$ if there is a function $\sigma$ from $Q$ into $Q'$, taking $X$ into $X'$, and a function $\tau$ from $Y'$ into $Y$, such that (i) if $x \in X$ defines defines the computational sequence $x = q_0, q_1, \ldots$ in $C$, then $\sigma(q_0), \sigma(q_1), \ldots$ is a subsequence of the computational sequence $x' = \sigma(x) = q'_0, \ldots$ in $P$; and (ii) if $C$ produces output $y$ from $x$, $P$ produces the output $y'$ from $x'$ where $\tau(y') = y$.

The notion of *pseudofunction* (Chapter 3), introduced by Walters in the context of distributive categories [Walters:1992b], corresponds exactly to that of an algorithm in the sense of Knuth (*prefunc-* *tional processors* [Khalil], on the other hand, correspond to computational methods). And we shall show that the notion of representation can be led to that of a *functor*. Thus, the study of computation in distributive categories can be viewed as a new, wide generalization of some basic ideas about computation and iteration which have been known for a long time. The scientific program carried out in this thesis deals in particular with the computational power of computation in distributive categories, and with the relations between iteration and refinement (implementation/representation).

Such a generalization has obviously to be compared with the existing general formalisms for the description of iteration. In particular, we would like to mention Heller's *recursion categories* [Heller:1990] and Elgot's *iteration theories* [Elgot:1975]. Heller's framework is rather different from ours, because he is interested in generalizing *recursion theory* rather than *computation theory*. For instance, all the objects of a recursion category are isomorphic, which is not true, for instance, as soon as the natural numbers are replaced by a generic ring $R$, as in Section 2.4, where the BSS model of computation will be compared with distributive categories. On the other hand, Heller proved that the *Turing completion* of certain subcategories of a (locally connected) iteration category are always recursion categories, in the sense that it is always possible to prove the existence of universal functions [Heller:1990].

The *feedback operator* of iteration theories, on the contrary, is very similar to the **call**$[-]$ operator introduced in Chapter 2. In particular, the normal form theorem exhibits some properties of **call**$[-]$ which are true also for the feedback operator [Bloom & Esik:1993]. However, iteration theories are restricted to consider sum of objects (an iteration theory is an $\boldsymbol{N}$-category, i.e., it is a category whose objects are the natural numbers), whereas in distributive categories both sums and products are allowed, thus allowing to describe data types (for instance, the "parallel product" of two functions is not expressible in an iteration theory, as it is not the related part of the normal form theorem).

Chapter 2 introduces in an elementary manner the definition of distributive computability, and gives the basic results which motivate further developments: first of all, the distributively computable functions over the successor and predecessor functions are exactly the recursive functions; second, the distributively computable functions over the operation of an ordered ring $R$ are exactly the functions computable in the BSS model (this result provides one of the few known equivalent formulations of the BSS computable functions).

Chapter 3 is devoted to the study of morphisms of automata. The main result proved here is a refinement theorem which says that an automaton on a free monoid implements (represents) the functions it computes, in the sense of Knuth's letter.

Chapter 4 dives deeply into category theory, in order to describe distributive computability in the greatest generality. The technical definition justifying the elementary treatment of Chapter 2 are given here, together with a series of theoretical results about distributive and extensive categories which allow to prove a theorem about the computational irrelevance of iteration (in a very general sense) under a simple syntactical assumption which corresponds to the absence of tests. Finally, the refinement theorem is lifted to the extensive categorical setting.

The last two chapters present the implementation of a series of tools (type-checker, compiler and interpreter) for programming in **IMP**$(G)$. The tools are based on a new representation of distributive data and programs (the first such representation was given in [Khalil & Walters:1993]) which allows to implement efficiently an interpreter.

Most of the material in this thesis is original. The equivalence theorems of Chapter 2 have been published in [Sabadini, Vigna & Walters:1996] and in [Vigna:1996] (the equivalence theorem for recursive functions appears here in a slightly stronger version). The basic ideas of Chapter 3 were introduced in [Sabadini, Vigna & Walters:1993a] and in [Sabadini, Vigna & Walters:1993b]; the presentation given here extends the results contained in the aforementioned references to generic monoids, and introduces some new counterexamples and theorems (in particular, Section 3.3).

The constructions given in Section 4.1.2 appear here for the first time, and starting from Section 4.2.1 the rest of Chapter 4 is original. The first definition of function computed by iteration was given by Khalil and Walters in terms of colimits; the definition was then simplified and restated in terms of sums in [Wagner, Khalil & Walters:1995] and [Sabadini, Vigna & Walters:1996]. In this thesis we use a new formulation, which is equivalent to the previous ones.

Finally, the material of the last two chapters has been entirely developed by the author, and partially presented in [Vigna:1995].

I would like to thank several people, without whom the making of this thesis would have been much harder, if not impossible. Paolo Boldi introduced me to the BSS model by means of his excellent survey, and commented the drafts. The material of Chapter 4 and of Chapter 6 has been heavily discussed with Steve Lack and Eelco Visser, respectively. In particular, the electronic communications with Steve have always been a useful source of inspiration about distributive and extensive categories, while Eelco answered all my (often stupid) questions about ASF+SDF[1] and suggested several improvements. The feedback received at the ASF+SDF95 workshop from the people of the University of Amsterdam was also a precious source of information. Danilo Bruschi and Gaetano Lanzarone kindly made the computing resources on which to run ASF+SDF available.

But the biggest "thanks" goes without any doubt to the people of the "Sala Dottorandi & Paolo Boldi" (Ph.D. room) of the University of Milan. For three years they have been not only brilliant researchers with whom interaction is always scientifically fruitful, but also friends whose relationships are warm and intense; the time spent together has been some of the best time of my life.

[1]The system ASF+SDF was used in order to realize the implementation described in the last two chapters.

# Chapter 2

# Distributive computability

In this chapter we shall introduce the concept of *distributive computability*. The definition we shall give of *function computable by iteration* is in the same spirit of Kleene's definition of recursive function [Kleene:1936], but it will be given in the environment of a countably extensive category, thus giving a wide and natural generalization of the concept of computable function.

Kleene's definitions builds inductively a set of functions from the powers of $\boldsymbol{N}$ to $\boldsymbol{N}$; starting from the zero function, the successor function and the projections, and closing by substitution, primitive recursion and minimization, we obtain a class of function which we call (partial) recursive.

In the definition which we shall give the domain and codomain of the functions will be obtained by using also the *sum* operator (disjoint union, in the category of sets). This fact will allow the definition of an operator of iteration which essentially incorporates both primitive recursion and minimization. Moreover, the rôle of the zero, successor and projection functions will be taken by an arbitrary set of basic functions; thus, the notion of distributive computability will be parametric, and will depend on the given basic set.

Recently, Blum, Shub and Smale proposed a generalization of the definition of computable function [Blum, Shub & Smale:1989] which is valid in an arbitrary ordered ring, rather than in the natural numbers. We will show how also this class of functions is generated by a particular instantiation of the model of distributive computability.

## 2.1 An elementary definition

In order to make this thesis accessible to the widest audience, we will first give a definition of distributively computable function in a special case, i.e., in the familiar category of sets. In other words, **Set** is now the "computational universe" (the rôle played by the category $\mathfrak{C}$ in Chapter 4).

For any pair of sets $X$, $Y$, the *sum* $X + Y$ is the disjoint union of $X$ and $Y$, i.e., the set $X \times \{0\} \cup Y \times \{1\}$. The *canonical injections* $\mathrm{inj}_1 : X \to X + Y$, $\mathrm{inj}_2 : Y \to X + Y$ are defined by

$$\begin{aligned} \mathrm{inj}_1(x) &= \langle x, 0\rangle \\ \mathrm{inj}_2(y) &= \langle y, 1\rangle. \end{aligned}$$

The *product* $X \times Y$ is the cartesian product of $X$ and $Y$, i.e., the set $\{\langle x, y\rangle \mid x \in X \wedge y \in Y\}$. The *canonical projections* $\mathrm{proj}_1 : X \times Y \to X$, $\mathrm{proj}_2 : X \times Y \to Y$ are defined by

$$\begin{aligned} \mathrm{proj}_1(\langle x, y\rangle) &= x \\ \mathrm{proj}_2(\langle x, y\rangle) &= y. \end{aligned}$$

Should ambiguities arise, we shall write $\mathrm{inj}_1^{X,Y}$ and so on.

The empty set $\varnothing$ and the singleton $I = \{*\}$ are the units of $+$ and $\times$ (up to isomorphism, i.e., up to bijections of sets; we shall not insist in making such isomorphisms explicit). For any set $X$, there are unique functions

$$\begin{aligned} !_X &: \varnothing \to X \\ ¡_X &: X \to I, \end{aligned}$$

called the *initial* and *terminal* maps[1].

We have also corresponding constructions on functions: for any pair of functions $f : X \to Z$ and $g : Y \to Z$, $(f \mid g) : X + Y \to Z$ is defined by

$$(f \mid g)(z) = \begin{cases} f(x) & \text{if } z = \langle x, 0\rangle \\ g(y) & \text{if } z = \langle y, 1\rangle \end{cases}$$

and for any pair of functions $f : Z \to X$ and $g : Z \to Y$, $(f, g) : Z \to X \times Y$ is defined by

$$(f, g)(x) = \langle f(x), g(x)\rangle.$$

We shall denote by $g \circ f$ the composition of the functions $f : X \to Y$ and $g : Y \to Z$.

Sums and products of sets are related by the *distributivity isomorphism*

$$\delta : X \times Y + X \times Z \xrightarrow{\cong} X \times (Y + Z),$$

which maps $\langle\langle x, w\rangle, k\rangle$ to $\langle x, \langle w, k\rangle\rangle$, with $k = 0$ or 1, and $w$ an element of $Y$ or $Z$, respectively. Moreover, any set $X$ has an associated identity function $\mathbf{1}_X : X \to X$.

We can now introduce some entities derived from the previous ones:

- $\Delta_X = (\mathbf{1}_X, \mathbf{1}_X) : X \to X \times X$;
- $\nabla_X = (\mathbf{1}_X \mid \mathbf{1}_X) : X + X \to X$;
- for each pair of functions $f : X \to X'$ and $g : Y \to Y'$, we have

  $$f \times g = (f \circ \mathrm{proj}_1, g \circ \mathrm{proj}_2) : X \times Y \to X' \times Y'$$

  and

  $$f + g = (\mathrm{inj}_1 \circ f \mid \mathrm{inj}_2 \circ g) : X + Y \to X' + Y'.$$

Note that $(f \mid g) = \nabla \circ (f + g)$ and $(f, g) = (f \times g) \circ \Delta$, so whatever we build using $(-, -)$ and $(- \mid -)$ can also be built using $\times$, $+$, $\Delta$ and $\nabla$; for instance, the canonical function $\delta : X \times Y + X \times Z \to X \times (Y + Z)$ we described previously is exactly

$$\delta = (\mathbf{1}_X \times \mathrm{inj}_1 \mid \mathbf{1}_X \times \mathrm{inj}_2),$$

and the isomorphisms of commutativity of sum and product, both denoted **twist** with an abuse of notation, can be expressed as

$$\mathbf{twist} : X \times Y \xrightarrow{(\mathrm{proj}_2, \mathrm{proj}_1)} Y \times X$$

and

$$\mathbf{twist} : X + Y \xrightarrow{(\mathrm{inj}_2 \mid \mathrm{inj}_1)} Y + X.$$

[1] In [Sabadini, Vigna & Walters:1996], initial and terminal maps are erroneously missing from the elementary definition.

Now, if we have a family of sets (the *basic sets*), consider all the sets that can be obtained from them and $\varnothing$, $I$ by repeated application of sums and products (these are the *derived sets*). Given a family of *basic functions* between derived sets, any function obtained by repeated use of $(- \mid -)$, $(-,-)$ and composition on the basic functions, injections, projections, identities, initial maps, terminal maps and $\delta^{-1}$ is again a function between derived sets: all functions defined in this way are called *derived* functions.

Finally, for each triple of sets $X, U, Y$, and any derived function

$$f : X + U \to U + Y$$

such that for each $x \in X$ there is an $n_x$ such that[2] $f^{n_x}(x) \in Y$ (i.e., the function *terminates* for each input) we define the *function which $f$ computes by iteration*, denoted by **call**$[\,X, U, Y, f\,]$ (or **call**$[\,f\,]$, if $X, U$ and $Y$ are clear from the context), as

$$\textbf{call}[\,X, U, Y, f\,](x) = f^{n_x}(x).$$

Note that if such an $n_x$ exists, it is unique, because we cannot iterate $f$ over an element of $Y$.

**Definition 1** The class of *distributively computable functions* (over the given family of basic functions) is exactly the class of functions of the form **call**$[\,f\,]$, with $f$ a derived function.

### 2.1.1 Some notation

We would like to establish some notation which will be useful throughout this thesis. A $+/\times$-algebra is an algebra having two constants 0, 1 and two binary operators $+$ and $\times$. A term of the free $+/\times$-algebra on $A$ is called a *distributive expression* on $A$.

The sets $\varnothing$ and $I$ are the *structural sets*. A *simple set* is either a basic set or a structural set. When the operations of sum and product and the constants 0 and 1 are interpreted in the category of sets, a distributive expression on the basic sets gives rise to a derived set. We shall frequently confuse the former and the latter.

A *structural function* is one of the injections, projections, identities, initial maps, terminal maps and inverse distributivity isomorphisms. A *simple function* is either a basic function or a structural function. Thus, derived functions are obtained by applying $(- \mid -)$, $(-,-)$ and composition to simple functions.

Given a function $f : X + Y \to Z$, we shall denote with $f_{|X}$ the restriction of $f$ to $X$.

Of course, analogous definitions can be given in an arbitrary distributive category by replacing “set” with “object” and “function” with “arrow”. Algebraically speaking, structural entities corresponds to constants and operators of a distributive category, while basic entities are generators. Retrictions are obtained by composition with an injection.

### 2.1.2 The normal form theorem

The theorem we are going to introduce, appeared for the first time in [Walters:1992b][3], is the distributive counterpart of Kleene’s normal form theorem, which says that in the construction of a recursive function it is possible to use the minimization operator just one time.

**Theorem 1** The following properties are true for **call**$[\,-\,]$ (we assume that **call**$[\,-\,]$ is defined whenever we mention it):

---

[2]We shall omit the natural number tagging the element of a sum whenever no confusion is possible, thus writing just $x$ instead of $\langle x, k\rangle$.

[3]Part (iv) of the theorem appeared for the fi rst time in [Sabadini & Walters:1993].

(i). For each function $f : X \to Y$,

$$f = \mathbf{call}[\,(\mathrm{inj}_2 \circ (f \mid !_Y) : X + \varnothing \to \varnothing + Y\,];$$

(ii). If $f : X + U \to U + Y$ and $g : Y + V \to V + Z$, then

$$\mathbf{call}[\,f\,] \circ \mathbf{call}[\,g\,] = \mathbf{call}[\,X, U + Y + V, Z,\ f + g\,].$$

(iii). If $f : X + U \to U + Y$ and $g : X' + U' \to U' + Y'$, then

$$\mathbf{call}[\,f\,] + \mathbf{call}[\,g\,] = \mathbf{call}[\,a \circ (f + g) \circ b\,],$$

where $a$ is the commutativity isomorphism

$$X + X' + U + U' \xrightarrow{\cong} X + U + X' + U',$$

and $b$ is the commutativity isomorphism

$$U + Y + U' + Y' \xrightarrow{\cong} U + U' + Y + Y';$$

moreover,

$$\mathbf{call}[\,f\,] \times \mathbf{call}[\,g\,] = \\ \mathbf{call}[\,d \circ (f_{|X} \times g_{|X'} \mid f_{|U} \times g_{|U'} \mid f_{|U} \times \mathrm{inj}_2^{U',Y'} \mid \mathrm{inj}_2^{U,Y} \times g_{|U'})\,],$$

where $d$ is the isomorphism

$$(U + Y) \times (U' + Y') \xrightarrow{\cong} UU' + UY' + YU' + YY'$$

induced by distributivity.

(iv). If $f : X + V + U \to U + V + Y$, then

$$\mathbf{call}[\,X, V, Y, \mathbf{call}[\,X + V, U, V + Y,\ f\,]\,] = \mathbf{call}[\,X, U + V, Y, i \circ f\,] : X \to Y,$$

where $i$ is the commutativity isomorphism

$$U + V + Y \xrightarrow{\cong} V + U + Y.$$

Note that since $(f \mid g) = \nabla \circ (f + g)$ and $(f, g) = (f \times g) \circ \Delta$, also ($\mathbf{call}[\,f\,] \mid \mathbf{call}[\,g\,]$) and ($\mathbf{call}[\,f\,], \mathbf{call}[\,g\,]$) can be expressed with a single use of $\mathbf{call}[\,-\,]$. We remark that the properties of $\mathbf{call}[\,-\,]$ which do not involve products are the properties of the *feedback* operator of iteration theories [Bloom & Esik:1993].

The immediate consequence of the normal form theorem is that the functions computed by iteration, when added to the family of basic functions, do *not* modify its computational power. Formally,

**Theorem 2** The class of distributively computable functions is the class of functions obtained by applying inductively $(- \mid -)$, $(-, -)$, composition *and* $\mathbf{call}[\,-\,]$ to simple functions.

**Proof.** We just have to show that, given a function $f$ defined using repeatedly **call**$[\,-\,]$, it is also possible to define it by applying **call**$[\,-\,]$ to a suitable derived function. But the normal form theorem show that any derived function used in the definition of $f$ is of the form **call**$[\,g\,]$, and that each time an operator is applied, **call**$[\,-\,]$ can be "extracted" in such a way to arrive to a definition of the form **call**[ **call**$[\cdots$ **call**$[\,g\,]\cdots]\,]$, with $g$ a derived function; now, part (iv) of the normal form theorem reduces all the applications of **call**$[\,-\,]$ to just one. ▮

A completely analogous analysis (as briefly discussed in Section 2.3) can be done on the **pcall**$[\,-\,]$ operator, which allows to iterate *any* $f : X + U \to U + Y$, producing possibly a partial function. This fact in turn yields immediately that the class of total distributively computable partial functions coincides with the class of distributively computable functions.

In order to prove the theorems of this chapter, it is now necessary to introduce some technical lemmata which show how it is possible to avoid the usage of initial maps in the definition of noninitial derived functions.

**Definition 2** Let $X$ be a derived set. The reduced form of $X$, named $\rho(X)$, is obtained inductively from $X$ as follows:

(i). $\rho(X) = X$ for simple sets;

(ii). $\rho(X \times \varnothing) = \rho(\varnothing \times X) = \varnothing$;

(iii). $\rho(X + \varnothing) = \rho(\varnothing + X) = \rho(X)$.

Clearly, $X \cong \rho(X)$, and we can fix this isomorphism (which we will call $\nu_X$) on the basis of the inductive definition ($X \times \varnothing \xrightarrow{\text{proj}_2} \varnothing$, and so on). Note that $X \cong \varnothing$ iff $\rho(X) = \varnothing$.

**Theorem 3** Let $f : X \to Y$ be a derived function, with $X \ncong \varnothing$. Then $f = \nu_Y^{-1} \circ f' \circ \nu_X$, where $f' : \rho(X) \to \rho(Y)$ is a derived function which is defined without using initial maps.

**Proof.** The claim is obvious for simple functions. Let us examine inductively the following cases:

- $f = (g \mid h) : X + X' \to Y$. If (without loss of generality) $h'$ is an initial map, $f' = g' \circ \text{inj}_1^{X,\varnothing}$ and $f = \nu_Y^{-1} \circ f' \circ \nu_X$. If instead neither $g'$ nor $h'$ are initial maps, $f' = (g' \mid h')$ and $f = \nu_Y^{-1} \circ f' \circ \nu_{X+X'}$.
- $f = (g, h) : X \to Y \times Y'$. In this case, the claim is obvious with $f' = (g', h')$, because neither $g$ nor $h$ can be initial.
- $f = h \circ g$. Again, the claim is obvious with $f' = h' \circ g'$ (recall that $f$ is initial iff $g$ is initial, and that if $h$ is initial then $g$ is initial). ▮

**Corollary 1** The class of functions obtained applying **call**$[\,-\,]$ to functions (between reduced derived sets) defined without using initial maps is equal to the class of distributively computable functions without initial maps.

At this point we can restrict our attention to functions defined without using initial maps. Indeed, the iteration of a function $f$ for which $f'$ is initial produces necessarily an initial map (if $X + U \cong \varnothing$, then $X \cong U \cong \varnothing$). Since initial maps do not compute anything (they have empty domain), in the proof of equivalence theorems we can forget them[4].

[4]Note that these considerations are true also at the categorical level, i.e., in the extensive categorical setting of Chapter 4.

## 2.2 Recursive functions

We now recall Kleene's definition of recursive function [Kleene:1936].

**Definition 3** The set of recursive functions from $\mathbf{N}^k$ to $\mathbf{N}$ is the smallest set of functions containing the 0 constant function, the successor function $\mathrm{succ} : \mathbf{N} \to \mathbf{N}$, the projections $\pi_j : \mathbf{N}^k \to \mathbf{N}$ and closed under the following constructions:

(i). (*substitution*) given $f : \mathbf{N}^k \to \mathbf{N}$ and $g_i : \mathbf{N}^j \to \mathbf{N}$ $(1 \le i \le k)$, form $f(g_1, g_2, \ldots, g_k) : \mathbf{N}^j \to \mathbf{N}$;

(ii). (*primitive recursion*) given $g : \mathbf{N}^k \to \mathbf{N}$, $h : \mathbf{N}^{k+2} \to \mathbf{N}$, form $f : \mathbf{N}^{k+1} \to \mathbf{N}$ by setting

$$f(\underline{m}, 0) = g(\underline{m})$$

and

$$f(\underline{m}, n+1) = h(\underline{m}, n,\ f(\underline{m}, n));$$

(iii). (*minimization*) given a function $f : \mathbf{N}^{k+1} \to \mathbf{N}$ such that for all $\underline{m}$ there is an $n_{\underline{m}}$ such that $f(\underline{m}, n_{\underline{m}}) = 0$, form $g : \mathbf{N}^k \to \mathbf{N}$ by setting

$$g(\underline{m}) = \mu_n f(\underline{m}, n) = \min\{n \in \mathbf{N} \mid f(\underline{m}, n) = 0\}.$$

If we drop the condition of existence of a solution in $n$ for $f(\underline{m}, n) = 0$, we have the definition of *partial* recursive function.

**Definition 4** The set of partial recursive functions from $\mathbf{N}^k$ to $\mathbf{N}$ is the smallest set of partial functions containing the 0 constant function, the successor function $\mathrm{succ} : \mathbf{N} \to \mathbf{N}$, the projections $\pi_j : \mathbf{N}^k \to \mathbf{N}$, and closed under substitution, primitive recursion, and under the following construction (*partial minimization*): given $f : \mathbf{N}^{k+1} \to \mathbf{N}$, form $g : \mathbf{N}^k \to \mathbf{N}$ by setting

$$g(\underline{m}) = \mu_n f(\underline{m}, n) = \min\{n \in \mathbf{N} \mid f(\underline{m}, n) = 0 \text{ and } f(\underline{m}, i) \text{ is defined for all } i < n\};$$

$g(\underline{m})$ is undefined when the aforementioned set is empty.

It is a well-known but non trivial fact that the class of recursive functions coincide with the class of total partial recursive functions. We shall give in the sequel a very simple proof of this fact using Theorem 2.

Now we instantiate the definition of distributive computability using as basic functions the following generalizations of the precedessor and successor functions:

$$\begin{aligned} p : \mathbf{N} &\to I + \mathbf{N} \quad (= \{*\} + \mathbf{N}) \\ n &\mapsto n - 1 \qquad \text{if } n > 0 \\ 0 &\mapsto * \end{aligned}$$

$$\begin{aligned} s : I + \mathbf{N} &\to \mathbf{N} \\ * &\mapsto 0 \\ n &\mapsto n + 1. \end{aligned}$$

Note that with these definitions $s$ and $p$ are inverse, and that the predecessor function allows to check if a number is equal to zero: indeed, by composing $p$ with a function of the form $f + g$, the computation will continue with $f$ or $g$ depending of the input of $p$ being equal to zero. We will denote with $\mathfrak{D}_{\mathbf{N}}$ the set of function distributively computable on $s$ and $p$.

### 2.2.1 Encodings

Now we would like to prove that the set $\mathfrak{D}_{\boldsymbol{N}}$ and the set of recursive function do coincide. Strictly speaking, this is impossible, because the domains and codomains of the first class of functions are taken from a much larger set.

However, we will exploit the fact that any derived set built starting from $\boldsymbol{N}$ and $I$ can be mapped injectively in $\boldsymbol{N}$, and that $\boldsymbol{N}$ can be mapped surjectively on any such derived set using standard arithmetic operations based upon primitive recursion only (see, for instance, [Manin:1977]); once we code the domain and codomain of the functions in $\mathfrak{D}_{\boldsymbol{N}}$ using this maps, we shall obtain exactly the class of recursive functions.

**Definition 5** The (bijective) encodings $\sigma_+ : \boldsymbol{N} + \boldsymbol{N} \to \boldsymbol{N}$ and $\sigma_\times : \boldsymbol{N} \times \boldsymbol{N} \to \boldsymbol{N}$ are defined by $\sigma_+(\langle n, k\rangle) = 2n + k$ and by the standard pairing function (for this latter and its inverse, see [Manin:1977]). Let $\mathbf{0} : I \to \boldsymbol{N}$ the constant zero function, and $\mathfrak{i}_{\boldsymbol{N}} : \boldsymbol{N} \to I$ the terminal map associated to $\boldsymbol{N}$. For each derived set $X$ we define the encodings $\sigma_X : X \to \boldsymbol{N}$ and $\sigma'_X : \boldsymbol{N} \to X$:

- if $X = I$, then $\sigma_X = \mathbf{0}$ and $\sigma'_X = \mathfrak{i}_{\boldsymbol{N}}$;
- if $X = \boldsymbol{N}$, then $\sigma_{\boldsymbol{N}} = \sigma'_N = \mathbf{1}_{\boldsymbol{N}}$;
- if $X = Y + Z$, then $\sigma_X = \sigma_+ \circ (\sigma_Y + \sigma_Z)$ and $\sigma'_X = (\sigma'_Y + \sigma'_Z) \circ \sigma_+^{-1}$;
- if $X = Y \times Z$, then $\sigma_X = \sigma_\times \circ (\sigma_Y \times \sigma_Z)$ and $\sigma'_X = (\sigma'_Y \times \sigma'_Z) \circ \sigma_\times^{-1}$.

Note that the encodings $\sigma$ and $\sigma'$ are related by the following property:

**Proposition 1** $\sigma'_X \circ \sigma_X = \mathbf{1}_X$.

**Proof.** A simple structural induction shows that

$$\sigma'_X \circ \sigma_X = (\sigma'_Y + \sigma'_Z) \circ \sigma_+^{-1} \circ \sigma_+ \circ (\sigma_Y + \sigma_Z) = \\ (\sigma'_Y + \sigma'_Z) \circ (\sigma_Y + \sigma_Z) = \sigma'_Y \circ \sigma_Y + \sigma'_Z \circ \sigma_Z = \mathbf{1}_X.$$

Analogously for the product. ▮

In what follows, given a function $f : X \to Y$ we will denote with $\widehat{f}$ the function $\sigma_Y \circ f \circ \sigma'_X$. The previous proposition shows that

**Corollary 2** $\widehat{g} \circ \widehat{f} = \widehat{g \circ f}$.

### 2.2.2 Equivalence theorems

Using the encodings we just defined, we can finally state the equivalence theorem:

**Theorem 4** Let $\mathcal{F} = \{f : \boldsymbol{N} \to \boldsymbol{N} \mid f \text{ is recursive}\}$ the set of recursive functions, and let $\mathcal{G} = \{\widehat{f} \mid f : X \to Y \in \mathfrak{D}_{\boldsymbol{N}}\}$ be the set of distributively computable functions whose domains and codomains have been coded in $\boldsymbol{N}$. Then $\mathcal{F} = \mathcal{G}$.

Note that we considered only recursive functions having $\boldsymbol{N}$ as domain: it is immediate to see that this restriction can be made without loss of generality.

**Proof.** We begin by showing that $\mathcal{F} \subseteq \mathcal{G}$, i.e., the recursive functions $\boldsymbol{N} \to \boldsymbol{N}$ are a subset of the distributively computable ones.

First of all note that the basic functions mentioned in the definition of recursive function are distributively computable (succ $= s \circ \mathrm{inj}_2$, and $0 = s \circ \mathrm{inj}_1$). Substitution can be obtained using $(-,-)$ and composition, so that the only things left to prove are primitive recursion and minimization.

**Primitive recursion.** Given functions $g : \boldsymbol{N}^k \to \boldsymbol{N}$ and $h : \boldsymbol{N}^{k+2} \to \boldsymbol{N}$ which define by primitive recursion a function $f : \boldsymbol{N}^{k+1} \to \boldsymbol{N}$, we want to build a function

$$\alpha : \boldsymbol{N}^k \times \boldsymbol{N} + (\boldsymbol{N}^k \times \boldsymbol{N} \times \boldsymbol{N} \times (I + \boldsymbol{N})) \to (\boldsymbol{N}^k \times \boldsymbol{N} \times \boldsymbol{N} \times (I + \boldsymbol{N})) + \boldsymbol{N}$$

such that $\mathbf{call}[\,\alpha\,] = f$.

Intuitively, the "middle part" of the state space is given by

$$\text{input} \times \text{last output} \times \text{recursion index} \times \text{counter}$$

(the recursion index is $n$ in Definition 3; the counter is necessary because it has to run backwards with respect to $n$).

We shall describe $\alpha$ by breaking it into the two component which acts on the summands of its domain; we shall understand the obvious injections in the codomain. The first part moves the data out of the initial state, decrements the counter, initialize the recursion index to 0 and applies at the same time the "initial data" function $g$:

$$\boldsymbol{N}^k \times \boldsymbol{N} \cong \boldsymbol{N}^k \times I \times \boldsymbol{N} \xrightarrow{(\mathbf{1},g)\times 0 \times p} \boldsymbol{N}^k \times \boldsymbol{N} \times \boldsymbol{N} \times (I + \boldsymbol{N}).$$

Note that

$$\boldsymbol{N}^k \times \boldsymbol{N} \times \boldsymbol{N} \times (I + \boldsymbol{N}) \cong (\boldsymbol{N}^k \times \boldsymbol{N} \times \boldsymbol{N}) + (\boldsymbol{N}^k \times \boldsymbol{N} \times \boldsymbol{N} \times \boldsymbol{N}),$$

so we can define the second part of $\alpha$ separately on the two summands. Clearly,

$$\boldsymbol{N}^k \times \boldsymbol{N} \times \boldsymbol{N} \xrightarrow{\mathrm{proj}_2} \boldsymbol{N}$$

projects the result on the final state. Indeed, when the counter arrives in the first part of $(I + \boldsymbol{N})$ we iterated the loop exactly the number of times required.

The following function defines the action on the second summand:

$$\begin{array}{rcl}
\boldsymbol{N}^k \times \boldsymbol{N} \times \boldsymbol{N} \times \boldsymbol{N} & \xrightarrow{\Delta \times \mathbf{1} \times \Delta \times \mathbf{1}} & \boldsymbol{N}^k \times \boldsymbol{N}^k \times \boldsymbol{N} \times \boldsymbol{N} \times \boldsymbol{N} \times \boldsymbol{N} \\
& \xrightarrow{\mathbf{1} \times \mathbf{1} \times \mathbf{twist} \times \mathbf{1} \times \mathbf{1}} & \boldsymbol{N}^k \times \boldsymbol{N}^k \times \boldsymbol{N} \times \boldsymbol{N} \times \boldsymbol{N} \times \boldsymbol{N} \\
& \xrightarrow{\mathbf{1} \times h \times \mathbf{1} \times \mathbf{1}} & \boldsymbol{N}^k \times \boldsymbol{N} \times \boldsymbol{N} \times \boldsymbol{N} \\
& \xrightarrow{\mathbf{1} \times \mathbf{1} \times (s \circ \mathrm{inj}_2) \times p} & \boldsymbol{N}^k \times \boldsymbol{N} \times \boldsymbol{N} \times (I + \boldsymbol{N}).
\end{array}$$

In order to show that $\mathbf{call}[\,\alpha\,] = f$, we shall compute explicitly the orbit induced by $\alpha$. For an input $(\underline{m}, n)$ we have

$$(\underline{m}, n) \mapsto (\underline{m}, g(\underline{m}), 0, n-1).$$

If $n = 0$, the next iteration will be

$$(\underline{m}, g(\underline{m}), 0) \mapsto g(\underline{m}) = f(\underline{m}, 0),$$

which computes correctly $f$. If $n > 0$,

$$\begin{array}{rcl}
(\underline{m}, g(\underline{m}), 0, n-1) & \mapsto & (\underline{m}, \underline{m}, g(\underline{m}), 0, 0, n-1) \\
& \mapsto & (\underline{m}, \underline{m}, 0, g(\underline{m}), 0, n-1) \\
& \mapsto & (\underline{m}, h(\underline{m}, 0, g(\underline{m})), 0, n-1) = (\underline{m}, f(\underline{m}, 1), 0, n-1) \\
& \mapsto & (\underline{m}, f(\underline{m}, 1), 1, n-2)
\end{array}$$

and, by induction, when $\alpha$ computes $f(\underline{m}, k)$ the counter is decremented to $n - k - 1$, thus moving the state to the final summand.

**Minimization.** The function

$$\beta : \boldsymbol{N}^k + (\boldsymbol{N}^k \times \boldsymbol{N}) \to (\boldsymbol{N}^k \times \boldsymbol{N}) + \boldsymbol{N}$$

which we shall use in order to emulate the minimization operator is defined on $\boldsymbol{N}^k$ by $\boldsymbol{N}^k \cong \boldsymbol{N}^k \times I \xrightarrow{\mathbf{1}\times 0} \boldsymbol{N}^k \times \boldsymbol{N}$, while on $\boldsymbol{N}^k \times \boldsymbol{N}$ it acts as follows:

$$\begin{array}{lll} \boldsymbol{N}^k \times \boldsymbol{N} & \xrightarrow{\qquad\Delta\qquad} & \boldsymbol{N}^k \times \boldsymbol{N} \times \boldsymbol{N}^k \times \boldsymbol{N} \\ & \xrightarrow{\quad \mathbf{1}\times\mathbf{1}\times f \quad} & \boldsymbol{N}^k \times \boldsymbol{N} \times \boldsymbol{N} \\ & \xrightarrow{\quad \mathbf{1}\times\mathbf{1}\times p \quad} & \boldsymbol{N}^k \times \boldsymbol{N} \times (I + \boldsymbol{N}) \\ & \xrightarrow{\qquad \delta^{-1} \qquad} & (\boldsymbol{N}^k \times \boldsymbol{N}) + (\boldsymbol{N}^k \times \boldsymbol{N} \times \boldsymbol{N}) \\ & \xrightarrow{\mathrm{proj}_2 + (\mathrm{proj}_1, \mathrm{proj}_2)} & \boldsymbol{N} + (\boldsymbol{N}^k \times \boldsymbol{N}) \\ & \xrightarrow{\qquad \mathbf{twist} \qquad} & (\boldsymbol{N}^k \times \boldsymbol{N}) + \boldsymbol{N} \\ & \xrightarrow{(\mathbf{1}\times(s\circ \mathrm{inj}_2))+\mathbf{1}} & (\boldsymbol{N}^k \times \boldsymbol{N}) + \boldsymbol{N}. \end{array}$$

Thus, if the result of $f$ is 0, the function lands in $\boldsymbol{N}$, returning the number of iterations; otherwise, it remains in $\boldsymbol{N}^k \times \boldsymbol{N}$, incrementing the iteration counter (the second factor of the product).

In more detail, for a given input $\underline{m}$ the first application of $\beta$ gives

$$\underline{m} \mapsto (\underline{m}, 0),$$

and then we have

$$\begin{array}{lll} (\underline{m}, 0) & \mapsto & (\underline{m}, 0, \underline{m}, 0) \\ & \mapsto & (\underline{m}, 0, f(\underline{m}, 0)) \\ & \mapsto & (\underline{m}, 0, f(\underline{m}, 0) - 1). \end{array}$$

Now, if $f(\underline{m}, 0) = 0$, we have simply

$$(\underline{m}, 0, *) \mapsto 0,$$

otherwise

$$(\underline{m}, 0, f(\underline{m}, 0) - 1) \mapsto (\underline{m}, 0) \mapsto (\underline{m}, 1),$$

so that $\beta$ keeps incrementing the loop counter until $f$ is evaluated to 0. Clearly, $\mathbf{call}[\,\beta\,] = \mu_n f(\underline{m}, n)$.

Let now $f : X \to Y$ be a distributively computable function. This function has been obtained by iterating a suitable derived function $g : X + U \to U + Y$. We shall denote with $h$ the loop $X + U + Y \to X + U + Y$ induced by $g$ (and defined formally by $h : X + U + Y \xrightarrow{(g \mid \mathrm{inj}_2)} U + Y \xrightarrow{\mathrm{inj}_2} X + (U + Y)$). Clearly, $f$ is the function obtained by applying $h$ to an element of $X$ until the generated orbit intersects $Y$ (all elements of $Y$ are fixed points of $h$).

Since by Corollary 2 we have $\widehat{h^k} = \widehat{h}^k$, if we shall be able to prove that for any loop $h : X + U + Y \to X + U + Y$ the function $\widehat{h}$ is recursive, we shall be able to use (as it happens typically when proving the computability of recursive functions using an imperative language) primitive recursion in order to compute $\widehat{h}^k$, and minimization in order to find out the number of iterations which are necessary to arrive in $Y$. At this point, primitive recursion will again enable us to actually iterate $\widehat{h}$ as much as it is necessary.

Thus, given a loop $h : (X + U) + Y \to (X + U) + Y$ (the grouping is chosen without loss of generality), consider the following recursion scheme:

$$\begin{cases} a(n, 0) = n - (n \bmod 4) \\ a(n, k + 1) = (\widehat{h} \circ a)(n, k). \end{cases}$$

Clearly, if $n$ is a multiple of four we have that $a(n, k) = \widehat{h}^k(n)$. Since the encoding $\sigma_{(X+U)+Y}$ sends the elements of $Y$ into odd numbers, $\mu_k(1 - (a(n, k) \bmod 2))$ is the least number of iterations after which $\widehat{h}$ "lands in $Y$" on input $n$, and

$$b(n) = a(n, \mu_k(1 - (a(n, k) \bmod 2)))$$

iterates $\widehat{h}$ exactly that number of times. Note that at the first iteration we ensure that $n$ is a multiple of four, and thus that it represents an element of $X$, making the application of $\mu$ possible (we do not have any guarantee of termination on $U$).

At this point we just have to precompose $b$ with multiplication by four (which corresponds to the inclusion of $X$ in $(X + U) + Y$) and postcompose it with integer division by two (which corresponds to the extraction of $Y$ from $(X + U) + Y$). The resulting recursive function is exactly $\widehat{f}$.

In order to complete the proof, we shall show by structural induction that all functions $\widehat{f}$, where $f$ is a derived function, are recursive. The following claims are immediately verified:

- the function $\widehat{s}$ sends even number in zero and odd numbers in $(n + 1)/2$;
- the function $\widehat{p}$ sends zero in zero and the other numbers in $2n - 1$;
- $\widehat{\mathrm{proj}_i} = \mathrm{proj}_i \circ \sigma_\times^{-1}$ $(i = 1, 2)$;
- $\widehat{\mathrm{inj}_i}(n) = 2n + i - 1$ $(i = 1, 2)$;
- $\widehat{\mathbf{1}_X} = \mathbf{1}_{\boldsymbol{N}}$;
- $\widehat{\mathrm{i}_X} = 0$;
- $\widehat{\delta^{-1}}(n) = 2\sigma_\times((\mathrm{proj}_1 \circ \sigma_\times^{-1})(n), \lfloor(\mathrm{proj}_2 \circ \sigma_\times^{-1})(n)/2\rfloor) + ((\mathrm{proj}_2 \circ \sigma_\times^{-1})(n) \bmod 2)$.

If $f = h \circ g$, then by Corollary 2 we have $\widehat{f} = \widehat{h} \circ \widehat{g}$. If $f = (g, h)$, $\widehat{f}$ is obtained by substituting $\widehat{g}$ and $\widehat{h}$ into the pairing function $\sigma_\times$. If $f = (g \mid h)$, $\widehat{f}(n) = f'(n, n \bmod 2)$, where $f'$ is obtained by primitive recursion as follows:

$$\begin{cases} f'(n, 0) = \widehat{g}(\lfloor n/2 \rfloor) \\ f'(n, k + 1) = \widehat{h}(\lfloor n/2 \rfloor). \end{cases} \blacksquare$$

What we just proved, and the considerations made at the beginning of this chapter, allows us to say that

**Lemma 1** The encodings $\sigma_X$, $\sigma'_X$ are distributively computable for each $X$.

The operation of encoding derived sets in $\boldsymbol{N}$ can be reversed: instead of restricting the distributively computable functions to the class $\mathcal{G}$ (the previous lemma implies $\mathcal{G} \subseteq \mathfrak{D}_{\boldsymbol{N}}$), we can enlarge $\mathcal{F}$ by composing its functions with the encodings. It is easy to show that

**Theorem 5** Let $\widehat{\mathcal{F}} = \{\sigma'_Y \circ f \circ \sigma_X \mid f : R^{n_X} \to R^{n_Y} \in \mathcal{F}\}$. Then $\widehat{\mathcal{F}} = \mathfrak{D}_{\boldsymbol{N}}$.

Note that this is the claim proved originally in [Sabadini, Vigna & Walters:1996].

**Proof.** Any function in $\widehat{\mathcal{F}}$ is also trivially in $\mathfrak{D}_{\boldsymbol{N}}$ (use lemma 1, the fact $\mathcal{F} = \mathcal{G} \subseteq \mathfrak{D}_{\boldsymbol{R}}$, and the closure of $\mathfrak{D}_{\boldsymbol{N}}$ with respect to composition). On the other hand, if $f : X \to Y \in \mathfrak{D}_{\boldsymbol{N}}$ then $\widehat{f} \in \mathcal{G} = \mathcal{F}$. But then

$$\widehat{\mathcal{F}} \ni \sigma'_Y \circ \widehat{f} \circ \sigma_X = \sigma'_Y \circ \sigma_Y \circ f \circ \sigma'_X \circ \sigma_X = f. \blacksquare$$

## 2.3 Partial functions

We shall now briefly outline the (single) modification necessary to handle partiality, and state the equivalence theorem between recursive and total partial recursive functions, which admits a very simple proof in this context.

**Definition 6** For each function $f : X + U \to U + Y$ we define the *partial* function

$$\textbf{pcall}[\, f\,] : X \to Y = \begin{cases} f^{n_x}(x) & \text{if there is an integer } n_x \text{ such that } f^{n_x}(x) \in Y \\ \perp \text{ (undefined)} & \text{otherwise.} \end{cases}$$

Substituting **pcall**[ – ] for **call**[ – ] in the proof of Theorem 4 (the operators on sets and functions should be redefined accordingly, in a standard manner) yields immediately a result analogous to Theorem 4 for partial functions. This fact, coupled with the normal form theorem, yields a simple proof of the well-known but non trivial

**Theorem 6** The class of total partial recursive functions coincides with the class of recursive functions.

**Proof.** Obvious, because the use of **call**[ – ] or **pcall**[ – ] in the definition of a recursive function can be reduced by Theorem 2 to a single, outermost application. ∎

## 2.4 Computability in an ordered ring

In [Blum, Shub & Smale:1989], Blum, Shub and Smale have defined a model of machine working on an arbitrary commutative ordered ring (or field) $R$. Their definition has given rise to a whole new theory of computability and computational complexity. In the *BSS model*, a machine has a finite control, given by a finite graph, and an unlimited number of registers, each capable of holding an element of $R$. The computational steps consist in computing polynomials, and possibly deciding the next step by comparing the result of an evaluation with 0. A pair of integer registers can be used as pointers in order to retrieve and set any register.

In this section we want to prove an equivalence result between the functions $R^l \to R^m$ (with $l, m$ finite) computable in the BSS model, and the functions distributively computable in the category of sets when the basic functions are the constants, sum, multiplication (possibly also inversion) and test for nonnegativity (note that we do not force a finiteness condition on the state space of BSS machines; this is unnecessary, as explained below). We remark that an extension of this result to the case of BSS machines with infinite input/output state space is possible by describing all spaces involved in the computation via suitable unbounded data types like stacks.

### 2.4.1 The finite dimensional BSS model

In [Blum, Shub & Smale:1989] two kinds of machine are defined: *finite* dimensional and *infinite* dimensional ones, the difference lying in the presence, in the second case, of an infinite number of registers in the input, output and state space, and of some machinery which is necessary in order to address such registers. We shall introduce the finite dimensional model, for a reason which will become clear below.

Recall that an *ordered ring* is a ring $R$ with a specified subset $P \subseteq R \setminus \{0\}$ such that:

1. if $\alpha, \beta \in P$ then $\alpha + \beta, \alpha\beta \in P$;

2. for all $\alpha \in R \setminus \{0\}$, either $\alpha \in P$ or $-\alpha \in P$, but not both.

In an ordered ring, the notation $\alpha > \beta$ stands for $\alpha - \beta \in P$.

In the rest of this section, the word "ring" will always mean "ordered commutative ring". We shall frequently use the term "polynomial function" for functions $R^i \to R^j$: this means that each of the $j$ components of the function is (induced by) a polynomial in $i$ variables. It will be always understood that if $R$ is a field such functions can also be rational functions. The elements of $R$ will be denoted by $\alpha, \beta, \ldots$, while the vectors of $R^n$ will be denoted by $\boldsymbol{\alpha} = \langle \alpha_1, \alpha_2, \ldots, \alpha_n \rangle, \boldsymbol{\beta}, \ldots$.

**Definition 7** A *finite dimensional machine $M$ over $R$* consists of three spaces: the input space $\bar{I} = R^l$, the output space $\bar{O} = R^m$ and the state space $\bar{S} = R^n$, together with a finite directed connected graph with node set $\bar{N} = \{1, 2, \ldots, N\}$ ($N > 1$) divided in four subsets: *input, computation, branch* and *output* nodes.

Node 1 is the only *input node*, having fan-in 0 and fan-out[5] 1; node $N$ is the only *output node*, having fan-out 0. They have associated linear functions (named $I(-)$ and $O(-)$), mapping respectively the input space to the state space and the state space to the output space. Any other node $k \in \{2, 3, \ldots, N-1\}$ can be of the following types:

(i). a *branching node*; in this case, $k$ has fan-out 2 and its two (distinguished) successors are $\beta^-(k)$ and $\beta^+(k)$; there is a polynomial function $h_k : \bar{S} \to R$ associated to $k$, and for a given state $\boldsymbol{\alpha} \in \bar{S}$, branching on $-$ or $+$ will depend upon whether or not $h_k(\boldsymbol{\alpha}) < 0$;

(ii). a *computation node*; in this case, $k$ has fan-out 1 and there is a polynomial function $g_k : \bar{S} \to \bar{S}$ associated with it.

We can view $M$ as a discrete dynamical system over the *full state space* $\bar{N} \times \bar{S}$. $M$ induces a *computing endomorphism* on the full state space:

$$\begin{array}{rcll}
\langle 1, \boldsymbol{\alpha} \rangle & \mapsto & \langle \beta(1), \boldsymbol{\alpha} \rangle & \\
\langle N, \boldsymbol{\alpha} \rangle & \mapsto & \langle N, \boldsymbol{\alpha} \rangle & \\
\langle k, \boldsymbol{\alpha} \rangle & \mapsto & \langle \beta(k), g_k(\boldsymbol{\alpha}) \rangle & \text{if } k \text{ is a computation node} \\
\langle k, \boldsymbol{\alpha} \rangle & \mapsto & \begin{cases} \langle \beta^-(k), \boldsymbol{\alpha} \rangle & \text{if } h_k(\boldsymbol{\alpha}) < 0 \\ \langle \beta^+(k), \boldsymbol{\alpha} \rangle & \text{if } h_k(\boldsymbol{\alpha}) \geq 0 \end{cases} & \text{if } k \text{ is a branching node.}
\end{array}$$

The *computation* of $M$ under input $\boldsymbol{\alpha}$ is the orbit generated in the full state space by the computing endomorphism starting from $\langle 1, I(\boldsymbol{\alpha}) \rangle$. If the orbit reaches a fixed point of the form $\langle N, \boldsymbol{\beta} \rangle$ for some $\boldsymbol{\beta} \in \bar{S}$ we say that the machine *halted*, and that its output is $O(\boldsymbol{\beta})$. The association $\boldsymbol{\alpha} \mapsto O(\boldsymbol{\beta})$ defines a partial function $\varphi_M$, which is called the *partial function computed by the machine $M$*. In what follows, we shall consider only machines computing *total* functions; the obvious extension to the partial case can be made by substituting the **pcall**[ − ] operator to the **call**[ − ] operator.

[5] If $k$ is a node with fan-out 1, then $\beta(k)$ denotes the "next" node in the graph after $k$.

When defining the infinite dimensional model, the state space—and possibly the input and output spaces—becomes $R^\infty$ (i.e., the space of infinite sequences of elements of $R$ in which only a finite number of components is nonzero) and nodes of *fifth type* are added, which allow to access any register of the state space. However, as suggested in [Blum, Shub & Smale:1989] (and proved in a different, particularly enlightening way in [Michaux:1989]), if the input and output space of an infinite dimensional machine $M$ are finite, then the function $\varphi_M$ is computed also by a finite dimensional machine. Thus, since in this paper we are discussing the class of computable functions $R^l \to R^m$ with $l, m$ finite, we can restrict our attention to finite dimensional machines without loss of generality.

## 2.4.2 The basic functions

For a given ring $R$, we shall now list the basic functions which will be used to generate the distributively computable functions (categorically speaking, we are defining the base distributive graph). The list includes the following functions:

$$\begin{array}{rcl} R \times R & \stackrel{*}{\to} & R \\ R \times R & \stackrel{+}{\to} & R \\ I & \xrightarrow{\ulcorner \alpha \urcorner} & R \text{ (for all } \alpha \in R) \\ R & \stackrel{\geq}{\to} & I + I. \end{array}$$

If $R$ is a field, we include also a function

$$R \xrightarrow{(-)^{-1}} R,$$

the intended semantics being: product and sum of two elements of the ring, constants, test for nonnegativity and inversion. Analytically,

$$\begin{array}{rcl} \langle \alpha, \beta \rangle & \stackrel{*}{\mapsto} & \alpha\beta \\ \langle \alpha, \beta \rangle & \stackrel{+}{\mapsto} & \alpha + \beta \\ * & \stackrel{\ulcorner \alpha \urcorner}{\mapsto} & \alpha \\ \alpha & \stackrel{\geq}{\mapsto} & \begin{cases} \langle *, 0 \rangle & \text{if } \alpha < 0 \\ \langle *, 1 \rangle & \text{if } \alpha \geq 0 \end{cases} \\ \alpha & \stackrel{(-)^{-1}}{\mapsto} & \alpha^{-1} \quad (\alpha \neq 0) \end{array}$$

Intuitively, we are claiming that we can multiply and sum (and possibly invert[6]) the elements of $R$, that we can generate any constant, and that we can compute the characteristic function of the positives[7]. We shall denote with $\mathfrak{D}_R$ the class of functions distributively computable starting from the functions above.

[6]In the BSS model it is assumed that a rational function will never be evaluated when the denominator is zero. Thus, we do not care too much about the defi nition of $0^{-1}$. It can be taken to be 0, for instance.

[7]We mention, without developing the theory, that the function $\geq$ can be replaced by a test for equality with zero, which allows to extend the equivalence theorem to unordered rings such as $\boldsymbol{C}$.

### 2.4.3 Encodings

As in the case of recursive functions, we must now establish an encoding for the derived sets built from $R$ and $I$. The solution previously proposed is no longer applicable, because it relied on the existence of a computable isomorphism $\boldsymbol{N} \to \boldsymbol{N} \times \boldsymbol{N}$, which is not in general available for an arbitrary ring $R$. Here we shall be contented of encoding sums into products.

**Definition 8** The encodings $\sigma(i, j) : R^i + R^j \to R \times R^i \times R^j$ and $\sigma'(i, j) : R \times R^i \times R^j \to R^i + R^j$ are defined by

$$\langle \boldsymbol{\alpha}, k\rangle \overset{\sigma(i,j)}{\mapsto} \begin{cases} \langle -1, \boldsymbol{\alpha}, \mathbf{0}\rangle & \text{if } k = 0 \\ \langle 0, \mathbf{0}, \boldsymbol{\alpha}\rangle & \text{if } k = 1 \end{cases} \qquad \langle \alpha, \boldsymbol{\beta}, \boldsymbol{\gamma}\rangle \overset{\sigma'(i,j)}{\mapsto} \begin{cases} \langle \boldsymbol{\beta}, 0\rangle & \text{if } \alpha < 0 \\ \langle \boldsymbol{\gamma}, 1\rangle & \text{otherwise.} \end{cases}$$

For any derived set $X$ we define inductively $\sigma_X : X \to R^{n_X}$ and $\sigma'_X : R^{n_X} \to X$ as follows:

- if $X = I$, then $n_I = 1$, $\sigma_I = 0$ and $\sigma'_I = *$;
- if $X = R$, then $n_R = 1$ and $\sigma_R = \sigma'_R = \mathbf{1}_R$;
- if $X = Y + Z$, then $n_X = n_Y + n_Z + 1$, $\sigma_X = \sigma(n_Y, n_Z) \circ (\sigma_Y + \sigma_Z)$ and $\sigma'_X = (\sigma'_Y + \sigma'_Z) \circ \sigma'(n_Y, n_Z)$;
- if $X = Y \times Z$ then $n_X = n_Y + n_Z$, $\sigma_X = \sigma_Y \times \sigma_Z$ and $\sigma'_X = \sigma'_Y \times \sigma'_Z$.

The encodings $\sigma_X$ and $\sigma'_X$ are related by the following fundamental property:

**Proposition 2** $\sigma'_X \circ \sigma_X = \mathbf{1}_X$.

**Proof.** We prove the claim by structural induction. The base case is obvious. If $X = Y + Z$, then we have that

$$\begin{aligned}\sigma'_X \circ \sigma_X &= (\sigma'_Y + \sigma'_Z) \circ \sigma'(n_Y, n_Z) \circ \sigma(n_Y, n_Z) \circ (\sigma_Y + \sigma_Z) \\ &= (\sigma'_Y + \sigma'_Z) \circ (\sigma_Y + \sigma_Z) = \sigma'_Y \circ \sigma_Y + \sigma'_Z \circ \sigma_Z = \mathbf{1}_X.\end{aligned}$$

Finally, if $X = Y \times Z$ then

$$\sigma'_X \circ \sigma_X = (\sigma'_Y \times \sigma'_Z) \circ (\sigma_Y \times \sigma_Z) = \sigma'_Y \circ \sigma_Y \times \sigma'_Z \circ \sigma_Z = \mathbf{1}_X. \blacksquare$$

In what follows, given a function $f : X \to Y$ we shall denote with $\widehat{f}$ the function $\sigma_Y \circ f \circ \sigma'_X$. Using the previous proposition, it is easy to prove the following

**Corollary 3** Let $f : X \to Y$, $g : X' \to Y'$. Then

- if $Y = X'$ then $\widehat{g} \circ \widehat{f} = \widehat{g \circ f}$;
- $(\widehat{f}, \widehat{g}) = \widehat{(f, g)}$;
- $(\widehat{f} \mid \widehat{g}) = \widehat{(f \mid g)} \circ \sigma(n_X, n_{X'})$;
- $(\widehat{f} \mid \widehat{g}) \circ \sigma'(n_X, n_{X'}) = \widehat{(f \mid g)}$.

### 2.4.4 Equivalence theorems

We firstly need a couple of technical lemmata:

**Lemma 2** The derived functions of the form $f : R^l \to R^m$ contain the polynomial functions. If $R$ is a field, they contain the rational functions.

**Proof.** We just have to show that the thesis holds for the functions $R^l \to R$ associated to polynomials; the general case can be obtained by applying the $(-,-)$ operator.

We work by structural induction on a polynomial $p$: if $p$ is a variable or a constant $\alpha$, the corresponding function is an identity or the function $I \xrightarrow{\ulcorner\alpha\urcorner} R$, respectively, precomposed with a suitable projection or terminal map. If $p = p'p''$, then by applying the operator $(-,-)$ to the functions associated to $p'$ and $p''$ and composing with $* : R^2 \to R$ we obtain the function associated to $p$. An analogous consideration can be made if $p = p' + p''$. If $R$ is a field any rational function can be easily obtained by inverting the denominator using $(-)^{-1}$. ∎

**Lemma 3** The derived functions contain the encodings.

**Proof.** Trivial, by structural induction. The base case is covered by identities, constant maps and projections. The maps $\sigma(i, j)$ and $\sigma'(i, j)$ can be easily shown to be derived functions, and the operations used in the inductive construction of $\sigma_X$ and $\sigma'_X$ are those allowed when building derived functions. ∎

We can now state and prove our main theorem. Since the domains and codomains of the functions computed in the BSS model are restricted to the powers of $R$, we have (as for recursive functions) to encode the derived sets appearing in the domains and codomains of functions in $\mathfrak{D}_R$ using the techniques developed in Section 2.4.3.

**Theorem 7** Let $\mathcal{F} = \{\varphi_M : R^l \to R^m \mid M \text{ is a BSS machine }\}$ be the set of functions computed by machines over $R$ in the BSS model, and $\mathcal{G} = \{\widehat{f} \mid f : P \to Q \in \mathfrak{D}_R\}$ be the set of distributively computable functions whose domains and codomains have been encoded into powers of $R$. Then $\mathcal{F} = \mathcal{G}$.

**Proof.** Let us prove that $\mathcal{F} \subseteq \mathcal{G}$, by building a derived function $f : R^l + N \cdot R^n \to N \cdot R^n + R^m$ which[8] emulates the behaviour of a given machine $M$ with node set $\bar{N}$. By Lemma 2 we know that we can freely use polynomial functions while building such a function.

[8]Note that by distributivity $\bar{N} \times R^n \cong N \cdot R^n = R^n + R^n + \cdots + R^n$ ($N$ times); we correspondingly use $N$-ary injections: they can be easily defined in terms of binary ones.

We define $f$ “piecewise” as follows (the sequence of arrows denotes functional composition):

$$
\begin{aligned}
f_0 &: R^l \xrightarrow{\mathrm{inj}_1 \circ I(-)} N \cdot R^n \\
f_1 &: R^n \xrightarrow{\mathrm{inj}_{\beta(1)}} N \cdot R^n \\
f_k &: R^n \xrightarrow{\mathrm{inj}_{\beta(k)} \circ g_k} N \cdot R^n \qquad \text{if } k \text{ is a computation node} \\
f_k &: R^n \xrightarrow{\Delta_{R^n}} R^n \times R^n \\
&\xrightarrow{h_k \times \mathbf{1}_{R^n}} R \times R^n \\
&\xrightarrow{\geq \times \mathbf{1}_{R^n}} (I + I) \times R^n \\
&\xrightarrow{\delta^{-1}} I \times R^n + I \times R^n \\
&\xrightarrow{\cong} R^n + R^n \\
&\xrightarrow{(\mathrm{inj}_{\beta^-(k)} \mid \mathrm{inj}_{\beta^+(k)})} N \cdot R^n \qquad \text{if } k \text{ is a branching node} \\
f_N &: R^n \xrightarrow{O(-)} R^m
\end{aligned}
$$

and then $f = (f_0 \mid f_1 \mid \cdots \mid f_N)$, understanding composition with the obvious injections into $N \cdot R^n + R^m$. A routine check shows that $f$ acts on $N \cdot R$ exactly as the computing endomorphism of $M$. Thus, since $\sigma_{R^l} = \mathbf{1}_{R^l}$ and $\sigma_{R^m} = \mathbf{1}_{R^m}$, we have **call**[ $f$ ]=$\varphi_M$, so $\mathcal{F} \subseteq \mathcal{G}$.

Let now $f : X \to Y$ be a function of $\mathfrak{D}_R$. As in the proof of Theorem 4, such a function has been obtained by iterating a suitable derived function $g : X + U \to U + Y$ (i.e., $f =$ **call**[ $g$ ]). We shall denote by $h$ the loop $X + U + Y \to X + U + Y$ induced by $g$. It is clear that $f$ is exactly the function obtained applying $h$ to an element of $X$ until the generated orbit intersects $Y$.

By Corollary 3, we have that $\widehat{h^k} = \widehat{h}^k$. Thus, if we can show that for any loop $h : X + U + Y \to X + U + Y$ there is a machine computing $\widehat{h}$, we can build a new machine $M$ with input space $R^{n_X}$, output space $R^{n_Y}$ and state space $R^{n_{X+U+Y}}$. The input map of $M$ immerges the element of $X$ encoded in $R^{n_X}$ into the encoding of the same element in $R^{n_{X+U+Y}}$; the output map extracts the encoding of an element of $Y$ from its encoding in $R^{n_{X+U+Y}}$ (such maps are obviously linear). The machine $M$ applies $\widehat{h}$ to its state space, and then checks for the element represented in $R^{n_{X+U+Y}}$ being in $Y$ (recall that such a check can be made simply by checking the nonnegativity of certain components of the vector encoding the element). If it is not, $M$ applies again $\widehat{h}$; otherwise, it moves to the output node. Clearly, $\varphi_M = \widehat{f}$.

In order to make complete the argument, we shall now show by structural induction that, for all derived functions $f : X \to Y$, $\widehat{f}$ can be computed by a suitable finite dimensional machine. We shall describe the machines informally; the details would just be cumbersome. Unless otherwise specified, the input will be represented by a vector $\boldsymbol{\alpha}$ (or a scalar $\alpha$).

It is not difficult to check that the base case is covered by the following machines (we shall not mention encodings when they are just the identity):

- the machines computing $* : R^2 \to R$ and $+ : R^2 \to R$ just output $\alpha_1\alpha_2$ and $\alpha_1 + \alpha_2$, respectively;

- the machine computing a constant $\ulcorner\alpha\urcorner \circ \sigma'_I : R \to R$ outputs that constant;

- the machine computing $\sigma_{I+I} \circ \geq : R \to R^3$ outputs $\langle -1, 0, 0\rangle$ if $\alpha < 0$, and $\langle 0, 0, 0\rangle$ if $\alpha \geq 0$;

- the machine computing a first (second) projection just outputs the variables corresponding to the first (second) component;
- the machine computing a first (second) injection outputs $\langle -1, \boldsymbol{\alpha}, \mathbf{0}\rangle$ ($\langle 0, \mathbf{0}, \boldsymbol{\alpha}\rangle$);
- the machine computing an identity outputs its input;
- the machine computing a terminal map outputs 0;
- the machine computing the inverse distributivity isomorphism on input $\langle\langle \alpha, \boldsymbol{\beta}, \boldsymbol{\gamma}\rangle, \boldsymbol{\delta}\rangle$ outputs $\langle \alpha, \langle \boldsymbol{\beta}, \boldsymbol{\delta}\rangle, \langle \boldsymbol{\gamma}, \boldsymbol{\delta}\rangle\rangle$.

If $f = h \circ g$, then by Corollary 3 $\widehat{h \circ g} = \widehat{h} \circ \widehat{g}$, so by connecting in series the machines corresponding to $\widehat{g}$ and $\widehat{h}$ we obtain a machine which computes $\widehat{f}$.

If $f = (g, h)$, with $g : X \to Y'$, $h : X \to Y''$ (so $Y = Y' \times Y''$), then we take a machine which duplicates its input, runs $\widehat{g}$ on the first copy and $\widehat{h}$ on the second one, and finally outputs the two results juxtaposed. Again by Corollary 3, this machine computes $\widehat{f}$.

If $f = (g \mid h)$, with $g : X' \to Y$, $h : X'' \to Y$ (so $X = X' + X''$), then we take a machine which on input $\langle \alpha, \boldsymbol{\beta}, \boldsymbol{\gamma}\rangle \in R \times R^{n_{X'}} \times R^{n_{X''}}$ checks if $\alpha < 0$: in this case, it outputs $\widehat{g}(\boldsymbol{\beta})$; otherwise, it outputs $\widehat{h}(\boldsymbol{\gamma})$. One more time, Corollary 3 guarantees that the computed function is $\widehat{f}$. This completes the proof. ▮

As in the case of recursive functions, instead of restricting the class of distributively computable functions we could enlarge $\mathcal{F}$ using the encodings. However, we can again prove that

**Theorem 8** Let $\widehat{\mathcal{F}} = \{\sigma'_Y \circ f \circ \sigma_X \mid f : R^{n_X} \to R^{n_Y} \in \mathcal{F}\}$. Then $\widehat{\mathcal{F}} = \mathfrak{D}_R$.

**Proof.** Any function in $\widehat{\mathcal{F}}$ is trivially in $\mathfrak{D}_R$ (use Lemma 3, the fact $\mathcal{F} = \mathcal{G} \subseteq \mathfrak{D}_R$, and closure of $\mathfrak{D}_R$ with respect to composition). On the other hand, if $f : X \to Y \in \mathfrak{D}_R$ then $\widehat{f} \in \mathcal{G} = \mathcal{F}$. But then

$$\widehat{\mathcal{F}} \ni \sigma'_Y \circ \widehat{f} \circ \sigma_X = \sigma'_Y \circ \sigma_Y \circ f \circ \sigma'_X \circ \sigma_X = f. \blacksquare$$

We shall now give an example by mapping to a derived function the machine described in [Blum, Shub & Smale:1989, p. 5] (in order to simplify the discussion we suppose that the machine works on the reals, and we use the square instead of the absolute value of $g^k(x)$). The machine has a built-in constant $c$, and evaluates iteratively a given polynomial $g$ on the input $x$ until $(g^k(x))^2 \geq c$.

There is one register and four nodes. The second node is a computation node which applies $g$ and moves to the third node, while the third node is a branching node which checks if the square of the register is greater than or equal to $c$, moving to the output node in this case, or to the second node otherwise.

Our mapping sends this machine to the function $f : R + 4 \cdot R \to 4 \cdot R + R$ defined as follows:

$$
\begin{aligned}
f_0 &: R \xrightarrow{\mathrm{inj}_1} 4 \cdot R \\
f_1 &: R \xrightarrow{\mathrm{inj}_2} 4 \cdot R \\
f_2 &: R \xrightarrow{\mathrm{inj}_3 \circ g} 4 \cdot R \\
f_3 &: R \xrightarrow{\Delta_R} R \times R \\
&\xrightarrow{((-)^2 - c) \times \mathbf{1}_R} R \times R \\
&\xrightarrow{\geq \times \mathbf{1}_R} (I + I) \times R \\
&\xrightarrow{\delta^{-1}} I \times R + I \times R \\
&\xrightarrow{\cong} R + R \\
&\xrightarrow{(\mathrm{inj}_2 \mid \mathrm{inj}_4)} 4 \cdot R \\
f_4 &: R \xrightarrow{\mathbf{1}_R} R
\end{aligned}
$$

An input $x$ generates the following orbit:

$$x \mapsto \langle x, 1\rangle \mapsto \langle x, 2\rangle \mapsto \langle g(x), 3\rangle .$$

Now $f_3$ works as follows:

$$g(x) \mapsto \langle g(x), g(x)\rangle \mapsto \langle g(x)^2 - c, g(x)\rangle \mapsto$$
$$\langle\langle *, k\rangle, g(x)\rangle \mapsto \langle\langle *, g(x)\rangle, k\rangle \mapsto \langle g(x), k\rangle \mapsto \langle g(x), 2k + 2\rangle ,$$

where $k$ is 0 if $g(x)^2 < c$, 1 otherwise. In the first case we restart from node 2, obtaining

$$\langle g(x), 2\rangle \mapsto \langle g^2(x), 3\rangle$$

and so on; in the second case, we apply $f_4$ and we output $g(x)$.

### 2.4.5 An application

We shall now apply Theorem 7 in order to obtain a new structural characterization à la Kleene of the functions $R^l \to R^m$ computable in the BSS model. The characterization is based on iteration rather than on recursion (a structural characterization based on recursion was given in [Blum, Shub & Smale:1989]), and it is clearly derived from the operations available in a distributive category.

**Definition 9** The set $\mathcal{K}_\nu$ of general $\nu$-recursive functions over $R$ is the smallest set of functions $R^l \to R^m$ containing the polynomial functions, and closed under the following constructions:

1. (*composition*) given $f : R^l \to R^m$ and $g : R^m \to R^n$, form $f \circ g$;
2. (*juxtaposition*) given $g_i : R^m \to R^{n_i}$ $(1 \leq i \leq k)$, form $(g_1, g_2, \ldots , g_k) : R^m \to R^{\sum_{1 \leq i \leq k} n_i}$;
3. (*cases*) given $f, g : R^m \to R^n$ form $h : R \times R^m \to R^n$, defined by

$$h(\alpha, \boldsymbol{\beta}) = \begin{cases} f(\boldsymbol{\beta}) & \text{if } \alpha < 0 \\ g(\boldsymbol{\beta}) & \text{if } \alpha \geq 0 \end{cases}$$

4. ($\nu$-*operator*) given $f : R \times R^n \to R \times R^n$, if for every $\boldsymbol{\alpha} \in R^n$ there is a least $n_{\boldsymbol{\alpha}}$ such that $f^{n_{\boldsymbol{\alpha}}}(-1, \boldsymbol{\alpha}) = \langle \rho_{\boldsymbol{\alpha}}, \boldsymbol{\beta}_{\boldsymbol{\alpha}} \rangle$ with $\rho_{\boldsymbol{\alpha}} \geq 0$, then form $\nu f : R^n \to R^n$ by setting $\nu f(\boldsymbol{\alpha}) = \boldsymbol{\beta}_{\boldsymbol{\alpha}}$.

If we consider also partial functions, we can relax the condition given on iteration, and for *any* $f : R \times R^n \to R \times R^n$ define $\nu f$ ($\nu f(\boldsymbol{\alpha})$ is undefined if there is no $n_{\boldsymbol{\alpha}}$ as in Definition 9). The other operations should be redefined as usual. We shall speak in this case of the partial $\nu$-recursive functions over $R$.

**Theorem 9** $\mathcal{K}_\nu = \mathcal{G}$.

**Proof.** We firstly prove left-to-right inclusion. We already know that all polynomials are derived functions. Composition and juxtaposition have corresponding categorical operators. Construction by cases on $f, g : R^m \to R^n$ can be obtained as follows:

$$R \times R^m \xrightarrow{\geq \times \mathbf{1}_{R^m}} (I + I) \times R^m \xrightarrow{\delta^{-1}} R^m + R^m \xrightarrow{(f \mid g)} R^n.$$

Given $f : R \times R^n \to R \times R^n$, consider the functions

$$\begin{aligned} g \quad &: \quad R^n \cong I \times R^n \xrightarrow{\ulcorner -1 \urcorner \times \mathbf{1}_{R^n}} R \times R^n \xrightarrow{\mathrm{inj}_1} R \times R^n + R^n \\ h \quad &: \quad R \times R^n \xrightarrow{f} R \times R^n \xrightarrow{\Delta_R \times \mathbf{1}_{R^n}} R \times R \times R^n \xrightarrow{\geq \times \mathbf{1}_{R \times R^n}} (I + I) \times R \times R^n \\ &\quad\quad \xrightarrow{\delta^{-1}} R \times R^n + R \times R^n \xrightarrow{\mathbf{1}_{R \times R^n} + \mathrm{proj}_2} R \times R^n + R^n. \end{aligned}$$

It is a matter of calculation to show that **call**$[\, R^n, R \times R^n, R^n, (g \mid h)\,] = \nu f$.

Some more work is required in order to prove that $\mathcal{G} \subseteq \mathcal{K}_\nu$. We apply structural induction again, following the lines of the proof of Theorem 7. The function computed by iterating the loop $h : X + U + Y \to X + U + Y$ induced by a derived function $g : X + U \to U + Y$ can be obtained by applying the $\nu$ operator to a suitable function $k : R \times R^{n_{X+U+Y}} \to R \times R^{n_{X+U+Y}}$. We shall define $k$ by juxtaposing the functions $k_1 : R \times R^{n_{X+U+Y}} \to R$ and $k_2 : R \times R^{n_{X+U+Y}} \to R^{n_{X+U+Y}}$ described below, which are clearly in $\mathcal{K}_\nu$[9]:

$$k_1(\alpha, \boldsymbol{\beta}) = \begin{cases} 0 & \text{if } \beta_1 \geq 0 \\ -2 & \text{otherwise} \end{cases}$$

$$k_2(\alpha, \boldsymbol{\beta}) = \begin{cases} \langle -1, -1, \beta_3, \ldots, \beta_{n_{X+U+Y}} \rangle & \text{if } \alpha = -1 \\ \widehat{h}(\boldsymbol{\beta}) & \text{otherwise.} \end{cases}$$

Note the usage of the "control variable" $\alpha$. When $k = (k_1, k_2)$ is applied, we check whether $\alpha = -1$ (first iteration). In this case, we set the first two coordinates of the encoding of $(X + U) + Y$ to $-1$: this forces the element encoded in $R^{n_{X+U+Y}}$ to be an element of $X$. At the same time, $\alpha$ is set to $-2$, so that the subsequent iterations will apply $\widehat{h}$ leaving $\alpha$ undisturbed, unless the element encoded in $\boldsymbol{\beta}$ becomes an element of $Y$ (i.e., $\beta_1 \geq 0$), in which case $\alpha$ is set to 0, stopping iteration (and necessarily $\boldsymbol{\beta}$ is left undisturbed by $\widehat{h}$). Since $g$ terminates for every element of $X$, and the iteration of $\widehat{h}$ necessarily starts from the encoding of an element of $X$, the termination condition on $g$ implies the termination condition required by the operator $\nu$, which can actually be applied to $k$. Precomposition and postcomposition of $\nu k$ with trivial maps preparing the input and extracting the output give as a result the function **call**$[\, g\,]$.

This machinery is actually necessary (similarly to what happened in the proof of Theorem 4): If we did not force the presence of the encoding of an element of $X$ at the first iteration, we could iterate over an element of $U$, over which we could not guarantee termination. The precomposition and postcomposition would in the end produce a (total) function $R^{n_X} \to R^{n_Y}$, but the $\nu$ operator would have been applied incorrectly.

[9] We are assuming, without loss of generality, that $X + U + Y$ is encoded as $(X + U) + Y$.

We are left to prove that for any derived function $f : X \to Y$ we have $\widehat{f} \in \mathcal{K}_\nu$. We use structural induction:

- the functions $* : R^2 \to R$, $+ : R^2 \to R$ and the constants are obviously representable as polynomials, and thus in $\mathcal{K}_\nu$;
- the function $\sigma_{I+I} \circ \geq : R \to R^3$ can be obtained by juxtaposition and cases:

$$(\sigma_{I+I} \circ \geq)(\alpha) = \begin{cases} \langle -1, 0, 0 \rangle & \text{if } \alpha < 0 \\ \langle 0, 0, 0 \rangle & \text{if } \alpha \geq 0; \end{cases}$$

- the projections can be obtained again by juxtaposition, discarding the variables which correspond to the discarded component;
- the first (second) injection outputs $\langle -1, \boldsymbol{\alpha}, \mathbf{0} \rangle$ ($\langle 0, \mathbf{0}, \boldsymbol{\alpha} \rangle$) on input $\boldsymbol{\alpha}$;
- the terminal map is the constant polynomial 0;
- the identities and the inverse distributivity isomorphism can be easily obtained by juxtaposition of polynomials.

The functions obtained by application of composition and $(-, -)$ are by definition inside $\mathcal{K}_\nu$; if $f = (g \mid h)$, with $g : X' \to Y$ e $h : X'' \to Y$ (so $X = X' + X''$), we define a function by cases using $\widehat{g} \circ \mathrm{proj}_1$, $\widehat{f} \circ \mathrm{proj}_2 : R^{n_{X'}} \times R^{n_{X''}} \to R^{n_Y}$. The resulting function is exactly $\widehat{f}$. This completes the proof. ▮

The obvious consequence of the previous theorem is that the $\nu$-recursive functions coincide with the functions computed in the BSS model:

**Corollary 4** $\mathcal{K}_\nu = \mathcal{F}$.

# Chapter 3

# Automata

In this chapter we shall introduce the notion of *morphism* of automata. Since any function $f : X + U \to U + Y$ can be seen as a one-letter automaton with state space $X + U + Y$, the operator **call**$[-]$ will allow to obtain, from automata whose state space has the form $X + U + Y$ and satisfying certain conditions, automata with state space $X + Y$, thus realizing an abstraction of the time and space necessary to compute a function. The notion of morphism will then enable us to "invert" the **call**$[-]$ operator, because the former automaton will represent the latter in the sense of [Knuth:1966].

As in the previous chapter, the description will be kept at an elementary level: only a small amount of category theory will be used, and for the time being we shall use as "computational universe" **Set**, the familiar category of sets and functions. We assume once for all that a set of basic functions (i.e., a distributive graph) has been chosen, and that all functions used are distributively computable.

Firstly we shall give the basic definitions, some of which are probably known to the reader. Then we shall prove a series of simple but nontrivial theorems about morphisms of automata, in order to give an intuition about their nature and to prepare the ground for the refinement theorem, which will be proved in its most general form in Chapter 4.

## 3.1 Basic definitions

**Definition 10** Let $M$ be a monoid and $\boldsymbol{X}$ an $M$-automaton; that is, a set $X$ endowed with an action of $M$

$$\alpha : M \times X \to X, \quad (m, x) \overset{\alpha}{\mapsto} m \cdot x;$$

the action is required to satisfy the usual axioms $m_2 \cdot (m_1 \cdot x) = (m_2 m_1) \cdot x$ and $1 \cdot x = x$. Define the category **Trans**$(\boldsymbol{X})$ (the *transition category of* $\boldsymbol{X}$) as follows:

(i). objects are states (that is, elements) of $X$;

(ii). arrows from $x$ to $y$ are state transitions; that is, elements $m \in M$ such that $m \cdot x = y$;

(iii). composition is monoid multiplication.

Note that since each arrow of a transition category is determined by an element $m \in M$ *and* by a domain and a codomain $x, y \in X$, many distinct arrows will be named by the same element of $M$. This fact should cause no confusion, but we shall reserve the possibility of writing arrows in the form $x \overset{m}{\to} y$ whenever necessary. If $M$ is the monoid of natural numbers $\boldsymbol{N}$, the action $\alpha$ is completely specified by a function $X \to X$, which we call again $\alpha$ for sake of semplicity.

We shall denote with $\mathrm{Ind}(M)$ the set of the *indecomposable elements* of $M$, i.e., the elements $m \in M$ such that $m = m'm''$ implies $m' = \varepsilon$ or $m'' = \varepsilon$. These are also all and only the minimal elements with respect to the prefix (or suffix) preordering on $M \setminus \{\varepsilon\}$ (in this preordering, $m \leq n$ iff $n = mm'$). Not amazingly, if $\boldsymbol{X}$ is an $M$-automaton we shall call the arrows of $\mathbf{Trans}(\boldsymbol{X})$ which are labelled by elements of $\mathrm{Ind}(M)$ *atomic*.

### 3.1.1 Pseudofunctions

Certain automata on $\boldsymbol{N}$ related to functions $f : X + U \to U + Y$ can compute functions by iteration. To this purpose, in [Khalil & Walters:1993] the notion of *pseudofunction* has been introduced:

**Definition 11** A *functional processor* or *pseudofunction* from $X$ to $Y$ is an $\boldsymbol{N}$-automaton with state space of the form $X + U + Y$ and action $\alpha$ such that

(i). for each $y$ in $Y$ we have $\alpha(y) = y$;

(ii). for each $x$ in $X$ there exists a natural number $n_x$ such that $\alpha^{n_x}(x) \in Y$.

A pseudofunction $\alpha$ defines a function $\mathbf{call}[\,\alpha\,] : X \to Y$ (by iteration) as follows:

$$\mathbf{call}[\,\alpha\,](x) = \alpha^{n_x}(x).$$

We shall call $U$ the set of *local* states of the pseudofunction $\alpha$, or the *hidden* states of the function $\mathbf{call}[\,\alpha\,]$.

Of course, any function of the form $f : X + U \to U + Y$ induces a pseudofunction

$$\varphi : X + U + Y \xrightarrow{(f \mid \mathrm{inj}_2)} U + Y \xrightarrow{\mathrm{inj}_2} X + (U + Y)$$

(something which we exploited in the proofs of several theorems of Chapter 2), and $\mathbf{call}[\,f\,] = \mathbf{call}[\,\varphi\,]$.

## 3.2 Morphisms

Our notion of morphism of automata [Sabadini, Vigna & Walters:1993a] is a generalization of the morphisms of automata given by a function on states which preserves the action up to substitution of a letter with a fixed word. Indeed, in our morphisms this substitution is *state-dependent*.

**Definition 12** A morphism of automata (or, in short, a *mapping*) from $\boldsymbol{X}$ to $\boldsymbol{Y}$, where $\boldsymbol{X}$ is an $M$-automaton and $\boldsymbol{Y}$ is an $N$-automaton, is a functor from $\mathbf{Trans}(\boldsymbol{X})$ to $\mathbf{Trans}(\boldsymbol{Y})$.

If $M = A^*$ and $N = B^*$ (i.e., $M$ and $N$ are free monoids) a mapping $F$ from $\boldsymbol{X}$ to $\boldsymbol{Y}$ is given by two functions (which we shall call ambiguously with the same name)

$$F : X \to Y$$

and

$$F : X \times A \to B^*$$

which satisfy the following condition: if $a \in A$ and $x \xrightarrow{a} x'$ is an arrow of $\mathbf{Trans}(\boldsymbol{X})$, then $F(x) \xrightarrow{F(x,a)} F(x')$ is an arrow of $\mathbf{Trans}(\boldsymbol{Y})$.

The following theorem partially characterizes isomorphic automata:

**Theorem 10** Let $\boldsymbol{X}$ and $\boldsymbol{Y}$ be automata on the monoids $M$ and $N$, and let $\boldsymbol{X}$ be isomorphic to $\boldsymbol{Y}$ in the sense of Definition 12. Then $|\operatorname{Ind}(M)| = |\operatorname{Ind}(N)|$, and if $\operatorname{Ind}(M)$ is a set of generators then the isomorphism is induced by a family of bijections $\{\varphi_x : \operatorname{Ind}(M) \to \operatorname{Ind}(N)\}_{x \in X}$.

**Proof.** Let $F : \boldsymbol{X} \to \boldsymbol{Y}$ be an isomorphism. Then $F$ is a bijection on objects, i.e., a bijection between the sets $X$ and $Y$, and a bijection between sets of arrows. Firstly, let us show that $F$ preserves and reflects atomicity. If $x \xrightarrow{m} x'$ is atomic, and $F(x \xrightarrow{m} x') = F(x) \xrightarrow{n_1} y \xrightarrow{n_2} F(x')$, then

$$x \xrightarrow{m} x' = F^{-1}(F(x \xrightarrow{m} x')) = F^{-1}(F(x) \xrightarrow{n_1} y \xrightarrow{n_2} F(x')) = x \xrightarrow{m_1} F^{-1}(y) \xrightarrow{m_2} x'$$

with

$$F(x \xrightarrow{m_1} F^{-1}(y)) = F(x) \xrightarrow{n_1} y \quad \text{and} \quad F(F^{-1}(y) \xrightarrow{m_2} x') = y \xrightarrow{n_2} F(x'),$$

so that necessarily $n_1 = \varepsilon$ or $n_2 = \varepsilon$. Thus, $F$ preserves atomicity. But symmetrically this is true for $F^{-1}$, so $F$ also reflects atomicity.

We shall denote with $\varphi_x$ the restriction of $F$ to the atomic arrows outgoing from $x$. Since $\boldsymbol{X}$ and $\boldsymbol{Y}$ are complete automata, $\varphi_x$ is a bijection between $\operatorname{Ind}(M)$ and $\operatorname{Ind}(N)$. In particular, for each indecomposable $m$ we have

$$F(x \xrightarrow{m} x') = F(x) \xrightarrow{\varphi_x(m)} F(x').$$

Finally, if $\operatorname{Ind}(M)$ is a set of generators then every arrow can be written as a composition of a finite number of atomic arrows, and the effect of $F$ can be computed by composition knowing just the bijections $\varphi_x$. ∎

**Corollary 5** Let $\boldsymbol{X}$ and $\boldsymbol{Y}$ be isomorphic automata on the free monoids $A^*$ and $B^*$. Then $|A| = |B|$.

Note that we cannot claim that $\boldsymbol{X}$ and $\boldsymbol{Y}$ are isomorphic as automata on $A^*$, in the classical sense. A simple counterexample follows:

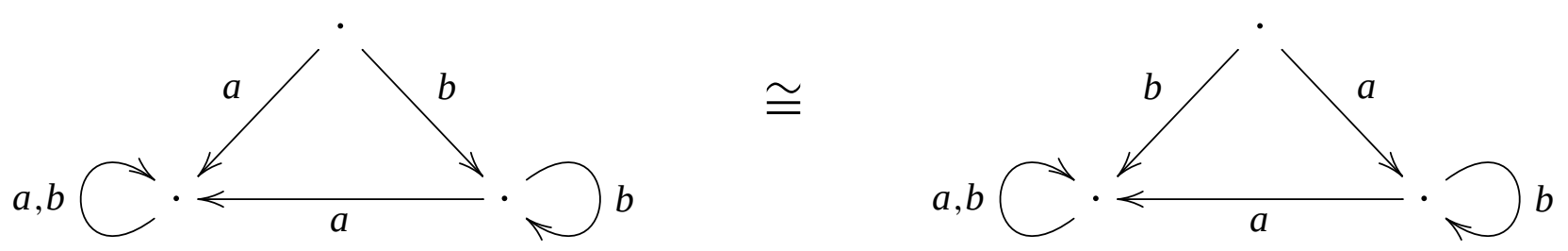

### 3.2.1 Refinement

In [Sabadini, Vigna & Walters:1993b] a particular case of mapping, named *refinement*, is studied. The notion of refinement plays a particularly important rôle in the theory of distributive computability because the automata built on distributively computable functions admit refinements into the automata built on the relative derived functions. In particular, we shall see that in a precise sense pseudofunctions are refinement of functions. The notion of refinement is also essential in any model of distributed and concurrent systems; the quoted reference gives some motivation to this purpose.

**Definition 13** A mapping $F : \boldsymbol{X} \to \boldsymbol{Y}$ is a *refinement* iff the functor $F : \mathbf{Trans}(\boldsymbol{X}) \to \mathbf{Trans}(\boldsymbol{Y})$ is a full inclusion.

In other words, in order to give a refinement of $\boldsymbol{X}$ one has to specify a bigger system $\boldsymbol{Y}$ which has a restriction to a system isomorphic to $\boldsymbol{X}$. If $M = A^*$ and $N = B^*$, the additional condition for $F$ being a refinement is that the function induced by $F$ between $\mathbf{Hom}(x, x')$ and $\mathbf{Hom}(F(x), F(x'))$ is a bijection, and that the function between the state spaces in injective.

Note that *non all* full inclusions in **Trans**($\boldsymbol{Y}$) can have as domain a category of the form **Trans**($\boldsymbol{X}$). For instance, consider the automaton

$$0 \xrightarrow{a} 1 \xrightarrow{a} 2 \circlearrowleft a$$

The full subcategory contaning the objects 0 and 1 has just one arrow $0 \xrightarrow{a} 1$; thus, it cannot be of the form **Trans**($\boldsymbol{X}$).

In order to give some intuition about the nature of refinements, we shall relate them with simpler properties. We start by giving the following

**Definition 14** An *expansive mapping* is an inclusion $F$ of **Trans**($\boldsymbol{X}$) in **Trans**($\boldsymbol{Y}$) such that if $F(x \xrightarrow{m} x') = F(x) \xrightarrow{n} F(x')$, where $x \xrightarrow{m} x'$ is atomic, there are no $x'' \in X$, $n' \in N$ such that $n'$ is a nontrivial suffix of $n$ and $F(x) \xrightarrow{n'} F(x'')$.

When an expansive mapping is applied to an atomic transition, the set of states spawned[1] by the resulting arrow lies entirely outside of the image of $\boldsymbol{X}$, except for the initial and final states (which are the image of the domain and of the codomain of the atomic transition). Indeed, if $M = A^*$ and $N = B^*$ we can restate Definition 14 as follows:

1. $Y = X + U$ for some set $U$;
2. if $F(x, a) = b_n \cdots b_1 \in B^*$, then $b_k \cdots b_1 \cdot x \in U$ for $k = 1, 2, \ldots, n-1$.

Expansiveness and fullness are related by the following

**Theorem 11** Let $\boldsymbol{X}$ and $\boldsymbol{Y}$ automata on $M$ and $N$, respectively. If a mapping $F : \boldsymbol{X} \to \boldsymbol{Y}$ is a refinement, then it is expansive.

**Proof.** Let $m$ be indecomposable, and $F(x \xrightarrow{m} x') = F(x) \xrightarrow{n} F(x')$. Let $n'$ be a suffix of $n$, and $F(x) \xrightarrow{n'} F(x'')$. Then $n = n''n'$, and $F(x'') \xrightarrow{n''} F(x')$. By fullness of $F$, there are $m', m'' \in M$ such that the arrows $x \xrightarrow{m'} x''$ and $x'' \xrightarrow{m''} x'$ have as image $F(x) \xrightarrow{n'} F(x'')$ and $F(x'') \xrightarrow{n''} F(x')$. By composition, we get

$$F(x \xrightarrow{m'} x'' \xrightarrow{m''} x') = F(x) \xrightarrow{n'} F(x'') \xrightarrow{n''} F(x') = F(x) \xrightarrow{n} F(x') = F(x \xrightarrow{m} x'),$$

and by faithfulness $m = m''m'$. Thus either $m' = m$ and $m'' = \varepsilon$, or $m' = \varepsilon$ and $m'' = m$. In both cases, $n'$ (which is the image of $m'$) is a trivial suffix of $n$. ▮

This theorem cannot be reversed. Consider the mapping between the following two automata

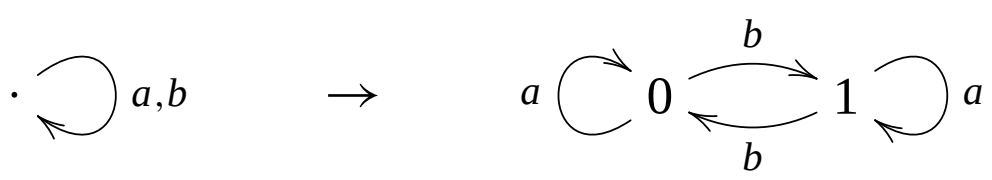

where $\cdot \mapsto 0$, $a \mapsto a$ and $b \mapsto b^2$. It is expansive, but not full ($0 \xrightarrow{bab} 0$ has no counterimage). There is however a relevant case in which we can invert Theorem 11:

**Theorem 12** Let $\boldsymbol{X}$ and $\boldsymbol{Y}$ be $\boldsymbol{N}$-automata. If a mapping $F : \boldsymbol{X} \to \boldsymbol{Y}$ is expansive, then it is a refinement.

[1] If the atomic decomposition of the image of an atomic transition is not unique, the set is not necessarily unique.

**Proof.** If $F$ is not a refinement, consider states $x, x' \in X$ and an arrow $F(x) \xrightarrow{k} F(x')$ which is not the image of an arrow from $x$ to $x'$. Assume without loss of generality that $k$ is minimal. Let $F(x) \xrightarrow{n} F(x'')$, $n \neq k$, be the image through $F$ of $x \xrightarrow{1} x''$. If $k > n$, then necessarily $F(x'') \xrightarrow{k-n} F(x')$ is not in the image of $\boldsymbol{X}$, which contradicts the minimality of $k$. But then $n > k$, which contradicts expansiveness. ▮

### 3.2.2 Pseudofunctions and refinement

The fundamental relation between computation by iteration and refinement is the fact that, given a pseudofunction $\alpha : X + U + Y \to X + U + Y$ with the property $\alpha^n(x) \in U + Y$ for all $n$ (without loss of generality), we can define an automaton $\boldsymbol{X}$ with state space $X + Y$ which computes **call**[ $\alpha$ ] in one step, and which admits a refinement in the automaton defined by $\alpha$, refinement which is induced by the inclusion $X + Y \to X + U + Y$ (the claim will be proved in Chapter 4, in the context of extensive categories). In a slogan: *the inverse of the application of* **call**[ − ] *is a refinement.* This fact generalizes to the following

**Theorem 13** Let $\boldsymbol{X}$ be an $A^*$-automaton whose action is defined using functions distributively computable on a basic set. Then $\boldsymbol{X}$ admits a refinement in an $A^*$-automaton whose action is defined using derived functions.

**Proof.** For each $a \in A$, $\alpha(a, -) : X \to X$ admits a refinement to a pseudofunction $\varphi_a : X + U_a + X \to X + U_a + X$. Now define

$$Y = I + X + \sum_{a \in A} U_a.$$

The action of $b \neq a$ on $U_a$ is the unique function $U_a \to I$, while on $X$ and $U_a$ is induced in an obvious way by $\varphi_a$. It is immediate to show that the inclusion $X \to Y$ induces a refinement. ▮

### 3.2.3 Examples

**Representations.** The existence of a refinement $\boldsymbol{X} \to \boldsymbol{Y}$, where $\boldsymbol{X}$ and $\boldsymbol{Y}$ are $\boldsymbol{N}$-automata, exhibits $\boldsymbol{Y}$ as a representation of $\boldsymbol{X}$ in the sense of [Knuth:1966].

**An IMP($G$) Interpreter.** The **IMP**($G$) interpreter described in [Khalil & Walters:1993] is easily seen to be a refinement of each particular **IMP**($G$) program.

**Independent actions are not necessarily parallel.** If we consider a fixed machine $\boldsymbol{Y}$, we may think of refinements of other machines $\boldsymbol{X}$ in $\boldsymbol{Y}$ as implementations of abstract machines $\boldsymbol{X}$ in a system $\boldsymbol{Y}$. The class of abstract machines has unbounded parallelism; however, in relation to $\boldsymbol{Y}$ it is possible to consider questions of resources. We can make the distinction between actions of $\boldsymbol{X}$ being "independent" and being "parallel". Actions are independent if they are parallel in the abstract machine $\boldsymbol{X}$. Actions are parallel if they are parallel in the system $\boldsymbol{Y}$. The following example can be expressed by saying that independent actions in an abstract machine may not be parallel in the implementation.

Given two automata $\boldsymbol{X}$, $\boldsymbol{Y}$, both on $\{a\}^*$, suppose that there are refinements of $\boldsymbol{X}$ to $\boldsymbol{X}'$ and $\boldsymbol{Y}$ to $\boldsymbol{Y}'$, where $X' = X + U$ and $Y' = Y + U$, the meaning being that the set $U$ is the state space of some temporarily used (and reset after use) resource like a scratch pad, or printer. Then the synchronous parallel product $\boldsymbol{X} \times \boldsymbol{Y}$ of $\boldsymbol{X}$ and $\boldsymbol{Y}$, which has state space $X \times Y$ and action of $\{a\}^*$ defined componentwise, may be refined to an automaton in which there is only one resource $U$ whose use is scheduled between $\boldsymbol{X}$ and $\boldsymbol{Y}$. The state space would be $Z = XY + UY + XY + XU$. The two letters $b, c$ would act respectively on the first two and on the last two summands, applying $\langle a, \mathbf{1}\rangle$ and $\langle \mathbf{1}, a\rangle$, respectively (a trap state would complete the automaton in the obvious way). The injection of $XY$ as first summand of $Z$ would then define a refinement, which would schedule the parallel action $\langle a, a\rangle$ to $bc$.

**Shutdown.** Consider a refined description of a system in which a new, destructive action can happen. This is a typical case of a sudden shutdown. We expect that the system can, at any time, be shut down, thus moving into a state which was impossible to reach before. In this case, the refinement space is formed by adding a single element, and a new letter to the alphabet; it sends to the new state any other state. The behaviour of the machine, if we ignore the shutdown state, is unmodified, which is exactly reflected in our definition of refinement.

### 3.2.4 Abstraction

Refinements introduce new information in the description of an automaton. In this section, on the contrary, we shall discuss how to reduce it.

**Definition 15** A mapping $F : \boldsymbol{X} \to \boldsymbol{Y}$ is an *abstraction* iff the functor $F : \mathbf{Trans}(\boldsymbol{X}) \to \mathbf{Trans}(\boldsymbol{Y})$ is surjective on objects and arrows.

An abstraction is essentially a quotient of automata. Let us consider the following example: for any automaton $\boldsymbol{X}$ on a free monoid, any partition of the states of $\boldsymbol{X}$ in two classes (linked by at least one transition) induces an abstraction $F$ having as codomain the automaton $\boldsymbol{Y}$ specified here:

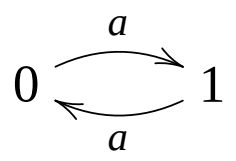

The abstraction maps the states in each class respectively to 0 and 1. All atomic transitions which do not change the class are mapped to the identity of 0 or 1, respectively, while an atomic transition which crosses the partition boundary is mapped to the suitable $a$-transition of $\boldsymbol{Y}$. As a result, each arrow of $\mathbf{Trans}(\boldsymbol{X})$ is mapped to an arrow of the form $x \xrightarrow{a^n} y$ (where $x, y = 0$ or 1); the exponent $n$ is equal to the number of changes of class that happened while the state transformations induced by the arrow were applied. In other words, $F$ forgets the information about the internals of the automaton, and allows us only to observe two states, and the transitions between those two states.

Note that *not all abstractions can be inverted by a refinement*. Indeed, an epimorphism $F : \mathbf{Trans}(\boldsymbol{X}) \to \mathbf{Trans}(\boldsymbol{Y})$ does not necessarily have a section which is a full inclusion. The following counterexample shows this fact:

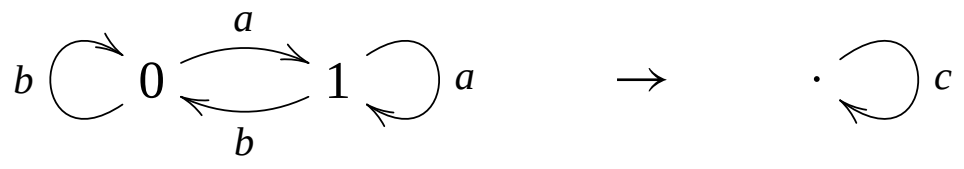

The mapping is defined by $a, b \mapsto c$. Any section of it has to send $c$ to $a$ or $b$. The inclusion cannot then be full because of the arrows $(ba)^n$ or $(ab)^n$, respectively.

The following example shows instead a case in which refinement of the abstraction is possible. Consider the mapping

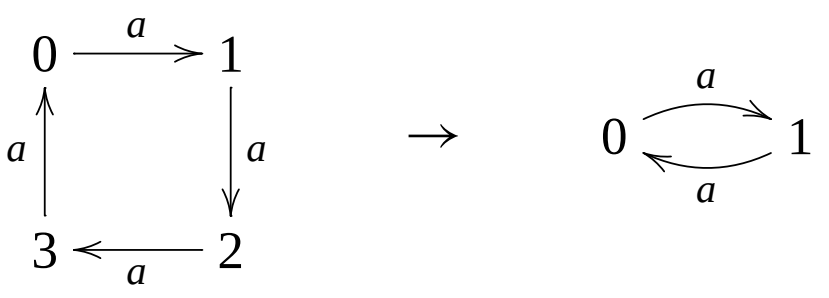

defined by $n \mapsto n/2$ and by

$$\begin{array}{ccc} 0 \xrightarrow{a} 1 & \mapsto & 0 \xrightarrow{\varepsilon} 0 \\ 1 \xrightarrow{a} 2 & \mapsto & 0 \xrightarrow{a} 1 \\ 2 \xrightarrow{a} 3 & \mapsto & 1 \xrightarrow{\varepsilon} 1 \\ 3 \xrightarrow{a} 0 & \mapsto & 1 \xrightarrow{a} 0. \end{array}$$

The refinement $x \mapsto 2x$, with

$$x \xrightarrow{a} y \quad \mapsto \quad 2x \xrightarrow{a^2} 2y \quad (x, y = 0 \text{ or } 1),$$

is a section of the given abstraction.

Note that having a refinement as a section is a property of a specific *abstraction*, not of the *automata* involved. Indeed, using the same two automata of the previous example, consider the abstraction $F$ defined by $n \mapsto (n \bmod 2)$, with

$$x \xrightarrow{a} y \quad \mapsto \quad (x \bmod 2) \xrightarrow{a} (y \bmod 2);$$

it is easy to see that there is no inversion through a refinement. Indeed, in order to invert $F$ we must send 0 and 1 in states $x$ and $y$ such that the shortest word in $\mathbf{Hom}(x, y)$ (or $\mathbf{Hom}(y, x)$) is $a^3$. Since $F$ preserves the word length, it is not possible to find a section for $0 \xrightarrow{a} 1$ ($1 \xrightarrow{a} 0$, respectively).

### 3.2.5 Input/Output

In this section we note that our notion of morphism allows to specify elegantly output of automata. Our basic assumption is that output is *state-dependent*, i.e., it depends *both* on the state *and* on the monoid element which is acting.

**Definition 16** An automaton on $M$ with output in $N$ is an $M$-automaton $\boldsymbol{X}$ endowed with a mapping $O_{\boldsymbol{X}} : \boldsymbol{X} \to N$ (where $N$ is seen as a one-state automaton).

The meaning of this definition is that for each state $x$ and each element $m$ of $M$ there is an element of $N$ which represents the output of the action of $m$ in the state $x$. The functoriality equations tell us that no output is possible if no action is specified, and that the output of the composition of actions is the composition of the output of the single actions.

## 3.3 Partial specification

Partially specified automata are automata on a free monoid whose action is defined, for each letter in the alphabet, only on a part of the state space. The non-completeness of information fits flawlessly in our theory, and it is particularly useful when specifying all details about trap states would be cumbersome.

**Definition 17** A partially specified automaton on $A^*$ is a set $X$, together with, for each $a \in A$, a decomposition $X = X_a + U_a$ and a function[2]

$$\alpha_a : X_a \to X.$$

These functions induce a partial action of $A^*$ on $X$. Again, define the category $\mathbf{Trans}(\boldsymbol{X})$ as follows:

[2]It could be argued that one could define a partial specification using functions $\alpha_a : X \to X + I$; but even in the general case we shall be working in an extensive category, and in such a category the two concepts coincide.

(i). objects are states (that is, elements) of $X$;

(ii). arrows from $x$ to $y$ are allowable state transitions, i.e., words $w \in A^*$ such that $w \cdot x = y$;

(iii). composition is monoid multiplication in $A^*$.

Note that we did not state the definition of partial specification for an automaton on an arbitrary monoid $M$. In that case, we have to specify a decomposition $X = X_m + U_m$ for each $m \in M$, but the decompositions are not free (if $m = m_1 m_2$, and the decompositions associated to $m_1$ and $m_2$ allow to apply them in sequence from a certain state, then the decomposition associated to $m$ has to include that state).

It is clear that the definition of mapping of automata given in the previous section is still perfectly valid (being built upon the notion of transition category). Thus, we can formulate our first

**Proposition 3** Each partially specified automaton $\boldsymbol{X}$ admits a refinement in an automaton $\boldsymbol{Y}$ on the same monoid.

**Proof.** Let $Y = X + I$, and define the action of $\boldsymbol{Y}$ extending the action of $\boldsymbol{X}$ via the unique map $U_\alpha + I \to I$ (i.e., we are adding a trap state to $\boldsymbol{X}$). The inclusion $X \to X + I$ then induces a full inclusion $\mathbf{Trans}(\boldsymbol{X}) \to \mathbf{Trans}(\boldsymbol{Y})$. ▮

We already showed that not all full subcategories of $\mathbf{Trans}(\boldsymbol{Y})$ are of the form $\mathbf{Trans}(\boldsymbol{X})$ for an automaton $\boldsymbol{X}$. However, the contrary is true for partially specified automata (and gives the main motivation for introducing them):

**Theorem 14** Let $\boldsymbol{Y}$ be a partially specified automaton on $B^*$. For each full subcategory $\mathfrak{C} \subseteq \mathbf{Trans}(\boldsymbol{Y})$ there is a partially specified automaton $\boldsymbol{X}$ such that $\mathfrak{C} \cong \mathbf{Trans}(\boldsymbol{X})$.

**Proof.** For each $y, y' \in \mathfrak{C}$ consider the set

$$N_{(y,y')} = \{y \xrightarrow{w} y' \mid \nexists\, y'' \in \mathfrak{C}, y \xrightarrow{w'} y'', y'' \xrightarrow{w''} y' \text{ such that } w = w''w'\},$$

where $w, w', w'' \neq \varepsilon$. Define a partially specified automaton $\boldsymbol{X}$ on the (not necessarily finite!) alphabet

$$A = \sum_{y,y' \in \mathfrak{C}} N_{(y,y')},$$

with an action on $X = |\mathfrak{C}|$ such that

$$X_{y \xrightarrow{w} y'} = \{y\}$$

and obviously $(y \xrightarrow{w} y') \cdot y = y'$. The mapping $F : \mathbf{Trans}(\boldsymbol{X}) \to \mathfrak{C}$ which is the identity on objects and sends an element $y \xrightarrow{w} y' \in A$ to the arrow $y \xrightarrow{w} y'$ of $\mathfrak{C}$ is trivially full. The crucial point of the proof is showing its faithfulness.

Let $x \xrightarrow{u} x'$, $x \xrightarrow{v} x'$ be arrows of $\mathbf{Trans}(\boldsymbol{X})$ which are mapped to the same arrow of $\mathfrak{C}$. This means that

$$u = (w_{m-1} w_{m-2} \cdots w_1 \cdot x \xrightarrow{w_m} x') \cdots (w_1 \cdot x \xrightarrow{w_2} w_2 w_1 \cdot x)(x \xrightarrow{w_1} w_1 \cdot x)$$

and

$$v = (z_{n-1} z_{n-2} \cdots z_1 \cdot x \xrightarrow{z_n} x') \cdots (z_1 \cdot x \xrightarrow{z_2} z_2 z_1 \cdot x)(x \xrightarrow{z_1} z_1 \cdot x),$$

with moreover

$$w_m w_{m-1} \cdots w_1 = z_n z_{n-1} \cdots z_1.$$

If any of the two arrows is an identity, so is the other one; otherwise, without loss of generality we can assume $z_1 = tw_1$. Since $\mathfrak{C}$ is full, it contains the arrow $w_1 \cdot x \xrightarrow{t} tw_1 \cdot x$. But then necessarily $t = \varepsilon$ and $w_1 = z_1$. The result follows by induction on $n + m$. ▮

The fullness hypothesis is necessary, as the following counterexample shows: consider the automaton

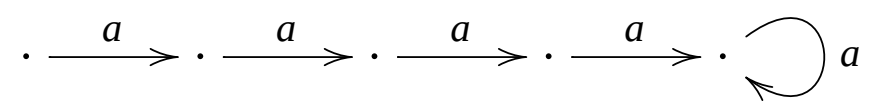

and the subcategory $\mathfrak{C}$ of its transition category generated by

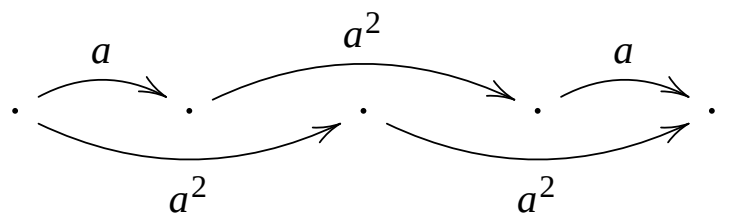

Suppose there is a partially specified automaton $\boldsymbol{X}$ such that $\mathbf{Trans}(\boldsymbol{X}) \cong \mathfrak{C}$. Since the generating arrows are atomic in $\mathfrak{C}$, the same must be true of their counterparts in $\mathbf{Trans}(\boldsymbol{X})$. But this implies that they are one-letter transitions; and the equation between the composition of the three upper arrows and the composition of the two lower arrows would imply that a three-letter word is equal to a two-letter word, which is clearly impossible.

Note that, as it happens for algebraic structures such as groups, a finitely generated category $\mathbf{Trans}(\boldsymbol{X})$ (with "finitely generated" we denote the finiteness of both $X$ and its alphabet $A$) can have (full) subcategories which are not isomorphic to any transition category of a finitely generated, partially specified automaton, even if $\boldsymbol{X}$ is completely specified. For instance, if we consider the following automaton, any partially specified automaton $\boldsymbol{X}$ such that $\mathbf{Trans}(\boldsymbol{X})$ is isomorphic to the full subcategory generated by 1 and 2 must have an infinite alphabet. (In this specific case $\boldsymbol{X}$ can be taken to be fully specified; more generally, whenever the cardinality of the set of letters enabled in $x$ is constant for all $x \in \boldsymbol{X}$, $\boldsymbol{X}$ can be always turned into an isomorphic, fully specified automaton.)

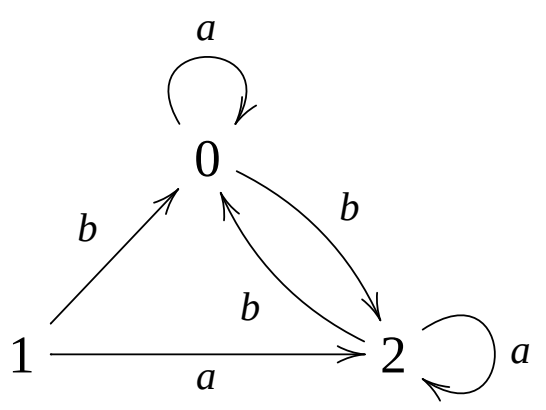

The full subcategory generated by 0 and 2 provides another interesting example, because it is isomorphic to a finitely generated, completely specified automaton. This does not happen in general, as the following example shows:

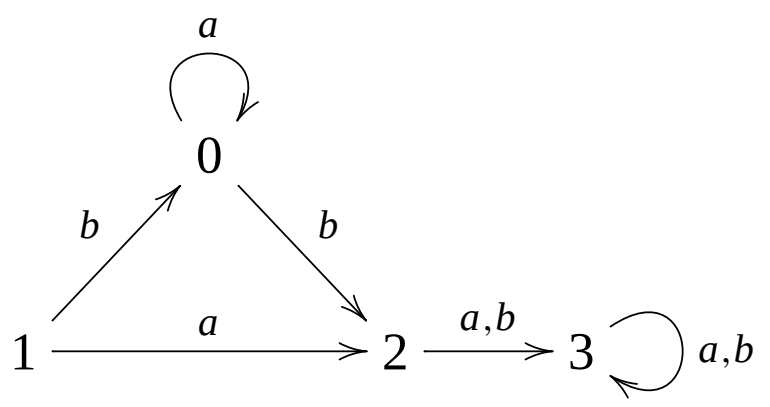

The full subcategory generated by 1 and 2 has as alphabet a countable set whose action is only defined on 1.

## 3.4 A Categorical Note

We would like to mention that what we exposed in this chapter can be embedded into a very general view. In what follows we shall use freely a series of categorical concepts.

It is clear that, given an $M$-automaton $\boldsymbol{X}$, there is a "forget domain and codomain" functor $\mathbf{Trans}(\boldsymbol{X}) \to M$ which sends an arrow $x \xrightarrow{m} x'$ to $m \in M$. The important idea, developed by F. W. Lawvere and A. Kock during seminars held in 1971 [Lawvere] is that this functor is not arbitrary, but it is rather a *discrete opfibration*. Indeed, *every discrete opfibration of the form* $F : \mathfrak{C} \to M$ *gives rise to an* $M$*-automaton* $\boldsymbol{X}$ *such that* $\mathbf{Trans}(\boldsymbol{X}) \cong \mathfrak{C}$. The unique lifting property tell us exactly that given a state (an object $x$ of $\mathfrak{C}$) and an arrow of $M$ (an element $m \in M$) there is a unique arrow going out of $x$ which is projected by $F$ on $m$. Since we have already seen that the output in $N$ of an automaton can be specified by a functor $O_{\boldsymbol{X}} : \mathbf{Trans}(\boldsymbol{X}) \to N$, an input/output automaton can be defined as a *span in* **Cat**, i.e., as a diagram

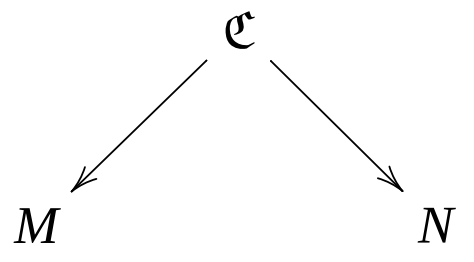

where the "left leg" of the span is a discrete opfibration. If $N = \{\varepsilon\}$, we get an automaton without output. Composition of automata with input/output is then defined using the following standard pullback construction:

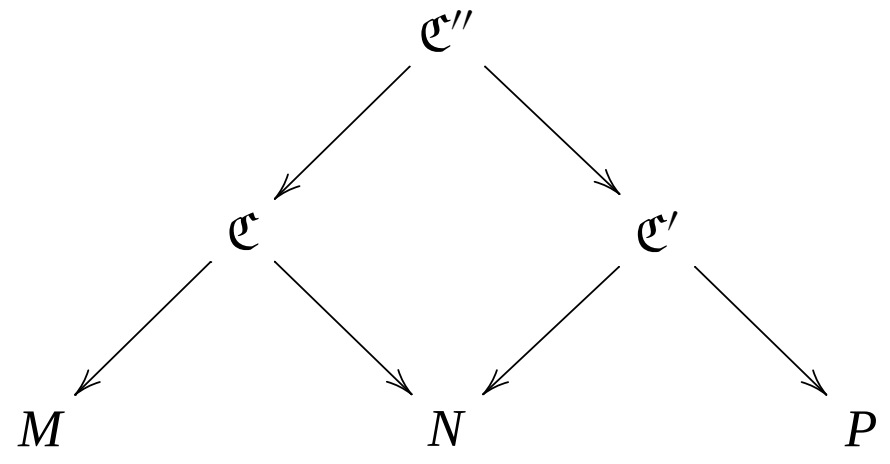

Functors between transition categories which commute with the legs of the respective spans form the 2-cells of a *bicategory of input/output automata*, whose objects are small categories (or monoids) and whose arrows are spans in **Cat** whose left leg is a discrete opfibration.[3] The mappings we defined are a generalization of such 2-cells, because no commutation condition is imposed.

If instead of monoids $M$ and $N$ we consider arbitrary small categories, we have that the generalized action of the "input category" is not defined everywhere, but rather each state of the automaton belongs to a certain class (where the classes are the object fibres of the left leg) on which a certain set of action is defined—i.e., partial specification. The obvious consideration that the identity is a discrete opfibration exhibits any partially specified automaton $\boldsymbol{X}$ as a span

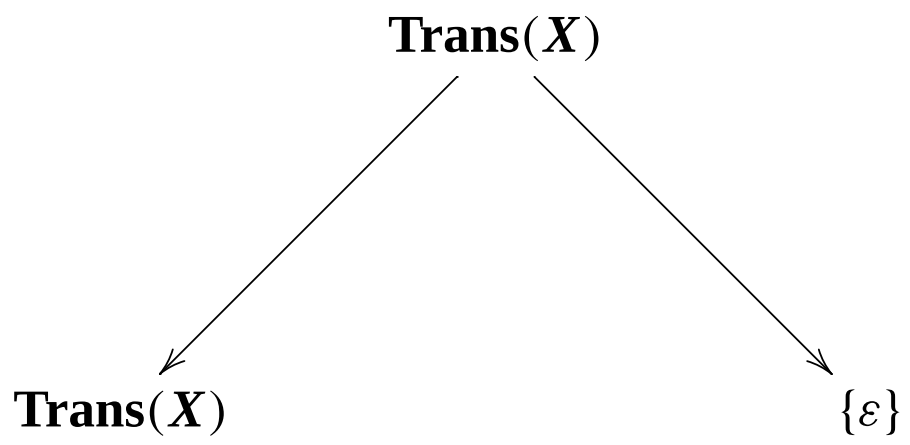

Thus, the restriction to free monoids for partially specified automata finds its natural lifting in the context of automata on *categories*, rather than monoids.

[3] Spans form a bicategory (see [Borceux:1994]), and discrete opfi brations are closed under composition and pullback along an arbitrary functor.

# Chapter 4

# The general theory

In this chapter we are going to describe the theory of distributive computability in its most general form. We shall assume a working knowledge of the basics of category theory (category, functor, equivalence, sum, product, and pullback); suitable references are [Mac Lane:1971] and [Borceux:1994].

The material is divided in three parts: an introduction to distributive categories, which contains a description of the free distributive category on a distributive graph (whose explicit construction and 2-categorical properties are given here for the first time); a review of the basics of extensive categories, including theorems of Carboni, Khalil, Lack and Walters, followed by a series of new results about iteration; finally, a proof of the refinement theorem (see Section 3.2.2) in the context of the automata internal to a given extensive category.

## 4.1 Distributive categories

**Definition 18** A *distributive category* is a category with finite sums (coproducts) and products such that the products distribute over the sums, i.e., such that the canonical arrow

$$\delta = (\mathbf{1}_X \times \mathrm{inj}_1 \mid \mathbf{1}_X \times \mathrm{inj}_2) : X \times Y + X \times Z \to X \times (Y + Z)$$

has an inverse. A *distributive functor* is a functor which preserves finite sums and products.

Distributive categories are essential, in our treatment, because they provide *categories of (free) syntax* for the definition of data types and programs in the language **IMP**$(G)$. We start by giving some examples of data types in a distributive category, and then we introduce the construction of the free distributive category on a distributive graph, which provides the abstract definition for an **IMP**$(G)$ program.

### 4.1.1 Data types

We present briefly here some examples from [Walters:1989a] which show how data types can be described using the syntax of a distributive category. Several more examples can be found in the quoted reference.

#### Stacks

The archetypical example of a data type is that of a *stack* of elements of some type (for instance, integers). Here we shall consider an object $X$, and we shall say that a *stack of $X$* is an object $S$ endowed with specified arrows

$$\text{push} : I + XS \to S$$
$$\text{pop} : S \to I + XS$$

which are inverses. Intuitively, the “push” arrow breaks into two arrows $1 \to S$ and $XS \to S$. The first one creates a new stack (in **Set**, an arrow $1 \to S$ is just an element of $S$), while the second one corresponds to the usual meaning of “pushing data onto an existing stack”. The “pop” arrow can have two outcomes: either $*$, the only element of $I$ (in **Set**), or a stack and an element of type $X$. The fact that the push and the pop arrows are inverse just tell us that popping a new stack gives an error, that pushing and popping gives back the same stack and the element just pushed.

Moreover, it is possible “solve” symbolically the equation $S \cong I + XS$, obtaining $S \cong (I - X)^{-1}$, which once expanded in a power series, gives $S \cong \sum_{k \geq 0} X^k$, which is a possible realization of a stack of type $X$ if our category has countable sums.

It is also important to note that $\sum_{k \geq 0} X^k$ is the formula which gives the free monoid on $X$ (for the definition of monoid in a category and the proof of the formula, see [Mac Lane:1971]). The free monoid is given in **Set** by the sequences of elements of $X$, were composition is concatenation. Thus, the algebraic and the combinatorial view of the data type coincide, in the sense that a possible algebraic realisation of the data type satisfies its equational definition.

**Binary trees**

Another fundamental data type is given by the *binary trees* labelled by $X$. In this case, we have an object $B$ and two arrows

$$\text{break} : B \to I + BXB$$
$$\text{make} : I + BXB \to B$$

which are again inverses. Intuitively, the “make” arrow builds a new binary tree from a label and given left and right subtrees, while the “break” arrow takes a binary tree and gives back its top label and its two subtrees. The properties of binary trees are derivable from the fact that these two arrows are inverses.

Once more, it is possible “solve” symbolically the equation $B \cong I + BXB$, obtaining $XB^2 - B + I \cong \varnothing$, which, once expanded in a power series, gives $B \cong \sum_{k \geq 0} C_k X^k$, where the $C_k$’s are the *Catalan numbers*, i.e., the number of binary trees with $k$ nodes.

We note that, analogously to the previous case, the formula $\sum_{k \geq 0} C_k X^k$ gives the free $X$-motor, an algebraic structure (a monoid with an action of $X$; for definition, properties, and a proof of the formula, see [Kasangian & Vigna:1991b]) which in **Set** is given by the trees labelled by elements of $X$ (composition is collation of the root); the same comments of the previous case apply.

### 4.1.2 The free distributive category on a distributive graph

Recall that a $+/\times$-algebra is an algebra having two constants 0, 1 and two binary operators $+$ and $\times$. A *distributive term* is a term of the free $+/\times$-algebra on $\{x\}$. For a distributive term $t$, we denote by $\varphi_t$ the number of occurrences of generators it contains (for “free places”). Given a set $A$, a distributive term $t$ and a list of $\varphi_t$ elements $a_i \in A$, by substitution (following the natural order inside the term) we obtain a term $t[a_1, a_2, \dots, a_{\varphi_t}]$ of the free $+/\times$-algebra on $A$, i.e., a distributive expression on $A$.

The *topos of distributive graphs* is the presheaf topos $\mathbf{Set}^{\mathfrak{G}}$. The objects of $\mathfrak{G}$ are given by $\{*\} + T^2$, where $T$ is the set of distributive terms. For each object $\langle t, u \rangle$ of $\mathfrak{G}$ there are $\varphi_t + \varphi_u$ arrows named $d_1, d_2, \dots, d_{\varphi_t}, c_1, c_2, \dots, c_{\varphi_u}$ which go from $\langle t, u \rangle$ to $*$.

Given a distributive graph (i.e., a functor) $G : \mathfrak{G} \to \mathbf{Set}$, $G(*)$ is called the set of *objects* of the graph (denoted by $A, B, \ldots$), while the sets $G(\langle t, u\rangle)$ are the sets of *arcs* (denoted by $a, b, \ldots$). Every arc $a \in G(\langle t, u\rangle)$ has an assigned "source" $t[G(d_1)(a), \ldots, G(d_{\varphi_t})(a)]$ and "target" $u[G(c_1)(a), \ldots, G(c_{\varphi_u})(a)]$, which are both distributive expressions on the objects of $G$. In other words, a distributive graph is given by a set of objects $A$ and by a directed graph whose vertices are the distributive expressions on $A$ (note that we admit multiple arcs and loops, following the notation of [Knuth:1973]). Morphisms (natural trasformations) are maps between sets of objects and between sets of arcs which preserve the source and target of an arc (the map on objects inducing by substitution the map on vertices). This shows that the topos $\mathbf{Set}^{\mathfrak{G}}$ is indeed the category of distributive graphs in the sense of [Walters:1992b].

**The essentially equational case.**

In order to keep the notation to a minimum, we shall first describe the construction in the context of distributive categories with chosen sums and products, with functors which strictly preserve the structure (the resulting algebraic theory is thus essentially equational). Then, we shall introduce the formal modifications which are necessary in order to handle the more general case.

If $\mathfrak{D}$ is a distributive category with chosen sums and products (we shall denote in general objects of a category with $X, Y, \ldots$ and arrows with $f, g, \ldots$) and $X_1, X_2, \ldots, X_{\varphi_t}$ are objects of $\mathfrak{D}$, the term $t[X_1, \ldots, X_{\varphi_t}]$ of the free $+/\times$-algebra on the objects of $\mathfrak{D}$ can be interpreted in $\mathfrak{D}$, giving an object $t^{\mathfrak{D}}[\![X_i, \ldots, X_{\varphi_t}]\!] \in |\mathfrak{D}|$. Moreover, if $F : \mathfrak{C} \to \mathfrak{D}$ is a distributive functor (in the sense that it preserves the chosen sums and products), we have

$$F(t^{\mathfrak{C}}[\![X_i, \ldots, X_{\varphi_t}]\!]) = t^{\mathfrak{D}}[\![F(X_i), \ldots, F(X_{\varphi_t})]\!].$$

Let now $\mathbf{Dist}^{\bullet}$ be the category of small distributive categories (with chosen sums and products, and with functors strictly preserving the structure). We describe a forgetful functor $\mathcal{U} : \mathbf{Dist}^{\bullet} \to \mathbf{Set}^{\mathfrak{G}}$ by the following equations:

(i). $\mathcal{U}(\mathfrak{D})(*) = |\mathfrak{D}|$;

(ii). $\mathcal{U}(\mathfrak{D})(\langle t, u\rangle) = \sum_{X_i, Y_j \in |\mathfrak{D}|} \mathfrak{D}(t^{\mathfrak{D}}[\![X_1, \ldots, X_{\varphi_t}]\!], u^{\mathfrak{D}}[\![Y_1, \ldots, Y_{\varphi_u}]\!])$;

(iii). $\mathcal{U}(\mathfrak{D})(d_i)(f : t^{\mathfrak{D}}[\![X_1, \ldots, X_{\varphi_t}]\!] \to u^{\mathfrak{D}}[\![Y_1, \ldots, Y_{\varphi_u}]\!]) = X_i$;

(iv). $\mathcal{U}(\mathfrak{D})(c_j)(f : t^{\mathfrak{D}}[\![X_1, \ldots, X_{\varphi_t}]\!] \to u^{\mathfrak{D}}[\![Y_1, \ldots, Y_{\varphi_u}]\!]) = Y_j$.

It is clear that if $F : \mathfrak{C} \to \mathfrak{D}$ is a distributive functor, it induces a natural transformation $\mathcal{U}(F) : \mathcal{U}(\mathfrak{C}) \to \mathcal{U}(\mathfrak{D})$. What we want to describe here is a left adjoint of $\mathcal{U}$, i.e., we want to build the *free distributive category on the distributive graph $G$*. Such a category will be denoted by $\mathcal{F}(G)$.

The *objects* of $\mathcal{F}(G)$ are given by the distributive expressions on $G(*)$. The *arrows* are given by the terms generated by the following (partial) operators:[1]

(i). if $f_1 : X_1 \to Y,\ f_2 : X_2 \to Y$, then $(f_1 \mid f_2) : X_1 + X_2 \to Y$;

(ii). if $f_1 : Y \to X_1,\ f_2 : Y \to X_2$, then $(f_1, f_2) : Y \to X_1 \times X_2$;

[1]We use the following shorthand: we understand the existence of two operators $d$ and $c$ which give the domain and codomain of an arrow; whenever we write $f : X \to Y$, we mean $d(f) = X$, $c(f) = Y$. When these equations appear as premises, they restrict the applicability of an operator (for instance, composition can by applied only if $c(f) = d(g)$), while when they appear as conclusions, they imply the addition of equations about $c$ and $d$ to the system (for instance, $c(f \circ g) = c(f)$ and $d(f \circ g) = d(g)$). Finally, all "constant" arrows depending on given objects (such as identities) are really operators which accepts objects and output arrows.

(iii). if $f : X \to Y$, $g : Y \to Z$, then $g \circ f : X \to Z$;

starting from the following constants:

(i). for each arc $a$ of $G$,

$$a : t[G(d_1)(a), \ldots, G(d_{\varphi_t})(a)] \to u[G(c_1)(a), \ldots, G(c_{\varphi_u})(a)];$$

(ii). for each object $X$, $\mathbf{1}_X : X \to X$, $!_X : \varnothing \to X$ and $\mathrm{i}_X : X \to I$;

(iii). for each pair of objects $X_1, X_2$, $\mathrm{proj}_i^{X_1,X_2} : X_1 \times X_2 \to X_i$ and $\mathrm{inj}_i^{X_1,X_2} : X_i \to X_1 + X_2$;

(iv). for each triple of objects $X, Y, Z$, $\delta^{-1}_{X,Y,Z} : X \times (Y + Z) \to X \times Y + X \times Z$;

The set of such terms is quotiented by the following equations ($i = 1, 2$):

(i). $(h \circ g) \circ f = h \circ (g \circ f)$;

(ii). $f \circ \mathbf{1}_X = \mathbf{1}_Y \circ f = f$;

(iii). if $f : \varnothing \to X$ then $f = !_X$;

(iv). if $f : X \to I$ then $f = \mathrm{i}_X$;

(v). $\delta_{X,Y,Z} \circ \delta^{-1}_{X,Y,Z} = \mathbf{1}_{(X\times Y)+(X\times Z)}$, and $\delta^{-1}_{X,Y,Z} \circ \delta_{X,Y,Z} = \mathbf{1}_{X\times(Y+Z)}$ ($\delta$ has been described in Definition 18);

(vi). if $f_i : X_i \to Y$, then

$$(f_1 \mid f_2) \circ \mathrm{inj}_i^{X_1,X_2} = f_i, \qquad f = (f \circ \mathrm{inj}_1^{X_1,X_2} \mid f \circ \mathrm{inj}_2^{X_1,X_2});$$

(vii). if $f_i : Y \to X_i$, then

$$\mathrm{proj}_i^{X_1,X_2} \circ (f_1, f_2) = f_i, \qquad f = (\mathrm{proj}_1^{X_1,X_2} \circ f, \mathrm{proj}_2^{X_1,X_2} \circ f).$$

It is straightforward to check that the operators $+$ and $\times$ define a sum and a product in $\mathcal{F}(G)$ for which the distributive law is true, and that there is a chosen sum and product for any pair of objects. Moreover, the sum (product) in $\mathcal{F}(G)$ is exactly the sum (product) of terms, in the sense that

$$t^{\mathcal{F}(G)}[\![A_1, \ldots, A_{\varphi_t}]\!] + u^{\mathcal{F}(G)}[\![B_1, \ldots, B_{\varphi_u}]\!] = (t+u)^{\mathcal{F}(G)}[\![A_1, \ldots, A_{\varphi_t}, B_1, \ldots, B_{\varphi_u}]\!];$$

this implies that $t[A_1, \ldots, A_{\varphi_t}] = t^{\mathcal{F}(G)}[\![A_1, \ldots, A_{\varphi_t}]\!]$.

**Theorem 15** $\mathcal{F}$ is left adjoint to $\mathcal{U}$.

**Proof.** The unit of the adjunction $\eta_G : G \to \mathcal{U}(\mathcal{F}(G))$ is defined by $\eta_G(*)(A) = A$ and $\eta_G(\langle t, u\rangle)(a) = a$; it is trivial to check that $\eta_G$ is indeed a morphism of distributive graphs.

Given a morphism $\alpha : G \to \mathcal{U}(\mathfrak{D})$, suppose the existence of a functor $\widehat{\alpha} : \mathcal{F}(G) \to \mathfrak{D}$ such that $\mathcal{U}(\widehat{\alpha}) \circ \eta_G = \alpha$; then for any term $t$ and any $A_1, \ldots, A_{\varphi_t} \in G(*)$ we have necessarily

$$\widehat{\alpha}(t[A_1, \ldots, A_{\varphi_t}]) = t^{\mathfrak{D}}[\![\widehat{\alpha}(A_1), \ldots, \widehat{\alpha}(A_{\varphi_t})]\!] = t^{\mathfrak{D}}[\![\alpha(*)(A_1), \ldots, \alpha(*)(A_{\varphi_t})]\!];$$

analogously, if for an arc of $a$ of $G$ we have

$$\alpha(\langle t, u\rangle)(a) = f,$$

we have necessarily also

$$\widehat{\alpha}(a : t[G(d_1)(a), \ldots, G(d_{\varphi_t})(a)] \to u[G(c_1)(a), \ldots, G(c_{\varphi_u})(a)]) = f.$$

The other constants have obvious images necessarily induced by the diagram choice, and the image of a generic term can then be obtained interpreting inductively the operators in $\mathfrak{D}$. So $\widehat{\alpha}$ is unique.

On the other hand, using the previous equations we can define by induction from $\alpha$ a functor $\widehat{\alpha}$, because all the equations satisfied by the arrows of $\mathcal{F}(G)$ are also satisfied in $\mathfrak{D}$. It is then trivial to show that $\widehat{\alpha}$ is distributive. This completes the proof. ▮

**The 2-categorical view.**

We now drop the condition about the presence of a choice of sums and products. Our treatment will follow the ideas described in [Walters:1989b].

Note first of all that a distributive expression on $A$ can be thought of as a binary tree whose internal nodes are labelled by $+$ and $\times$, and whose leaves are labelled by 0, 1 or by an element of $A$.

If $\mathfrak{D}$ is a distributive category and $X_1, X_2, \ldots, X_{\varphi_t}$ are objects of $\mathfrak{D}$, there will be in general many objects $X$ of $\mathfrak{D}$ which interpret the distributive expression $t[X_1, \ldots, X_{\varphi_t}]$, in the sense that each such object is equipped with with a suitable choice of products and coproducts diagrams related to the operations described by $t$. The injections and projections of such diagrams can be thought of as labelling the arcs of $t$, in such a way that a node labelled by $+$ ($\times$, respectively) is linked to its children by arcs labelled by the suitable injections (projections, respectively). This labelled tree will be denoted by $t^{\mathfrak{D}}_X\{X_1, \ldots, X_{\varphi_t}\}$: it essentially "witnesses" how $X$ has been built using the $X_i$'s. Note that there is a degree of ambiguity here, in that there can be several different labellings which adapt to the given data; when using this notation, we shall always have in mind a specific tree, which will be clear from the context. Given trees $t^{\mathfrak{D}}_X\{X_1, \ldots, X_{\varphi_t}\}$, $t^{\mathfrak{D}}_Y\{Y_i, \ldots, Y_{\varphi_t}\}$, and arrows $f_i : X_i \to Y_i$, we shall write $t^{\mathfrak{D}}_{X,Y}[\![f_1, \ldots, f_{\varphi_t}]\!]$ for the obvious arrow from $X$ to $Y$ built from the $f_i$'s using the operations described in $t$.

If $F : \mathfrak{C} \to \mathfrak{D}$ is a distributive functor, it can be "applied" to a tree $t^{\mathfrak{C}}_X\{X_1, \ldots, X_{\varphi_t}\}$ by taking the image through $F$ of all labels (i.e., injections and projections) and all objects. The resulting tree witnesses the construction in $\mathfrak{D}$ of $F(X)$ by means of the $F(X_i)$'s. Finally, we note that given trees $t^{\mathfrak{D}}_X\{X_1, \ldots, X_{\varphi_t}\}$ and $t^{\mathfrak{D}}_Y\{X_1, \ldots, X_{\varphi_t}\}$, there is exactly one isomorphism from $X$ to $Y$ which commutes with all the diagrams of the two trees, in the appropriate sense.

Let now **Dist** be the 2-category of small distributive categories, and consider the 2-category $\mathbf{Cat}(\mathbf{Set}^{\mathfrak{G}}) \simeq \mathbf{Cat}^{\mathfrak{G}}$ of the internal categories of $\mathbf{Set}^{\mathfrak{G}}$. Then, $\mathcal{U} : \mathbf{Dist} \to \mathbf{Cat}(\mathbf{Set}^{\mathfrak{G}})$ is a forgetful 2-functor defined by:

(i). $\mathcal{U}(\mathfrak{D})(*) = \mathfrak{D}$;

(ii). $\mathcal{U}(\mathfrak{D})(\langle t, u\rangle)$ is a category having as objects triples

$$\langle f : X \to Y, t^{\mathfrak{D}}_X\{X_1, \ldots, X_{\varphi_t}\}, u^{\mathfrak{D}}_Y\{Y_1, \ldots, Y_{\varphi_u}\}\rangle;$$

arrows (to a "primed" triple) are $\varphi_t + \varphi_u$-tuples of arrows $f_i : X_i \to X'_i$, $g_j : Y_j \to Y'_j$ of $\mathfrak{D}$ which satisfy

$$u^{\mathfrak{D}}_{Y,Y'}[\![g_1, \ldots, g_{\varphi_u}]\!] \circ f = f' \circ t^{\mathfrak{D}}_{X,X'}[\![f_1, \ldots, f_{\varphi_t}]\!];$$

(iii). $\mathcal{U}(\mathfrak{D})(d_i)(f : X \to Y, t_X^{\mathfrak{D}}\{X_1, \dots, X_{\varphi_t}\}, u_Y^{\mathfrak{D}}\{Y_1, \dots, Y_{\varphi_u}\}) = X_i$, and

$$\mathcal{U}(\mathfrak{D})(c_j)(f : X \to Y, t_X^{\mathfrak{D}}\{X_1, \dots, X_{\varphi_t}\}, u_Y^{\mathfrak{D}}\{Y_1, \dots, Y_{\varphi_u}\}) = Y_j;$$

analogously on arrows.

It is clear that if $F : \mathfrak{C} \to \mathfrak{D}$ is a distributive functor, it induces a functor $\mathcal{U}(F) : \mathcal{U}(\mathfrak{C}) \to \mathcal{U}(\mathfrak{D})$ ($\mathcal{U}(F)(*) = F$; for the rest, just apply $F$ to every element of a triple and to every arrow of a tuple). The same happens for natural transformations.

Let us define the "unit of the adjunction" $\eta_G : G \to \mathcal{U}(\mathcal{F}(G))$ by $\eta_G(*)(A) = A$ and

$$\eta_G(\langle t, u\rangle)(a) = \langle a : t^{\mathcal{F}(G)}[\![A_1, \dots, A_{\varphi_t}]\!] \to u^{\mathcal{F}(G)}[\![B_1, \dots, B_{\varphi_u}]\!], \\ t^{\mathcal{F}(G)}\{A_1, \dots, A_{\varphi_t}\}, u^{\mathcal{F}(G)}\{B_1, \dots, B_{\varphi_u}\}\rangle,$$

where $A_i = G(d_i)(a)$, $B_j = G(c_j)(a)$ and the trees are given by the chosen sums and products in $\mathcal{F}(G)$. The universal property we are interested in is

**Theorem 16** $\mathcal{U}(-) \circ \eta_G$ is an equivalence between $\mathbf{Dist}(\mathcal{F}(G), \mathfrak{D})$ and $\mathbf{Cat}(\mathbf{Set}^{\mathfrak{G}})(G, \mathcal{U}(\mathfrak{D}))$.

**Proof.** Given a functor $F : G \to \mathcal{U}(\mathfrak{D})$ (note that we see $G$ as a discrete internal category of $\mathbf{Set}^{\mathfrak{G}}$), we define a functor $\widehat{F} : \mathcal{F}(G) \to \mathfrak{D}$ as follows: by choosing for each pair of objects $X, Y$ of $\mathfrak{D}$ a product and a coproduct diagram, we can define for any term $t$ and any $A_1, \dots, A_{\varphi_t} \in G(*)$

$$\widehat{F}(t[A_1, \dots, A_{\varphi_t}]) = t^{\mathfrak{D}}[\![F(*)(A_1), \dots, F(*)(A_{\varphi_t})]\!];$$

analogously, if for an arc of $a$ of $G$ we have

$$F(\langle t, u\rangle)(a) = \langle f : X \to Y, \\ t_X^{\mathfrak{D}}\{F(*)(A_1), \dots, F(*)(A_{\varphi_t})\}, u_Y^{\mathfrak{D}}\{F(*)(B_1), \dots, F(*)(B_{\varphi_u})\}\rangle, \quad (4.1)$$

where by functoriality again $A_i = G(d_i)(a)$ and $B_j = G(c_j)(a)$, we set

$$\widehat{F}(a) = \beta_a \circ f \circ \alpha_a : \widehat{F}(t[A_1, \dots, A_{\varphi_t}]) \to \widehat{F}(u[B_1, \dots, B_{\varphi_u}]),$$

where $\alpha_a : \widehat{F}(t[A_1, \dots, A_{\varphi_t}]) \to X$ and $\beta_a : Y \to \widehat{F}(u[B_1, \dots, B_{\varphi_u}])$ denote the unique isomorphisms which commute with the diagrams. The other constants have obvious images induced by the diagram choice, and the image of a general term can then be obtained interpreting inductively the operators in $\mathfrak{D}$.

In order to show that this association induces an equivalence, we note that $\mathcal{U}(\widehat{F}) \circ \eta_G$ is given by $(\mathcal{U}(\widehat{F}) \circ \eta)(*)(A) = F(*)(A)$, and by

$$(\mathcal{U}(\widehat{F}) \circ \eta)(\langle t, u\rangle)(a) = \langle \beta_f \circ f \circ \alpha_f, \\ t^{\mathfrak{D}}\{F(*)(A_1), \dots, F(*)(A_{\varphi_t})\}, u^{\mathfrak{D}}\{F(*)(B_1), \dots, F(*)(B_{\varphi_u})\}\rangle,$$

where $f : X \to Y$ is the first component of $F(\langle t, u\rangle)(a)$; then, the isomorphisms $\alpha_{(-)}$ and $\beta_{(-)}$ induce a natural isomorphism $F \cong \mathcal{U}(\widehat{F}) \circ \eta$.

The last step is proving that $\mathcal{U}(-) \circ \eta_G$ is full and faithful. Given functors $F, F' : G \to \mathcal{U}(\mathfrak{D})$, a natural transformation $\psi : F \Rightarrow F'$ is a family of arrows indexed by the objects and by the arcs of $G$ in such a way that for each object there is a morphism $\psi_A : F(*)(A) \to F'(*)(A)$ of $\mathfrak{D}$, and for each arc $a \in G(\langle t, u\rangle)$ there is a morphism

$$\psi_a : F(a) \to F'(a),$$

which amounts (using the notation of (4.1), and "priming" the tuple for $F'(a)$) to a $\varphi_t + \varphi_u$-tuple of morphisms $\psi_a^{A_i} : F(*)(A_i) \to F'(*)(A_i)$, $\psi_a^{B_j} : F(*)(B_j) \to F'(*)(B_j)$ satisfying

$$u^{\mathfrak{D}}_{Y,Y'}[\![\psi_a^{A_1}, \ldots, \psi_a^{A_{\varphi_t}}]\!] \circ f = f' \circ t^{\mathfrak{D}}_{X,X'}[\![\psi_a^{B_1}, \ldots, \psi_a^{B_{\varphi_u}}]\!].$$

But such a family of tuples is exactly a natural transformation $\mathcal{U}(F) \circ \eta_G \Rightarrow \mathcal{U}(F') \circ \eta_G$ (the last equation is the commutativity condition for all the basic arrows; the condition extends immediately to a generic arrow by structural induction). ∎

### 4.1.3 The language IMP($G$)

Given a distributive graph $G$, an (isolated) program in the language **IMP**($G$), introduced by Walters in [Walters:1992b], is simply an arrow

$$f : X \to X$$

of the free distributive category on $G$. The arcs and the objects of $G$ are used in order to specify the signature of the basic operations allowed in an **IMP**($G$) program (for instance, the graph of recursive functions has set of objects $\{\boldsymbol{N}\}$ and two arcs $p : \boldsymbol{N} \to I + \boldsymbol{N}$ and $s : I + \boldsymbol{N} \to \boldsymbol{N}$). Then, the free distributive category on $G$ provides the abstract syntax of all **IMP**($G$) programs. The notation used in the previous section provides instead a concrete syntax, and it will be used in Chapter 6 when describing the formal tools which allow to manipulate **IMP**($G$) programs.

Given a distributive category $\mathfrak{D}$, a distributive graph morphism $\alpha : G \to \mathcal{U}(\mathfrak{D})$ is defined by assigning to each object of $G$ an object of $\mathfrak{D}$ (these are the *basic objects* of Chapter 2) and to each arc of $G$ an arrow of $\mathfrak{D}$ (these are the *basic functions*). Then, the morphism extends to a distributive functor $\mathcal{F}(G) \to \mathfrak{D}$ which allows to interpret the arrows of $\mathcal{F}(G)$ in $\mathfrak{D}$. The image of an arrow of $\mathcal{F}(G)$ is a *derived function*, because it is obtained by interpreting in $\mathfrak{D}$ a term built using exactly the constants and operations described in Section 2.1.

Following the same lines of the elementary introduction given in Section 2.1, it is now natural to consider arrows of the form $f : X + U \to U + Y$, and the possibility of iterating such arrows. However, in a distributive category we have no way of expressing the property that the iteration of such an arrow *terminates*, giving as result an arrow $X \to Y$. This is the motivation for the introduction of *extensive categories*.

## 4.2 Extensive categories

**Definition 19** A category is said to be *extensive* if the canonical functor

$$\mathfrak{E}/A \times \mathfrak{E}/B \to \mathfrak{E}/(A + B)$$

is an equivalence, for each pair of objects $A, B$ in $\mathfrak{E}$.

Countably extensive categories (i.e., categories which have countable sums and enjoy the extensivity property for countable sums) provide *semantical universes* in which we can interpret a free distributive category. The peculiarity of countable extensivity is that it allows to define what is an *arrow computed by iteration*.

In an extensive category we can restrict and corestrict at the same time arrows which land into a sum. If we have an arrow $f : C \to A + B$, we can pull $f$ back along the injections of $A$ and $B$ into $A + B$ and obtain arrows $f_{\lceil A} : f^{-1}(A) \to A$ and $f_{B\rceil} : f^{-1}(B) \to B$ such that $f_{\lceil A} + f_{B\rceil} = f$. In other words, whenever we have an arrow into a sum, we can break the source of the arrow into the part which maps into the first summand, and the part which maps into the second summand (we shall use the set-like notation $f^{-1}(A)$ for the part of the domain which lands in $A$; if $X \cong A+B \cong A'+B'$ we shall use the notation $A \cap A'$ for the pullback of the injection of $A$ along the injection of $A'$).

This property derives from the following theorem [Carboni, Lack & Walters:1993]:

**Theorem 17** A category with finite sums is extensive iff it has pullbacks along injections, and for each commutative diagram

$$\begin{array}{ccccc} f^{-1}(A) & \longrightarrow & C & \longleftarrow & f^{-1}(B) \\ \downarrow & & \downarrow{\scriptstyle f} & & \downarrow \\ A & \underset{\mathrm{inj}_1}{\longrightarrow} & A+B & \underset{\mathrm{inj}_2}{\longleftarrow} & B \end{array}$$

the top row is a coproduct diagrams exactly when the two squares are pullbacks.

Another important property of extensive categories we shall use is the *cancellation lemma* [Lack:1995]:

**Theorem 18** If $A + B \cong A + C$, by an isomorphism commuting with the injections of $A$, then $B \cong C$.

Consider now an arrow $f : X + U \to U + Y$ in a countably extensive category. Intuitively, $f$ describes a dynamical system with initial states in $X$. By applying $f$, we obtain either a final state in $Y$, or a *local* state in $U$; in the latter case, we can apply again $f$. We can define the *arrow which $f$ computes by iteration*, denoted by **call**[ $X, U, Y, f$ ] (or **call**[ $f$ ], if $X$, $U$ and $Y$ are clear from the context), as follows: let

$$f_0 = f_{|X} : X \to U + Y$$
$$f_{k+1} = f_{|U} \circ f_{k\lceil U} : f_k^{-1}(U) \to U + Y.$$

Then $f_k^{-1}(Y)$ is the part of $X$ which lands in $Y$ after exactly $k+1$ iterations. The iteration of $f$ *terminates* if $\sum_{k\geq 0} f_k^{-1}(Y) \cong X$ (where the isomorphism is given by the injections $f_k^{-1}(Y) \to X$), in which case we can define **call**[ $f$ ] : $X \to Y$ as

$$\mathbf{call}[\, f\,] = \left(f_{0Y\rceil} \mid f_{1Y\rceil} \mid \cdots \mid f_{kY\rceil} \mid \cdots\right) = \nabla \circ \sum_{k\geq 0} f_{kY\rceil}.$$

This framework allows us to define, for any subcategory with finite sums $\mathfrak{C}$ of a countably extensive category $\mathfrak{E}$, the (for the time being a) set **call**[ $\mathfrak{C}$ ] of the *arrows computable by iteration in* $\mathfrak{C}$, which is formed by the arrows computed by arrows $f : X + U \to U + Y$ (in $\mathfrak{C}$). The normal form theorem can be now restated as follows:

**Theorem 19** Let $\mathfrak{E}$ be a countably extensive category, and $\mathfrak{C}$ a subcategory of $\mathfrak{E}$ with finite sums.

(i). For each terminating arrow[2] $f : X \to Y$ of $\mathfrak{C}$,

$$f = \mathbf{call}[\, \mathrm{inj}_2 \circ (f \mid \mathbf{1}_Y) : X + \varnothing \to \varnothing + Y\,].$$

(ii). If $f : X + U \to U + Y$ and $g : Y + V \to V + Z$, then

$$\mathbf{call}[\, f\,] \circ \mathbf{call}[\, g\,] = \mathbf{call}[\, X, U + Y + V, Z, f + g.\,]$$

[2]Note that in the future we shall avoid to quote explicitly the termination condition; each time **call**[ − ] will be used, we shall understand the termination condition for the arrows it is applied to.

(iii). If $f : X + U \to U + Y$ and $g : X' + U' \to U' + Y'$, then

$$\mathbf{call}[\, f\,] + \mathbf{call}[\, g\,] = \mathbf{call}[\, a \circ (f + g) \circ b\,]$$

where $a$ is the commutativity isomorphism

$$X + X' + U + U' \stackrel{\cong}{\to} X + U + X' + U',$$

and $b$ is the commutativity isomorphism

$$U + Y + U' + Y' \stackrel{\cong}{\to} U + U' + Y + Y'.$$

(iv). If $f : X + V + U \to U + V + Y$, then

$$\mathbf{call}[\, X, V, Y, \mathbf{call}[\, X + V, U, V + Y, f\,]\,] = \mathbf{call}[\, X, U + V, Y, i \circ f\,] : X \to Y,$$

where $i$ is the commutativity isomorphism

$$U + V + Y \to V + U + Y.$$

Moreover, if $\mathfrak{E}$ has finite products, $\mathfrak{C}$ is distributive, $f : X + U \to U + Y$ and $g : X' + U' \to U' + Y'$, then

$$\mathbf{call}[\, f\,] \times \mathbf{call}[\, g\,] = \mathbf{call}[\, d \circ (f_{|X} \times g_{|X'} \mid f_{|U} \times g_{|U'} \mid f_{|U} \times \mathrm{inj}_2^{U',Y'} \mid \mathrm{inj}_2^{U,Y} \times g_{|U'})\,],$$

where $d$ is the isomorphism

$$(U + Y) \times (U' + Y') \stackrel{\cong}{\to} UU' + UY' + YU' + YY'$$

induced by distributivity.

We can interpret the previous claims as follows:

(i). $\mathbf{call}[\,\mathfrak{C}\,]$ contains the arrows of $\mathfrak{C}$;

(ii). $\mathbf{call}[\,\mathfrak{C}\,]$ is closed under composition (thus, it is a subcategory of $\mathfrak{E}$ which contains $\mathfrak{C}$);

(iii). $\mathbf{call}[\,\mathfrak{C}\,]$ has finite sums;

(iv). $\mathbf{call}[\,\mathbf{call}[\,\mathfrak{C}\,]\,] = \mathbf{call}[\,\mathfrak{C}\,]$ (i.e., $\mathbf{call}[\,-\,]$ is idempotent).

Moreover, if $\mathfrak{E}$ has finite products and $\mathfrak{C}$ is distributive then $\mathbf{call}[\,\mathfrak{C}\,]$ is distributive. The proof of the normal form theorem for extensive categories has been given (in terms of functional processors) in [Khalil:1992b], [Khalil:1992a] and [Khalil]. We shall refer from now to the category $\mathbf{call}[\,\mathfrak{C}\,]$ as to the *iterative closure of* $\mathfrak{C}$ *in* $\mathfrak{E}$.

### 4.2.1 Functions and functional processors

The way in which a terminating arrow and the relative computed arrow have been defined here differs from the way it was defined in [Wagner, Khalil & Walters:1995] and [Sabadini, Vigna & Walters:1996]. The form described here is simpler to manage in equivalence proofs, and gives nicer formulae in some (but not all) parts of the normal form theorem.

The two definitions are of course equivalent, but we wish to formulate this fact more precisely. Given a prefunctional processor $\varphi : X + U + Y \to X + U + Y$, i.e., a loop which satisfies $\varphi \circ$ $\mathrm{inj}_2^{X+U,Y} = \mathbf{1}_Y$ and which factors[3] through the injection $U + Y \to X + U + Y$, we can obtain an arrow $\psi : X + U \to U + Y$ by factoring and restricting to $X + U$. On the other hand, given an arrow $f : X + U \to U + Y$, we shall call

$$h : X + U + Y \xrightarrow{(f \mid \mathrm{inj}_2^{U,Y})} U + Y \xrightarrow{\mathrm{inj}_2^{X,U+Y}} X + U + Y$$

the *loop induced by* $f$, which is trivially a prefunctional processor.

**Theorem 20** The prefunctional processor $h : X + U + Y \to X + U + Y$ associated to a terminating arrow $f : X + U \to U + Y$ is a functional processor in the sense of [Wagner, Khalil & Walters:1995]; moreover, both $h$ and $f$ compute the same arrow.

**Proof.** We set up some notation following [Wagner, Khalil & Walters:1995]. Let $i$, $j$ and $k$ be the injections of $X$, $U$ and $Y$, respectively, in $X + U + Y$ (the fact that $k$ is also a summation index should cause no confusion). The following pullbacks

$$\begin{array}{ccccc}
A_n & \xrightarrow{a_n} & X_n & \xrightarrow{h_n} & Y \\
{\scriptstyle a'_n}\downarrow & & \downarrow{\scriptstyle i_{X_n}} & & \downarrow{\scriptstyle k} \\
X'_{n-1} & \xrightarrow[i_{X'_{n-1}}]{} & X & \xrightarrow[h^n \circ i]{} & X + U + Y
\end{array}$$

where $X_n = (h^n \circ i)^{-1}(Y)$ and $X'_n = (h^n \circ i)^{-1}(X + U)$ (so $X_n + X'_n \cong X$) exhibit

$$A_n = (h^n \circ i)^{-1}(Y) \cap (h^{n-1} \circ i)^{-1}(X + U)$$

as the part of $X$ which lands in $Y$ after exactly $n$ iterations. The arrow $h$ is a functional processor iff the arrows $i_{X_n} \circ a_n : A_n \to X$ form a coproduct diagram. Then, the arrow computed by iteration by $h$ is defined by the arrows $h_n \circ a_n$.

Since $f$ is a terminating arrow, we have a decomposition $\sum_{k \geq 0} f_k^{-1}(Y) \cong X$ induced by the iteration of $f$, and $f_n^{-1}(U) \cong \sum_{k>n} f_k^{-1}(Y)$ by the cancellation lemma. The idea behind the proof is that the correspondence between the iteration of $f$ and the iteration of $h$ is described by the following equations:

(i). $A_{n+1} \cong f_n^{-1}(Y)$;

(ii). $X_{n+1} \cong \sum_{k=0}^{n} f_k^{-1}(Y)$;

(iii). $X'_{n+1} \cong \sum_{k>n} f_k^{-1}(Y)$;

(iv). $h_{n+1} = (f_{0Y\urcorner} \mid f_{1Y\urcorner} \mid \cdots \mid f_{nY\urcorner})$.

If we can prove that these equations hold, clearly $h$ will be a functional processor computing the arrow **call**[ $f$ ].

Since $X_0 = \varnothing$ and $X'_0 = X$, for $n = 0$ we have

$$X_1 = (h \circ i)^{-1}(Y) = h_{|X}^{-1}(Y) = f_{|X}^{-1}(Y) = f_0^{-1}(Y),$$

[3]The factoring condition is not imposed in [Wagner, Khalil & Walters:1995], but it will be essential when discussing the inversion of iteration through refinement, and it can be imposed without loss of generality; for if a prefunctional processor does not factor through $U + Y$, one can enlarge the local state space $U$ to $V = X + U$, injecting in the obvious manner the input state space $X$ into the new local state space $V$ without affecting the computed arrow. The resulting loop on $X + V + Y$ obviously factors through the injection $V + Y \to X + V + Y$.

so $A_1 = X_1 = f_0^{-1}(Y)$. On the other hand,

$$X_1' = (h \circ i)^{-1}(X+U) = (\mathrm{inj}_2^{X,U+Y} \circ f_{|X})^{-1}(X+U) = f_0^{-1}(U) \cong \sum_{k>0} f_k^{-1}(Y).$$

Now a tedious but trivial diagram chasing argument shows that the following diagram commutes for $n > 0$:

$$\begin{array}{ccccc}
\sum_{k>n} f_k^{-1}(Y) & \longrightarrow & X & \longleftarrow & \sum_{k=0}^{n} f_k^{-1}(Y) \\
{\scriptstyle \mathrm{inj}_2^{X,U} \circ f_{n\lceil U}} \downarrow & & \downarrow {\scriptstyle h^{n+1} \circ i} & & \downarrow {\scriptstyle (f_{0Y\rceil} \mid f_{1Y\rceil} \mid \cdots \mid f_{nY\rceil})} \\
X+U & \underset{(i \mid j)}{\longrightarrow} & X+U+Y & \underset{k}{\longleftarrow} & Y
\end{array}$$

But the top row is by definition a coproduct diagram. By Theorem 17 the two squares are pullbacks. Thus, also the following diagram

$$\begin{array}{ccccc}
f_n^{-1}(Y) & \longrightarrow & \sum_{k=0}^{n} f_n^{-1}(Y) & \xrightarrow{(f_{0Y\rceil} \mid \cdots \mid f_{nY\rceil})} & Y \\
\downarrow & & \downarrow & & \downarrow {\scriptstyle k} \\
\sum_{k\geq n} f_k^{-1}(Y) & \longrightarrow & X & \underset{h^{n+1} \circ i}{\longrightarrow} & X+U+Y
\end{array}$$

(where all the unspecified arrows are the obvious injections) is formed by two pullback squares (being $f_n^{-1}(Y) = \sum_{k=0}^{n} f_n^{-1}(Y) \cap \sum_{k\geq n} f_k^{-1}(Y)$). This completes the proof. ▮

Of course, also the following holds:

**Theorem 21** The arrow $\psi : X+U \to U+Y$ associated to a functional processor $\varphi : X+U+Y \to X+U+Y$ by factorization through $U+Y$ and restriction to $X+U$ is terminating, and both $\psi$ and $\varphi$ compute the same arrow.

**Proof.** Since $\varphi$ is the loop induced by $\psi$, we have that, as is the previous theorem, the following diagram commutes for all $n \geq 0$

$$\begin{array}{ccccc}
\psi_n^{-1}(U) & \longrightarrow & X & \longleftarrow & \sum_{k=0}^{n} \psi_k^{-1}(Y) \\
{\scriptstyle \mathrm{inj}_2^{X,U} \circ \psi_{n\lceil U}} \downarrow & & \downarrow {\scriptstyle \varphi^{n+1} \circ i} & & \downarrow {\scriptstyle (\psi_{0Y\rceil} \mid \psi_{1Y\rceil} \mid \cdots \mid \psi_{nY\rceil})} \\
X+U & \underset{(i \mid j)}{\longrightarrow} & X+U+Y & \underset{k}{\longleftarrow} & Y
\end{array}$$

and being the top row a coproduct diagram, the two squares are pullbacks (note that we cannot claim that $U \cong \sum_{k>n} \psi_k^{-1}(Y)$, for $\psi$ is not known to be terminating). But $\varphi$ is a functional processor, and this implies that the following two squares are pullbacks for all $n > 0$:

$$\begin{array}{ccccc}
\sum_{k>n} A_k & \longrightarrow & X & \longleftarrow & \sum_{k=1}^{n} A_k \\
\downarrow & & \downarrow {\scriptstyle \varphi^{n} \circ i} & & \downarrow {\scriptstyle \varphi_n} \\
X+U & \underset{(i \mid j)}{\longrightarrow} & X+U+Y & \underset{k}{\longleftarrow} & Y
\end{array}$$

which implies $A_n \cong \psi_{n-1}^{-1}(Y)$ and $\varphi_n = (\psi_{0Y\rceil} \mid \cdots \mid \psi_{n-1Y\rceil})$. ▮

### 4.2.2 The semantics of IMP($G$) programs

As it is shown in [Carboni, Lack & Walters:1993], an extensive category with finite products is distributive. Thus, given a distributive graph $G$ and a countably extensive category with finite products $\mathfrak{E}$, as in Section 4.1.3 a distributive graph morphism $\alpha : G \to \mathcal{U}(\mathfrak{E})$ extends to the free distributive category on $G$, resulting in a semantic assignment for all **IMP**($G$) programs. However, we have now the possibility to close by iteration the image of $\mathcal{F}(G)$ in $\mathfrak{E}$. In this way we assign an arrow of $\mathfrak{E}$ to all arrows computed by iteration; moreover, by the normal form theorem the extension of $G$ and $\alpha$ by such an arrow does not change the iterative closure.

### 4.2.3 Technical lemmata

**Lemma 4** Let $\mathfrak{E}$ and $\mathfrak{F}$ be a extensive categories, and $S : \mathfrak{E} \to \mathfrak{F}$ a functor which preserves finite sums. Then $S$ preserves pullbacks along injections.

**Proof.** If we have a pullback of $f : A \to B + C$ along an injection (of $B$ or $C$), we can get the following diagram by pulling back along the other injection:

$$\begin{array}{ccccc}
f^{-1}(B) & \longrightarrow & A & \longleftarrow & f^{-1}(C) \\
\big\downarrow{\scriptstyle f_{\restriction B}} & & \big\downarrow{\scriptstyle f} & & \big\downarrow{\scriptstyle f_{C\rceil}} \\
B & \xrightarrow[\mathrm{inj}_1]{} & B + C & \xleftarrow[\mathrm{inj}_2]{} & C
\end{array}$$

and then by extensivity the top row is a coproduct diagram. But $S$ preserves finite sums, so that the following diagram

$$\begin{array}{ccccc}
S(f^{-1}(B)) & \longrightarrow & S(A) & \longleftarrow & S(f^{-1}(C)) \\
\big\downarrow{\scriptstyle S(f_{\restriction B})} & & \big\downarrow{\scriptstyle S(f)} & & \big\downarrow{\scriptstyle S(f_{C\rceil})} \\
S(B) & \xrightarrow[\mathrm{inj}_1]{} & S(B) + S(C) & \xleftarrow[\mathrm{inj}_2]{} & S(C)
\end{array}$$

has a coproduct diagram as first row. As a consequence, the two squares are pullbacks, and $S(f)_{\restriction S(B)} \cong S(f_{\restriction B})$. ∎

**Definition 20** An arrow $f : X + U \to U + Y$ of an extensive category $\mathfrak{E}$ is said to be *finitely iterable* if ultimately $f_k^{-1}(Y) \cong \varnothing$. Given a subcategory with sums $\mathfrak{C}$ of $\mathfrak{E}$, $\mathfrak{C}$ is said to be finitely iterable in $\mathfrak{E}$ if all its arrows are finitely iterable.

**Lemma 5** If the free distributive category generated by a distributive graph is extensive, then it is finitely iterable within itself.

**Proof.** We have to prove that, given an arrow $f : X + U \to U + Y$ of $\mathcal{F}(G)$, ultimately $f_k^{-1}(Y) \cong \varnothing$. Let $v : |\mathcal{F}(G)| \to \boldsymbol{N}$ be the $+/\times$-algebra morphism which sends each object of $G$ to 1. Then $v(X) = v(Y)$ for all isomorphic $X, Y \in |\mathcal{F}(G)|$, and $v(X) = 0$ iff $X \cong \varnothing$.

Since $\sum_{0 \le k \le n} f_k^{-1}(Y)$ is a complemented subobject of $X$ for any $n$ and the only sums in a free distributive category are the syntactic ones, the upper bound

$$\sum_{0 \le k \le n} v(f_k^{-1}(Y)) = v\Bigg(\sum_{0 \le k \le n} f_k^{-1}(Y)\Bigg) \le v\Bigg(\sum_{0 \le k \le n} f_k^{-1}(Y)\Bigg) + v(f_n^{-1}(U)) = v(X)$$

implies that $f_k^{-1}(Y) \not\cong \varnothing$ only for a finite number of $k$'s. ∎

Whenever $\mathfrak{C}$ is finitely iterable, we can define the iterative closure **call**[ $\mathfrak{C}$ ] even if $\mathfrak{E}$ does not possess countable sums, because the infinite sums appearing in the definition of **call**[ − ] all reduce to finite sums. This is particularly important in the case of the previous lemma, because then $\mathfrak{C} = \mathfrak{E}$, and then obviously **call**[ $\mathfrak{E}$ ] = $\mathfrak{E}$, where the iterative closure is taken within $\mathfrak{E}$ itself.

We now study the effect of sum preserving functors on iterative closures.

**Lemma 6** Let $\mathfrak{E}$ and $\mathfrak{F}$ be extensive categories, $S : \mathfrak{E} \to \mathfrak{F}$ a functor which preserves finite sums, and $f : X + U \to U + Y$ a terminating, finitely iterable arrow. Then $S(f)$ is a terminating, finitely iterable arrow, and $S(\textbf{call}[\, X, U, Y, f\,]) = \textbf{call}[\, S(X), S(U), S(Y), S(f)\,]$. If $\mathfrak{E}, \mathfrak{F}$ are countably extensive and $S$ preserves countable sums the statement is true for any terminating arrow.

**Proof.** Let us prove that $S(f_k) = S(f)_k$. Clearly $S(f_0) = S(f)_0$. By induction,

$$S(f)_{k+1} = S(f)_{|S(U)} \circ S(f)_{k\lceil S(U)} = S(f_{|U}) \circ S(f_{k\lceil U}) = S(f_{|U} \circ f_{k\lceil U}) = S(f_{k+1}).$$

Then

$$\begin{aligned} S(\textbf{call}[\, f\,]) &= S\left(f_{0Y\rceil} \mid f_{1Y\rceil} \mid \cdots \mid f_{kY\rceil} \mid \cdots\right) \\ &= \left(S(f_{0Y\rceil}) \mid S(f_{1Y\rceil}) \mid \cdots \mid S(f_{kY\rceil}) \mid \cdots\right) \\ &= \left(S(f_0)_{S(Y)\rceil} \mid S(f_1)_{S(Y)\rceil} \mid \cdots \mid S(f_k)_{S(Y)\rceil} \mid \cdots\right) \\ &= \left(S(f)_{0S(Y)\rceil} \mid S(f)_{1S(Y)\rceil} \mid \cdots \mid S(f)_{kS(Y)\rceil} \mid \cdots\right) \\ &= \textbf{call}[\, S(f)\,]. \end{aligned}$$

Analogously for the finite case. ∎

**Corollary 6** Let $\mathfrak{E}$ and $\mathfrak{F}$ be extensive categories, $S : \mathfrak{E} \to \mathfrak{F}$ a functor which preserves finite sums, and $\mathfrak{C}$ a finitely iterable subcategory with sums of $\mathfrak{E}$. Then $S(\textbf{call}[\,\mathfrak{C}\,]) \subseteq \textbf{call}[\, S(\mathfrak{C})\,]$. If $\mathfrak{E}, \mathfrak{F}$ are countably extensive and $S$ preserves countable sums, the statement is true for any subcategory with finite sums.

**Proof.** The two categories involved have trivially the same class of objects. Moreover, Lemma 6 says that $S(\textbf{call}[\,\mathfrak{C}\,]) \subseteq \textbf{call}[\, S(\mathfrak{D})\,]$. ∎

It is now natural to ask the following question: when is the inclusion of the previous corollary an equality? This situation is enough important to deserve a name:

**Definition 21** A sum preserving functor $S : \mathfrak{E} \to \mathfrak{F}$, where $\mathfrak{E}, \mathfrak{F}$ are countably extensive categories, is said to *preserve the iteration* of a subcategory with finite sums $\mathfrak{C}$ of $\mathfrak{E}$ iff

$$S(\textbf{call}[\,\mathfrak{C}\,]) = \textbf{call}[\, S(\mathfrak{C})\,].$$

The essential reason which keeps an arbitrary functor from preserving iteration is that even if $S(f)$ is terminating, $f$ could be nonterminating. For instance, consider the arrow

$$f : A + B + A + B \xrightarrow{(\mathrm{inj}_1 \mid \mathrm{inj}_2 \mid \mathrm{inj}_3 \mid \mathrm{inj}_2)} A + B + A,$$

which is clearly not terminating on $A + B$. Any sum preserving functor $S$ satisfying $S(B) \cong \varnothing$ will bring $f$ into an arrow

$$S(f) : S(A) + S(A) \to S(A) + S(A)$$

such that $\textbf{call}[\, S(f)\,] = \mathbf{1}_{S(A)}$. We shall now give three sufficient conditions for the iteration of a subcategory being preserved.

**Theorem 22** A countable sum preserving functor which is conservative (i.e., which reflects isomorphisms) preserves the iteration of any subcategory.

**Proof.** If $S(f) : S(X) + S(U) \to S(U) + S(Y)$ is terminating, as in the proof of Lemma 6 we have

$$S(X) \cong \sum_{k\geq 0} S(X)_k \cong \sum_{k\geq 0} S(X_k) \cong S\Big(\sum_{k\geq 0} X_k\Big),$$

where all the isomorphisms are given by the obvious injections. This implies the the inclusion $\sum_{k\geq 0} X_k \to X$ is an isomorphism, by the conservativity of $S$. ∎

**Theorem 23** If $\mathfrak{C}$ is closed under subobjects, any countable sum preserving functor $S : \mathfrak{C} \to \mathfrak{F}$ preserves the iteration of $\mathfrak{C}$.

**Proof.** Let $S(f)$ be a terminating arrow. If we restrict $f$ to $\sum_{k\geq 0} X_k + U$, we obtain an arrow $f'$ which is terminating. Moreover,

$$S(f') = S(f_{|\sum_{k\geq 0} X_k + U}) = S(f)_{|S(\sum_{k\geq 0} X_k + U)} = S(f)_{|S(X)+S(U)} = S(f),$$

so **call**$[\, S(f') \,] =$ **call**$[\, S(f) \,]$. ∎

Theorems 22 and 23 are obviously true also for finitely iterable subcategories and finite sum preserving functors.

**Theorem 24** Any finite sum preserving functor $S$ satisfying $S(X) \cong \varnothing \implies X \cong \varnothing$ preserves iteration of finitely iterable subcategories. If $\mathfrak{C}$ has complemented subobjects and $S$ preserves countable sums, it preserves iteration of any subcategory.

**Proof.** With the same notation of the previous two theorems, we have $\sum_{k\geq 0} X_k + X' \cong X$, where $X'$ is obtained by extensivity in the finite case and by complementation of $\sum_{k\geq 0} X_k$ in the countable case. Thus we have

$$S\Big(\sum_{k\geq 0} X_k\Big) + S(X') \cong S(X),$$

which by the cancellation lemma implies $S(X') \cong \varnothing$; by hypothesis $X' \cong \varnothing$. This means that the termination of $S(f)$ implies the termination of $f$. ∎

### 4.2.4 The control function theorem

**Definition 22** The *normal form* of a distributive expression on a set $A$ is obtained by interpreting the distributive expression in the free semiring on $A$, and reducing it to a sum of monomials. An arc of a distributive graph is called a *control function* if the normal form of its target is a sum of at least two monomials.

Note that in $\mathcal{F}(G)$ there is an isomorphism between the interpretation of a distributive expression and the interpretation of its normal form.

**Theorem 25** Let $G$ be a distributive graph which contains no control functions; then $\mathcal{F}(G)$ is extensive.

**Proof.** We shall prove[4] by structural induction that if $f : X \to Y$ is a morphism of $\mathcal{F}(G)$, and $Y \cong V + W$, then there are functions $f_{\lceil V} : f^{-1}(V) \to V$, $f_{W\rceil} : f^{-1}(W) \to W$ such that $f_{\lceil V} + f_{W\rceil} = f$. This proves that the canonical functor $+$ is full and essentially surjective. Faithfulness follows trivially from the monicity of the injections of a distributive category.

It is clear that by hypothesis if $f$ is the image of an arc of $G$ then either $V \cong \varnothing$ or $W \cong \varnothing$; hence, either $f_{\lceil V} = f$ or $f_{W\rceil} = f$. If $f$ is an identity the case is trivial. If $f$ is an injection, it is easy to check that $f_{\lceil V}$ and $f_{W\rceil}$ are still injections. Finally, if $f$ is a projection or the inverse distributivity isomorphism the decomposition of the codomain induces trivially the decomposition of the domain.

Suppose now $f = h \circ g$, with $g : X \to Z$. Then $f$ satisfies the induction hypothesis, so that $Z \cong f^{-1}(V) + f^{-1}(W)$. But now $g$ in turn can be broken into $g_{\lceil h^{-1}(V)}$ and $g_{h^{-1}(W)\rceil}$, so that $h = h_{\lceil V} \circ g_{\lceil h^{-1}(V)} + h_{W\rceil} \circ g_{h^{-1}(W)\rceil}$.

If $f = (h \mid g)$, then we have $f_{\lceil V} = (h_{\lceil V} \mid g_{\lceil V})$ and $f_{W\rceil} = (h_{W\rceil} \mid g_{W\rceil})$. Clearly, $h^{-1}(V) + g^{-1}(V) + h^{-1}(W) + g^{-1}(W) \cong X$.

If $f = (g, h) : X \to A \times B$, then by distribution we can write in normal form $A \cong \sum_i A_i$ and $B \cong \sum_j B_j$, where the $A_i$'s and the $B_j$'s are monomials, and we can break $g$ into $g_i : g^{-1}(A_i) \to A_i$, with $g = \sum_i g_i$; the same can be done for $h$. Note that if we define $C_{ij} = g^{-1}(A_i) \cap h^{-1}(B_j)$, then $\sum_{i,j} C_{ij} \cong X$.

Since $A \times B \cong V + W$, there is a partition of the pairs $\langle i, j\rangle$ in two sets $K$, $L$ such that

$$\sum_{\langle i,j\rangle \in K} A_i B_j \cong V \qquad \sum_{\langle i,j\rangle \in L} A_i B_j \cong W.$$

But this implies that if we define

$$f_{\lceil V} = \sum_{\langle i,j\rangle \in K} (g_{i\,|C_{ij}}, h_{j\,|C_{ij}}) \qquad f_{W\rceil} = \sum_{\langle i,j\rangle \in L} (g_{i\,|C_{ij}}, h_{j\,|C_{ij}})$$

then

$$f_{\lceil V} + f_{W\rceil} = \sum_{\langle i,j\rangle \in K} (g_{i\,|C_{ij}}, h_{j\,|C_{ij}}) + \sum_{\langle i,j\rangle \in L} (g_{i\,|C_{ij}}, h_{j\,|C_{ij}}) = \sum_{i,j} (g_{i\,|C_{ij}}, h_{j\,|C_{ij}}) = (g, h). \blacksquare$$

The following theorem says that if no "control structure" (in the form of a control function) is present in the distributive base graph of a free distributive category, then closure by iteration does not yield any new function.

**Theorem 26** Let $G$ be a distributive graph, $\mathfrak{E}$ an extensive category and $S : \mathcal{F}(G) \to \mathfrak{E}$ a distributive functor. If no arc of $G$ is a control function, then

$$S(\mathcal{F}(G)) = \mathbf{call}[\, S(\mathcal{F}(G))\,].$$

**Proof.** Since $\mathcal{F}(G)$ is extensive, we have $\mathcal{F}(G) = \mathbf{call}[\,\mathcal{F}(G)\,]$. But then by Theorem 23

$$\mathbf{call}[\, S(\mathcal{F}(G))\,] = S(\mathbf{call}[\,\mathcal{F}(G)\,]) = S(\mathcal{F}(G)). \blacksquare$$

[4]A different proof has been suggested by R.F.C. Walters. Under the given hypotheses, it is possible to build a distributive graph $G'$ all whose arcs start from a product of objects and end on a single object, and such that $\mathcal{F}(G) \cong \mathcal{F}(G')$. Then $\mathcal{F}(G')$ is just the free category with sums on the free category with products on $G'$; but a free category with sums is always extensive.

## 4.3 Refinement in lextensive categories

We are now in the position to prove the refinement theorem (stated in Section 3.2.2) in an extensive setting. For this purpose, however, we shall need to internalize the notion of transition category to the extensive universe we are using. This requires the existence of finite limits (pullbacks are necessary in order to define internal categories, while products are necessary in order to define monoids and monoid actions), and a category which is extensive and has finite limits is called *lextensive* (or *countably lextensive*, in case extensivity is enjoyed for countable sums).

We begin by recalling the definition of internal category:

**Definition 23** Let $\mathfrak{C}$ be a category with pullbacks. An *internal category* $\mathfrak{X}$ of $\mathfrak{C}$ is given by

(i). objects $X_0$, $X_1$ of $\mathfrak{C}$, called the *object of objects* and the *object of arrows*;

(ii). arrows $\partial_0, \partial_1 : X_1 \to X_0$ of $\mathfrak{C}$, called *domain* and *codomain*;

(iii). an arrow $i : X_0 \to X_1$, called *identity*;

(iv). an arrow $c : X_1 \times_{X_0} X_1 \to X_1$ called *composition*, where $X_1 \times_{X_0} X_1$ is the pullback of $\partial_0$ and $\partial_1$ (with projections $\pi_1$ and $\pi_0$, respectively—note the reversed indices);

satisfying the following equations:

(i). $\partial_0 \circ i = \partial_1 \circ i = \mathbf{1}_{X_0}$;

(ii). $\partial_1 \circ \pi_1 = \partial_1 \circ c$ and $\partial_0 \circ \pi_0 = \partial_0 \circ c$;

(iii). $c \circ (\mathbf{1}_{X_1}, i \circ \partial_0) = c \circ (i \circ \partial_1, \mathbf{1}_{X_1}) = \mathbf{1}_{X_1}$;

(iv). $c \circ (\mathbf{1}_{X_1} \times_{X_0} c) = c \circ (c \times_{X_0} \mathbf{1}_{X_1})$.

A *internal functor* $F$ between internal categories $\mathfrak{X}$ and $\mathfrak{Y}$ of $\mathfrak{C}$ is given by a pair of arrows $F_0 : X_0 \to Y_0$, $F_1 : X_1 \to Y_1$ which satisfies the following equations:

(i). $\partial_0 \circ F_1 = F_0 \circ \partial_0$ and $\partial_1 \circ F_1 = F_0 \circ \partial_1$;

(ii). $F_1 \circ i = i \circ F_0$;

(iii). $F_1 \circ c = c \circ (F_1 \times_{F_0} F_1)$.

An internal functor $F$ is *injective on objects* if $F_0$ is a mono; it is *faithful* if $F_1$ is a mono; it is *full* if the factorization of $F_1$ through $\partial_0^{-1}(F_0) \cap \partial_1^{-1}(F_0)$ is an epi. A *full inclusion* is a full and faithful functor which is injective on objects.

Note that the equations which must be satisfied by the domain, codomain and identity arrows imply that the indicated arrows into the pullback $X_1 \times_{X_0} X_1$ do exist (for more details, see [Borceux:1994]). The definition of fullness requires the construction of the pullback diagrams

$$\begin{array}{ccccc} X_1 & & & & \\ & \overset{j_q}{\searrow} & & \overset{F_1}{\searrow} & \\ & & \partial_q^{-1}(F_0) & \longrightarrow & Y_1 \\ & \underset{\partial_q}{\searrow} & \downarrow & & \downarrow \partial_q \\ & & X_0 & \xrightarrow{F_0} & Y_0 \end{array}$$

for $q = 0, 1$ and of the factorization $F_1'$ of $F_1$ through the following pullback:

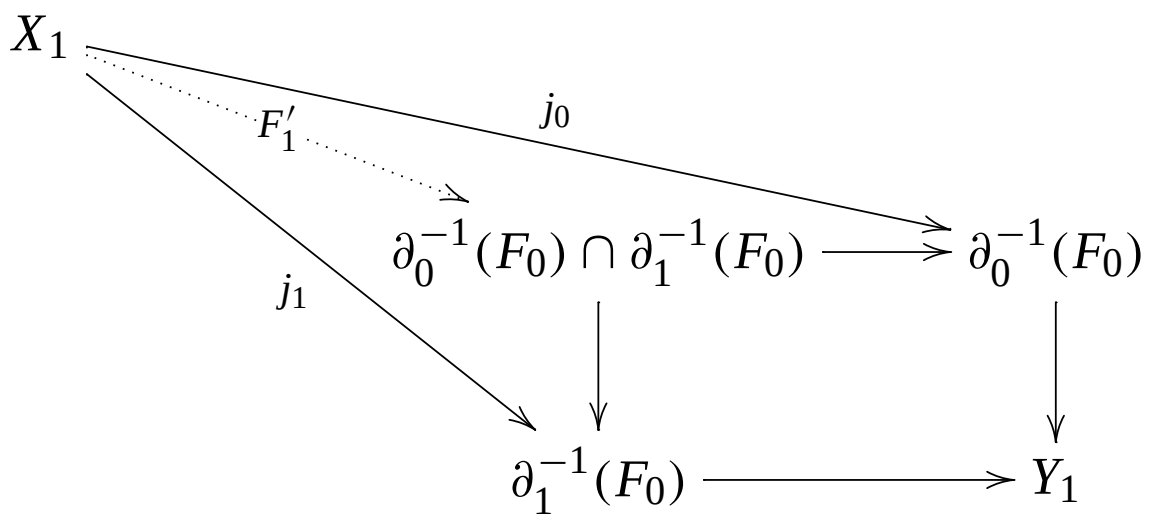

**Definition 24** Let $\langle M, \mu, \varepsilon\rangle$ be a monoid in a category with finite limits $\mathfrak{C}$; an automaton $\boldsymbol{X}$ on $M$ is given by an object $X$ of $\mathfrak{C}$ together with a monoid action $\alpha : M \times X \to X$. The transition category of $\boldsymbol{X}$ is the internal category $\mathbf{Trans}(\boldsymbol{X})$ of $\mathfrak{C}$ defined as follows:

(i). the objects of $\mathbf{Trans}(\boldsymbol{X})$ are given by $X$, the arrows by $MX$;

(ii). $i : X \xrightarrow{(\varepsilon \circ \mathrm{i}_X, \mathbf{1}_X)} MX$;

(iii). $\partial_0 : MX \xrightarrow{\mathrm{proj}_2} X$;

(iv). $\partial_1 : MX \xrightarrow{\alpha} X$;

(v). $c : (MX) \times_X (MX) \xrightarrow{(\pi_1, \pi_0)} MX \times MX \xrightarrow{(\mathrm{proj}_1, \mathrm{proj}_3, \mathrm{proj}_4)} MMX \xrightarrow{\mu \times \mathbf{1}_X} MX.$

Rewording simply our definitions, we shall say that a morphism of automata in $\mathfrak{C}$ is an *internal* functor between the respective transition categories, and that a *refinement* is a functor which is a full injection.

**Theorem 27** The data of Definition 24 satisfy the equations of an internal category of $\mathfrak{C}$.

**Proof.** The equations about the identity arrow $i$ are trivially satisfied because $\mathrm{proj}_2 \circ (\varepsilon \circ \mathrm{i}_X, \mathbf{1}_X) = \mathbf{1}_X$ and because of the unit axiom of $M$. Both $\partial_0 \circ \pi_0$ and $\partial_0 \circ c$ are just projections on the fourth factor of $(MX) \times_X (MX)$, and

$$\begin{aligned}\partial_1 \circ c &= \alpha \circ c\\ &= \alpha \circ (\mu \times \mathbf{1}_X) \circ (\mathrm{proj}_1, \mathrm{proj}_3, \mathrm{proj}_4) \circ (\pi_1, \pi_0)\\ &= \alpha \circ (\mathbf{1}_M \times \alpha) \circ (\mathrm{proj}_1, \mathrm{proj}_3, \mathrm{proj}_4) \circ (\pi_1, \pi_0)\\ &= \alpha \circ (\mathrm{proj}_1 \circ \pi_1, \alpha \circ \pi_0)\\ &= \alpha \circ (\mathrm{proj}_1 \circ \pi_1, \mathrm{proj}_2 \circ \pi_1)\\ &= \alpha \circ \pi_1,\end{aligned}$$

because of the associativity of the action of $M$ and of the commutativity of the pullback diagram of $(MX) \times_X (MX)$. We prove the first of the two equations regarding composition with the identity:

$$\begin{aligned}c \circ (\mathbf{1}_{MX}, i \circ \partial_0) &= (\mu \times \mathbf{1}_X) \circ (\mathrm{proj}_1, \mathrm{proj}_3, \mathrm{proj}_4) \circ (\pi_1, \pi_0) \circ (\mathbf{1}_{MX}, i \circ \partial_0)\\ &= (\mu \times \mathbf{1}_X) \circ (\mathrm{proj}_1, (\varepsilon \circ \mathrm{i}_X, \mathbf{1}_X) \circ \mathrm{proj}_2)\\ &= (\mu \times \mathbf{1}_X) \circ (\mathbf{1}_M \times (\varepsilon \circ \mathrm{i}_X, \mathbf{1}_X))\\ &= (\mu \times \mathbf{1}_X) \circ (\mathbf{1}_M \times \varepsilon \times \mathbf{1}_X) \circ (\mathbf{1}_M \times (\mathrm{i}_X, \mathbf{1}_X))\\ &= (\mathrm{proj}_1 \times \mathbf{1}_X) \circ (\mathbf{1}_M \times (\mathrm{i}_X, \mathbf{1}_X))\\ &= \mathbf{1}_M \times \mathbf{1}_X = \mathbf{1}_{MX},\end{aligned}$$

the other one having a similar proof. Finally, the associativity equation derives immediately from the associativity of $M$. ∎

As we noticed at the beginning of this chapter, in a distributive category with countable sums it is possible to build certain algebraic structures. In particular, the free monoid on $I$ will be denoted by $\boldsymbol{N} = I + I + \cdots$. Given a loop $f$ on $X$, there is an obvious $\boldsymbol{N}$-automaton with state space $X$, and action defined by

$$\alpha : \boldsymbol{N}X \cong X + X + \cdots \xrightarrow{(\mathbf{1}_X \mid f \mid f^2 \mid \cdots)} X.$$

We shall describe such automata by giving just the loop $f$. With these premises, we can finally state and prove the refinement theorem:

**Theorem 28** Let $f : X + U \to U + Y$ be a terminating arrow of a lextensive category $\mathfrak{C}$. Let $\boldsymbol{X}$ be the $\boldsymbol{N}$-automaton

$$X + Y \xrightarrow{\mathrm{inj}_2^{X,Y} \circ (\mathbf{call}[\, f\,] \mid \mathbf{1}_Y)} X + Y,$$

and $\boldsymbol{X}'$ be the $\boldsymbol{N}$-automaton on $X + U + Y$ given by the functional processor $h$ induced by $f$. Then, there is a refinement from $\boldsymbol{X}$ to $\boldsymbol{X}'$ whose object component is the inclusion $F_0 : X+Y \to X+U+Y$.

**Proof.** We have to define an arrow

$$F_1 : \boldsymbol{N}(X+Y) \to \boldsymbol{N}(X+U+Y)$$

which satisfies, together with the inclusion $F_0 : X + Y \to X + U + Y$, the equations of an internal functor. By distributivity $\boldsymbol{N}(X+Y) \cong \boldsymbol{N}X + \boldsymbol{N}Y$, and $F_{1|\boldsymbol{N}Y}$ is just the inclusion in $\boldsymbol{N}(X+U+Y) \cong \boldsymbol{N}X + \boldsymbol{N}U + \boldsymbol{N}Y$. For what matters $\boldsymbol{N}X$, we set

$$\sum_{i \geq 0} X_i \overset{g}{\cong} X$$

on the basis of the iteration of $f$, and we define for $j > 0$

$$F_1^{j,i} : X_i \xrightarrow{g_i} X \xrightarrow{\iota_{i+j}} \boldsymbol{N}X,$$

where $\iota_{i+j}$ is the injection in the $i + j$-th copy of $X$. Then $F_1^j : X \xrightarrow{(F_1^{j,0} \mid F_1^{j,1} \mid \cdots)} \boldsymbol{N}X$ and

$$F_{1|\boldsymbol{N}X} : \boldsymbol{N}X \xrightarrow{(F_1^0 \mid F_1^1 \mid \cdots)} \boldsymbol{N}X \to \boldsymbol{N}X + \boldsymbol{N}U + \boldsymbol{N}Y \cong \boldsymbol{N}(X+U+Y),$$

where $F_1^0$ is the injection of $X$ into the first component of $\boldsymbol{N}X$.

We have now to show that the functorial equations hold. The domain equation $\partial_0 \circ F_1 = F_0 \circ \partial_0$ is trivially satisfied. The codomain equation, restricted to $\boldsymbol{N}Y$, is also trivially satisfied because in both $\boldsymbol{X}$ and $\boldsymbol{X}'$ the action on that component is the identity. We now restrict the codomain equation to $X_i \xrightarrow{g_i} X \xrightarrow{\iota_{i+j}} \boldsymbol{N}X$, obtaining for $j > 0$

$$\begin{aligned} \partial_1 \circ F_1^{j,i} &= \alpha' \circ \iota_{i+j} \circ g_i \\ &= h^{i+j} \circ i \circ g_i \\ &= k \circ (f_{0\lceil Y} \mid \cdots \mid f_{i+j-1\lceil Y}) \circ g_i \\ &= k \circ f_{i\lceil Y} \\ &= F_0 \circ \mathrm{inj}_2^{X,Y} \circ f_{i\lceil Y}, \end{aligned}$$

where we used the commutativity of the diagrams in the proof of Theorem 20. Thus, $\partial_1 \circ F_1^j = F_0 \circ \mathrm{inj}_2^{X,Y} \circ \mathbf{call}[\,f\,]$, and (by the definition of $F_1^0$)

$$\begin{aligned}\partial_1 \circ F_{1|\boldsymbol{N}X} = F_0 \circ (\mathrm{inj}_1^{X,Y} \mid \mathrm{inj}_2^{X,Y} \circ \mathbf{call}[\,f\,] \mid \mathrm{inj}_2^{X,Y} \circ \mathbf{call}[\,f\,] \mid \cdots) = \\ F_0 \circ (\mathrm{inj}_1^{X,Y} \mid \mathrm{inj}_2^{X,Y} \circ \mathbf{call}[\,f\,] \mid \mathrm{inj}_2^{X,Y} \circ (\mathbf{call}[\,f\,] \mid \mathbf{1}_Y) \circ \mathrm{inj}_2^{X,Y} \circ \mathbf{call}[\,f\,] \mid \cdots) = \\ F_0 \circ \alpha'_{|\boldsymbol{N}X} = F_0 \circ \partial_{1|\boldsymbol{N}X}.\end{aligned}$$

The identity equations are trivial, for both $F_1 \circ i$ and $i \circ F_0$ are the inclusion of $X + Y$ into the first summand of $\boldsymbol{N}(X + U + Y)$ (note also that $i = F_1^0$).

Finally, we show that $F$ is full and faithful. Fidelity is immediate once we notice that $F_1$ is built using injections (which are monic in any distributive category) and the codomains of all injections are disjoint. Indeed, the codomain of two inclusions $X_i \xrightarrow{g_i} X \xrightarrow{\iota_{i+j}} \boldsymbol{N}X$ lies in different copies of $X$ when $j$ varies, and in different summands of $X$ when $i$ varies.

Now we note that $\partial_0^{-1}(F_0) \cong \boldsymbol{N}(X + Y)$ and (letting $V = \partial_{1|\boldsymbol{N}U}^{-1}(X + Y)$ and $V' = \partial_{1|\boldsymbol{N}U}^{-1}(U)$)

$$\partial_1^{-1}(F_0) \cong X + \sum_{j>0} \sum_{i<j} X_i + V + \boldsymbol{N}Y.$$

This can be seen by building the following diagram, whose commutativity can be easily derived from the commutativity of the diagrams in the proof of Theorem 20:

$$\begin{array}{ccccc} X + \sum_{j>0} \sum_{i<j} X_i + V + \boldsymbol{N}Y & \longrightarrow & \boldsymbol{N}(X + U + Y) & \longleftarrow & \sum_{j>0} \sum_{i \geq j} X_i + V' \\ \downarrow & & \downarrow \partial_1 & & \downarrow \\ X + Y & \xrightarrow[F_0]{} & X + U + Y & \longleftarrow & U \end{array}$$

The vertical arrows are obtained by suitably restricting and corestricting the action $\alpha'$. The first row is a coproduct, so the two squares are pullback. As a result,

$$\partial_0^{-1}(F_0) \cap \partial_1^{-1}(F_0) \cong X + \sum_{j>0} \sum_{i<j} X_i + \boldsymbol{N}Y.$$

Since $F_1$ if given by $(F_1^0 \mid F_1^1 \mid \cdots) + \mathbf{1}_{\boldsymbol{N}Y}$, the codomain of $F_1^0$ is $X$ and $F_1^{j,i}$ is the injection into the component $X_i$ which lies in the $j$-th summand of the sum $\sum_{j>0} \sum_{i<j} X_i$, we have that the factorization of $F_1$ through $\partial_0^{-1}(F_0) \cap \partial_1^{-1}(F_0)$ gives $\sum_{j>0} \sum_{i<j} \mathbf{1}_{X_i}$, which implies that $F_1$ is an isomorphism. ▮

## 4.4 Some consideration on partiality

The general theory described here is based on *total* functions. The iterative closure $\mathbf{call}[\,\mathfrak{C}\,]$ is built using only terminating arrows in order to preserve the good categorical properties (distributivity, for instance) of $\mathfrak{C}$.

In the **Set**-based case we mentioned that it possible to consider a $\mathbf{pcall}[\,-\,]$ operator returning partial functions. As long as suitable set-theoretical definitions are used in the definition of the operators $(-, -)$ and $(- \mid -)$, the theory can be indeed rebuilt without any modification. The categorical extension, however, is more problematic, because the category of sets and partial functions does not have products.

In [Wagner, Khalil & Walters:1995] a definition is proposed of *fix-point solution* of a prefunctional processor $\varphi : X + U + Y \to X + U + Y$ (called there a *program*) in a distributive category $\mathfrak{D}$; it is an arrow $\sigma : X + U + Y \to Y$ such that $\sigma \circ \varphi = \sigma$ and $\sigma \circ k = \mathbf{1}_Y$ (where $k$ is the injection $Y \to X + U + Y$). The *fix-point semantics* $\bar{\varphi}$ of $\varphi$ is then defined as $\sigma_{|X}$. Given a distributive subcategory $\mathfrak{C} \subseteq \mathfrak{D}$, one can consider the category $\mathfrak{C}^\dagger$ of the fix-point semantics of (not necessarily functional) processors of $\mathfrak{D}$, which is distributive. Thus, $\mathfrak{C}^\dagger$ should generalize the operator **call**$[-]$ to the case of arrows without termination condition (we say "generalize" because it is proved in [Wagner, Khalil & Walters:1995] that a pseudofunction has a unique fix-point semantics which coincides with the function it computes by iteration).

However, it is easy to see that for any objects $X, Y$ of $\mathfrak{C}$ the identity on $X + Y$ is an everywhere nonterminating prefunctional processor. Since the definition of fix-point solution implies that $\sigma$ can be defined arbitrarily on the part of the input and local state space on which $\varphi$ does not terminate, we have that *any arrow* from $X$ to $Y$ is a fixed-point semantics of the identity on $X + Y$, so $\mathfrak{C}^\dagger$ is just the full subcategory of $\mathfrak{D}$ generated by $\mathfrak{C}$.

In order to convince the reader that unwanted functions are included in $\mathfrak{C}^\dagger$ even in nondegenerate cases, we exhibit another example: let $\mathfrak{C} \subset$ **Set** be the distributive subcategory generated by the successor and predecessor functions of Section 2.2; consider a recursively enumerable, nonrecursive subset $A \subset \boldsymbol{N}$. It is immediate to build a processor with codomain $I + I$ which outputs in the first copy of $I$ on the elements of $A$ and it is undefined elsewhere. But then the characteristic function of $A$ is a fix-point semantics of such a processor, and it is by construction nonrecursive.

Thus, in the opinion of the author the correct generalization of the idea of iterative computation to nonterminating arrows remains an open problem.

# Chapter 5

# A new coding for IMP$(G)$

In this chapter we shall introduce a new coding for **IMP**$(G)$ data and programs which is particularly suited to implementation. We shall show how the elements of a derived set can be described as a suitable *forest*, and that it is possible to associate inductively to a derived arrow a function on forests.

When discussing such strictly syntactical issues, two possibilities arise: a wordy description, which highlights the main ideas and skips the syntactical details, or a very formal development, which usually turns the reader away. Since the coding will be used in the formal specification of a series of **IMP**$(G)$ tools (see Chapter 6), in this chapter we shall discuss the main mathematical and algorithmical ideas informally; a fully formal (even executable!) description of the same ideas will be given in the specification.

## 5.1 A new efficient coding for IMP$(G)$ data

In [Khalil & Walters:1993] a coding for the data of **IMP**$(G)$ programs has been proposed which represents an element of a set $X$ derived from the basic sets $A$, $B$, ... as a word, using the fact that $X$ is a "subset" of $(A + B + \cdots + I)^*$. Correspondingly, a specific coding has been given for the terms built using a certain set of operators which come from the structure of a distributive category. The coding transforms a term in a series of rewriting rules, whose application emulates the behaviour of the function described by the term on a given word (representing an element of the domain of the function).

We are going to propose a new, more efficient coding which is at the basis of the tools described in the next chapter. The basic idea behind the coding is that, in analogy with standard programming, *products represent structures and sums represent union types*. Thus, each time we shall need to represent an element of a sum, we shall *tag* it with the index of its summand.

### 5.1.1 The old coding

Let $D$ be the sum of all the basic sets plus $\{*, e\}$. The *length* $|X|$ of a derived set $X$ is inductively given by $|\varnothing| = |I| = 1$ and $|X + Y| = |X \times Y| = |X| + |Y|$ (if $X$ is seen as a tree, it is just its number of leaves). An element of $X$ is represented by a word over $D$:

(i). the only element of $I$ is represented by $*$ ($\varnothing$ has no elements, but we assign it the word $e$);

(ii). the elements of a basic set are represented by themselves (as one-letter words);

(iii). the elements of $X \times Y$ are represented by words $uv$, and the elements of $X + Y$ by words $ue^{|V|}$ or $e^{|U|}v$, where $u$ and $v$ represent elements of $X$ and $Y$, respectively.

The distinguished element $e$ is used in order to tell which components of a derived set "do hold" a value. For instance, if $X = A + B$ then the elements of $X$ can be elements of $A$ or elements of $B$; correspondingly, a representation of either of the form $ae$, or of the form $eb$. It is immediate to see that the length of the word representing an element of $X$ is $|X|$.

### 5.1.2 The new coding

The new coding we are going to propose is based on ordered forests whose leaves are labelled by elements of the basic sets and whose nodes are labelled by natural numbers. We shall identify the empty forest with the only element of $I$, the one-node forests with the elements of the basic sets, denote by juxtaposition the ordered concatenation of forests and by $\lhd n, f \rhd$ the creation of a tree with a root labelled by $n \in \boldsymbol{N}$ and having as first-level children the trees of the forest $f$. The precise definition and properties of such a structure, the *free $\boldsymbol{N}$-motor on the sum of the basic sets*, have been studied in [Pair & Quère:1968, Vigna:1991, Kasangian & Vigna:1991a]. We shall describe the manipulation of forest informally; the discussion can be recast in term of algebraic operations and initiality in the category of $\boldsymbol{N}$-motors.

We shall define our coding on the basis of a normal form for the derived sets. We shall consider $+$ and $\times$ as *operators of variable arity* (something which is allowed by associativity). A *flat term (or object)* is defined inductively as:

(i). a basic set;

(ii). a list of flat terms surrounded by "$+($" and "$)$", or "$\times($" and "$)$" (a $+$-list and a $\times$-list, respectively).

We confuse again a flat term and and its obvious interpretation as a set, calling both of them *flat objects*.

Now we define a normal form (obtained by associative merge) on the flat objects which optimizes the storage requirements. A *flat object in normal form* is inductively defined as:

(i). a basic set

(ii). a $+$-list of length different from 1, whose items are basic sets and $\times$-lists in normal form;

(iii). a $\times$-list of length different from 1, whose items are basic sets and $+$-lists in normal form.

In other words, flat objects in normal form do not contain lists with just one element, and it never happens that a sublist can be merged in the containing one by associativity.

For each derived set (or flat object) $X$ there is exactly one corresponding flat object in normal form, obtained by list merging after having applied the mapping $\nu$ inductively defined by

$$\begin{aligned} \nu(I) &= \times() \\ \nu(\varnothing) &= +() \\ \nu(A) &= A \\ \nu(X+Y) &= +(\nu(X)\,\nu(Y)) \\ \nu(X\times Y) &= \times(\nu(X)\,\nu(Y)). \end{aligned}$$

We shall denote the normal form of a flat object $X$ by $\bar{X}$. By abuse of notation, we shall write $\bar{X}$ for $\overline{\nu(X)}$ when $X$ is a derived set. Of course, many derived sets (or flat objects) correspond to a normal form. Note also that there is a canonical isomorphism between a derived set $X$, $\nu(X)$ and $\overline{\nu(X)}$.

Now the representation of an element $x \in X$, where $X$ is a flat object, is defined as follows:

(i). if $X$ is a basic set, $x$ represents itself;

(ii). if $X = \times(X_1\, X_2\, \cdots\, X_p)$, then $x$ is a tuple $\langle x_1, x_2, \ldots, x_p\rangle$, and if $u_k$ represents $x_k$ as an element of $X_k$ the representation of $x$ as an element of $X$ is $u_1 u_2 \cdots u_p$;

(iii). if $X = +(X_1\, X_2\, \cdots\, X_p)$, then $x$ is a pair $\langle u, k\rangle$, where $0 \leq k < p$ and $u$ represents $x$ as an element of $X_k$, and the representation of $x$ as an element of $X$ is $\triangleleft k, u\triangleright$.

Given an element $x$ of $X$, we shall write $\circ x$ for its representation as a forest (following the notation of [Khalil & Walters:1993]); when $X$ is a derived set, we shall write $\circ x$ for the representation of $x$ as an element of $\nu(X)$. It is immediate that

**Proposition 4** For all $x, x' \in X$, $\circ x = \circ x'$ implies $x = x'$.

Note that for sake of notational simplicity we shall write the $+(X)$ (or $\times(X)$) for the simple object or $\times$-list in normal form $X$ (for the simple object or $+$-list in normal form $X$, respectively). The notation is not ambiguous, since there are no $+/\times$-lists in normal form with just one element, and it allows us to describe in one shot as $+(X_1\, \cdots\, X_p)$ a $+$-list, when $p > 1$, a $\times$-list or simple object, when $p = 1$, or the empty set, when $p = 0$ (analogously for $\times(X_1\, \cdots\, X_p)$).

Once we reduce to flat objects in normal form, it is immediate, for instance, to note that the space necessary for storing $n \cdot X^n$ is $O(n^2)$ with the old coding and $O(n)$ with the new one. In general, the storage required for the sum of two flat objects is the sum of the storage for $X$ and $Y$ in the old case, and the maximum (plus one) of the storage for $X$ and $Y$ in the new case.

### 5.1.3 Normalization and associativity

The new coding we just described has a disadvantage: while the associativity isomorphisms related to the product, i.e., the maps

$$X \times (Y \times Z) \xrightarrow{a} (X \times Y) \times Z$$

derived from the projections, satisfy $\circ x = \circ a(x)$ for every element $x$ of $X$, and moreover $\circ x$ is exactly the representation of $x$ as an element of $\times(X\ Y\ Z)$, we have that if

$$X + (Y + Z) \xrightarrow{b} (X + Y) + Z$$

is the analogous isomorphism for the sum and $z$ is an element of $Z$, $\circ z = \triangleleft 1, \triangleleft 1, z\triangleright\triangleright$ but $\circ b(z) = \triangleleft 1, z\triangleright$. Moreover, the representation of $z$ as an element of $+(X\ Y\ Z)$ is $\triangleleft 2, z\triangleright$. Thus, the associativity isomorphisms of product induce the identity at the representation level, but this does not happen for the sum.

However, there is a simple way of overcoming this problem. For a forest $u$, let $\bar{u}$ denote its normal form with respect to the rewrite rule

$$\triangleleft j, \triangleleft k, u\triangleright\triangleright = \triangleleft j + k, u\triangleright.$$

It is straightforward to prove the following

**Proposition 5** Let $X$, $Y$ and $Z$ be flat objects such that $X = +(Y\ Z)$; let $\bar{Y} = +(Y_1\, \cdots\, Y_q)$ and $\bar{Z} = +(Z_1\, \cdots\, Z_r)$, and suppose $q, r > 0$. Then if $x_1$ is an element of $Y$, with representation $u_1$ in $\bar{Y}$, the representation of $\mathrm{inj}_1(x_1)$ in $\bar{X}$ is $\overline{\triangleleft 0, u_1\triangleright}$; analogously, if $x_2$ is an element of $Z$, with representation $u_2$ in $\bar{Z}$, the representation of $\mathrm{inj}_2(x_2)$ in $\bar{X}$ is $\overline{\triangleleft q, u_2\triangleright}$. If $q = 0$ (or $r = 0$) the representation of $\mathrm{inj}_2(x_2)$ (or $\mathrm{inj}_1(x_1)$) is $u_2$ ($u_1$, respectively).

This proposition has several consequences: first of all, it explains how to obtain the representation of an element of $X$ as an element of $\bar{X}$; second, it defines the effect of injections on representations; third, it allows to reverse the effect of the injections:

**Corollary 7** With the same notation of Proposition 5, suppose $q, r > 1$; then an element of $Y$ (or $Z$) injected in $\bar{X}$ is coded as $\lhd n, u \rhd$, with $0 \leq n < q$ ($q \leq n < q + r$, respectively). If $q = 1$ ($r = 1$, respectively) its coding as an element of $\bar{Y}$ is $u$, otherwise it is $\lhd n, u \rhd$ ($\lhd n - q, u \rhd$, respectively).

The categorical result underlying the previous discussion is the *coherence theorem* for monoidal categories (see [Mac Lane:1971]). The corollary of interest here is that if we choose a normal form with respect to associativity of $+$ and $\times$, there is a canonical way of making the interpretation of the original term and of the normal form isomorphic.

Thus, we shall suppose that all objects are given in flat normal form (some preprocessing can take care of this fact) and that all arrows are precomposed and postcomposed with the canonical isomorphisms from and to the normal forms (in practice, we shall work in an equivalent category where $+$ and $\times$ happen to be strict). This has the major advantage that all the monoidal isomorphisms (associativity and unit axioms) will induce identities at the representation level.

## 5.2 Coding IMP$(G)$ programs

Recall that an **IMP**$(G)$ program is an arrow in a free distributive category. Our goal is to transform it into an operator on forests, which can be actually computed by a real machine.

### 5.2.1 Coding simple functions

We now introduce the coding of the simple functions, which is given by stating the effect of the coding on normal form representations. For each basic function $a$ we denote with $\bar{a}$ the function that $a$ induces on the representations (and defined in the obvious way: if $u = \circ x$ then $\bar{a}(u) = \circ a(x)$).

$$
\begin{array}{rcl}
\circ a : u & \mapsto & \bar{a}(u) \\
\circ \mathrm{inj}_1^{+(X_1 \cdots X_p), +(Y_1 \cdots Y_q)} : u & \mapsto & \overline{\lhd 0, u \rhd} \\
\circ \mathrm{inj}_2^{+(X_1 \cdots X_p), +(Y_1 \cdots Y_q)} : u & \mapsto & \overline{\lhd p, u \rhd} \\
\circ \mathrm{proj}_1^{\times(X_1 \cdots X_p), \times(Y_1 \cdots Y_q)} : u_1 \cdots u_p v_1 \cdots v_q & \mapsto & u_1 \cdots u_p \\
\circ \mathrm{proj}_2^{\times(X_1 \cdots X_p), \times(Y_1 \cdots Y_q)} : u_1 \cdots u_p v_1 \cdots v_q & \mapsto & v_1 \cdots v_q \\
\circ \mathrm{i}_X : u & \mapsto & \varepsilon \\
\circ \mathbf{1}_X : u & \mapsto & u
\end{array}
$$

Note that all operations on the right side produce automatically forests in normal form, except for the injections, for which the required normalization is shallow: one just need to check that the rewrite rule $\lhd j, \lhd k, u \rhd \rhd = \lhd j + k, u \rhd$ is applicable at the outermost level. We supposed $p, q > 0$, for otherwise the injections trivialize to identities or initial maps.

The distributivity isomorphism has a more complex description; Corollary 7 tell us how to "unpack" the representation of an element of a sum, and, depending on $q$ or $r$ being equal to one, we have the following four cases:

$$\begin{aligned}\circ\delta^{-1}_{\times(X_1\cdots X_p),+(Y_1\cdots Y_q),+(Z_1\cdots Z_r)} : u\lhd 0, v\rhd &\mapsto \lhd 0, uv\rhd \text{ if } q=1\\ u\lhd k, v\rhd &\mapsto \lhd 0, u\lhd k, v\rhd\rhd \text{ if } q>1, k<q\\ u\lhd q, v\rhd &\mapsto \lhd n, uv\rhd \text{ if } r=1\\ u\lhd k, v\rhd &\mapsto \lhd n, u\lhd k-q, v\rhd\rhd \text{ if } r>1, k\geq q\end{aligned}$$

where we assumed $p, q, r > 0$ and

$$\overline{\times(X_1\cdots X_p + (Y_1\cdots Y_q))} = +(S_1\cdots S_n);$$

note that $n > 0$, and that $n > 1$ implies $p, q = 1$, $Y_1 = *()$ and $X_1 = +(S_1 \cdots S_n)$. If $p$, $q$ or $r$ are equal to zero, $\delta^{-1}$ is the identity or an injection.

### 5.2.2 Coding derived functions

To each operator used in the construction of derived arrows we associate inductively an operator on forests:

$$\begin{aligned}\circ(f,g) : u &\mapsto \circ f(u)\circ g(u)\\ \circ g\circ f : u &\mapsto (\circ g\circ\circ f)(u).\end{aligned}$$

In the case of the $(-\mid -)$ operator, as for the distributivity isomorphism, we must take some more care. Suppose $\overline{\mathbf{dom}(f)} = +(X_1\cdots X_p)$, $\overline{\mathbf{dom}(g)} = +(Y_1\cdots Y_q)$ and $p, q > 0$. Then, following Corollary 7 we have

$$\begin{aligned}\circ(f\mid g) : \lhd 0, u\rhd &\mapsto \circ f(u) \text{ if } p=1\\ \lhd k, u\rhd &\mapsto \circ f(\lhd k, u\rhd) \text{ if } p>1, k<p\\ \lhd p, u\rhd &\mapsto \circ f(u) \text{ if } q=1\\ \lhd k, u\rhd &\mapsto \circ f(\lhd k-p, u\rhd) \text{ if } q>1, k\geq p.\end{aligned}$$

The relevant result here is the following

**Theorem 29** Let $f$ be a derived function. Then $\circ f(\circ x) = \circ f(x)$.

**Proof.** A straightforward but tedious exercise in structural induction and case-by-case analysis. ▮

The reader should notice that $\circ-$ sends arrows to (simple) operations of forests. This will allow to precompile an arrow into a kind of p-code which is easily executed.

## 5.3 A type-checking algorithm for IMP$(G)$ programs

In the practical description of **IMP**$(G)$ programs, we certainly want to forget the associativity isomorphisms, and at the same time to specify structural arrows in a polymorphic form (as we did, for instance, in the proof of Theorem 4). It would indeed be cumbersome to have to write $X\times Y \xrightarrow{\text{proj}_1^{X,Y}} X$ instead of just $X\times Y \xrightarrow{\text{proj}_1} X$. This implies that injections and projections should have constraints relaxed with respect to the definition given in Section 2.1: we should admit $X+Y \xrightarrow{\text{inj}_1} X+(Y+Z)$ or $X\times(Y\times Z) \xrightarrow{\text{proj}_1} X\times Y$. Note that we should admit *also* $X \xrightarrow{\text{inj}_1} X+(Y+Z)$ or $X\times(Y\times Z) \xrightarrow{\text{proj}_1} X$.

When specifying an **IMP**$(G)$ program, we shall build an arrow using the usual operators and constant of a distributive category (see Section 4.1.2), but with two important differences:

(i). each structural arrow can be also specified in a polymorphic form, e.g., $\mathrm{proj}_2$;

(ii). the operators can be applied to any pair of arrows, regardless of the standard equational constraints (such as the domain of $g$ being equal to the codomain of $f$ when building $g \circ f$).

Note that we shall use "arrow" in this relaxed sense in the rest of this chapter, and in the next one.

The second point is imposed by the first one, because the equational conditions on domains and codomains which restrain the application of an operator cannot be checked for polymorphic arrows. Thus, there is no guarantee that we did not compose uncomposable functions, or that we did not write $(f, g)$ when $f$ and $g$ did not have the same domain. The *type-checker* of **IMP**$(G)$ is an algorithm which takes an arrow with specified domain and codomain, and outputs "yes" if there is a way of instatiating the polymorphic arrows contained in the arrow in such a way to satisfy the usual compositional constraint. (Note that we do not claim that if the type-checker answers "no" no instantiation is possible.)

Polymorphism is essential because it allows to use a friendly syntax, similar to the one which can be found in mathematics papers. On the other hand, polymorphism makes type-checking much more difficult, because suddenly there is a possibility that the (co)domain of an arrow is not computable, in the following sense: We say that **dom**$(f)$ (or **cod**$(f)$) is computable for a simple arrow $f$ if $f$ is a basic arrow, or a completely specified structural arrow; moreover, **dom**$(!)$ are **cod**$(\mathrm{i})$ computable; if $f$ is a derived arrow, its (co)domain is computable if the standard equations of category theory allow to infer it from the (co)domains of the simple arrows it is built from. More details can be found in Section 6.5.

Since the definition of a new arrow will be given in the form $X \xrightarrow{f} Y$, we shall check that $f$ has $X$ as domain and $Y$ as codomain *up to associativity (and unity) isomorphisms*. This means that we cannot assume that the syntactic structure of the (co)domain will follow the structure of the arrow. For instance, if $f : X \to Y$ and $g : X \to Z \times W$ then

$$X \xrightarrow{(f,g)} (Y \times Z) \times W$$

should pass the type check, even if we have to modify the abstract syntax tree of the codomain to $Y \times (Z \times W)$.

The approach we take is to transform the objects found in the program (i.e., the distributive expressions), which are described using binary sums and products, into flat objects in normal form. With this premise, the type-check of simple, polymorphic arrows consists just in verifying that the form of the domain and codomain satisfies certain simple rules; for instance, domain and codomain of an identity must be the same, while a first projection must to start on $\times(X_1 \cdots X_p)$ and end on $\times(X_1, \cdots, X_{p'})$, with $0 < p' < p$. A simple arrow which is not polymorphic can obviously be type-checked directly.

Then, each time we have to type-check an arrow of the form $X \xrightarrow{(f,g)} Y$, we try to split $\bar{Y}$, seen as a $\times$-list, into two sublists, and we check that $f$ and $g$ are well-typed with respect to the domain $X$ and to the codomain given by the first and second sublist, respectively (a dual behaviour is kept with respect to $(- \mid -)$). Of course, all splittings have to be tried. For instance, if $f : X + X \to Z$ and $g : Y \to Z$, we can write

$$(f \mid g) : X + (X + Y) \to Z.$$

When the distributive expressions will be turned in flat normal form the domain will become $+(X\ X\ Y)$, and subsequently the rule for $(- \mid -)$ will successfully try the splitting $+(X\ X)/+(Y)$ (the latter item being reducible to the normal form $Y$).

This mechanism can lead to ambiguous arrows. For instance, definining the arrow **twist** (commutativity isomorphism) as

$$X + (X + X) \xrightarrow{(\mathrm{inj}_2 \mid \mathrm{inj}_1)} (X + X) + X$$

does not make clear whether it is required to twist around the first or the second + sign (remember that associativity is irrelevant). In these cases, it is necessary to specify completely the arrows, as in

$$X + (X + X) \xrightarrow{(\mathrm{inj}_2^{X+X,X} \mid \mathrm{inj}_1^{X+X,X})} (X + X) + X.$$

List splitting, however, cannot take care of *composition*. If we have to check if $X \xrightarrow{g \circ f} Z$, we need to establish that $\mathbf{cod}(f) = \mathbf{dom}(g)$. However, it is not possible in general to compute $\mathbf{cod}(f)$ or $\mathbf{dom}(g)$ (also because $f$ and $g$ could be not well-typed). Nonetheless, we expect that arrows with a unique semantics such as $\mathrm{inj}_1 \circ \mathrm{i}_X : X \to I + I$ should be checkable.

Whenever we compose two arrow, we should solve a set of *type equations* over objects. Some theory is necessary in order to establish under which conditions a system of such equations has a unique solution, and how to compute it. As a simple example, consider the arrow

$$X \xrightarrow{\mathrm{proj}_2 \circ f \circ \mathrm{inj}_1} Y'$$

with $f : X + X' \to Y \times Y'$. The domain of $f$ and the codomain of $\mathrm{inj}_1$ have to be the same, and by definition of injection we write the equation $X + \mathbf{V} = X + X'$ (where $\mathbf{V}$ is a variable representing the unknown part of the codomain of $\mathrm{inj}_1$) which admits the unique solution $\mathbf{V} = X'$, because objects form a cancellative monoid with respect to both sum and product. For analogous reasons we have $Y \times Y' = \mathbf{V}' \times Y'$, which gives $\mathbf{V}' = Y$. In this simple case distinct variables appear in distinct equations, but this does not happen in general.

While this is the "best" solution to the problem of type-checking composition of arrows, it has big computational disadvantages. Moreover, it can be argued that simpler methods can lead to more affordable type-checkers whose power is comparable to the previous case, in the sense that the arrows which are no longer checkable with the simpler method are extremely artificial.

We want to describe here such a method. The basic idea is that when type-checking $g \circ f$ we just need a *reasonable guess* for a $Y$ such that $X \xrightarrow{f} Y \xrightarrow{g} Z$. If $f$ and $g$ are well-typed with a certain choice of $Y$, we have certainly $Y = \mathbf{cod}(f) = \mathbf{dom}(g)$. The simplest way of building such a $Y$ is to compute the domain of $g$ (or the codomain of $f$) *assuming that $f$ and $g$ are well-typed*. This allows to use that standard equation of category theory, such as $\mathbf{cod}((f, g)) = \mathbf{cod}(f) \times \mathbf{cod}(g)$. It is easy to see, for instance, that in the case of $\mathrm{inj}_1 \circ \mathrm{i}_X : X \to I + I$ the algorithm would try to compute the domain of $\mathrm{inj}_1$, which is unknown, but would then try to compute the codomain of $\mathrm{i}_X$, which is $I$, and then the type-checking of $X \xrightarrow{\mathrm{i}_X} I$ and $I \xrightarrow{\mathrm{inj}_1} I + I$ would be successful.

More generally, it is not difficult to show that

**Theorem 30** Let $f_0, f_1, \dots, f_n$ be a sequence of arrows. If for each $k = 1, 2, \dots, n$ we have that $\mathbf{cod}(f_{k-1})$ or $\mathbf{dom}(f_k)$ are computable, then $f_0 \circ f_1 \circ \cdots \circ f_n$ can be type-checked.

Getting back to our example

$$X \xrightarrow{\mathrm{proj}_2 \circ f \circ \mathrm{inj}_1} Y',$$

the type checker would perform as follows (supposing $\circ$ is left associative): $\mathbf{dom}(\mathrm{proj}_2)$ is not computable, but $\mathbf{cod}(f \circ \mathrm{inj}_1) = \mathbf{cod}(f) = Y \times Y'$ is, and $Y \times Y' \xrightarrow{\mathrm{proj}_2} Y'$ type-checks correctly. When checking $f \circ \mathrm{inj}_1$, $\mathbf{cod}(\mathrm{inj}_1)$ cannot be built, but $\mathbf{dom}(f) = X + X'$, and finally $X \xrightarrow{\mathrm{inj}_1} X + X'$ and $X + X' \xrightarrow{f} Y \times Y'$ type-check correctly.

It is of course possible to build artificial examples such as

$$X \times Z \xrightarrow{\mathrm{inj}_1 \circ \mathrm{proj}_1} X + Y.$$

In principle, we could establish (if $X$, $Y$ and $Z$ are simple objects) that the codomain of $\text{proj}_1$ is $X$, which coincides with the domain of $\text{inj}_1$. Such arrows, however, are almost unreadable by humans for the same reasons which make it difficult to manage them on a computer.

## 5.4 Some remarks on implementation

We would like to remark that the new coding proposed here can be easily and efficiently implemented on a standard computing machine as follows: each tree is represented by a pointer (either to some basic data representation, or to a forest tagged with a natural number), and a forest is represented by an array of such pointers. It is not difficult to see that this is the standard way of storing variant record, for instance in object-oriented languages with dynamic typing. Thus, the coding presented here is an essential step towards an efficient implementation of distributive programs on standard machines.

It could be argued that the old coding could be "compressed" by representing sequences of the special $e$ symbol as tagged natural numbers. Thus, for instance, the representation of an element of $n \cdot X^n$ would be of the form $e^{jn}x_1x_2 \cdots x_n e^{kn}$, which requires $O(n)$ space. However, this would impose the application of a very complicated "packing/unpacking mechanism" every time a derived arrow has to be computed. Moreover, we would always need, for instance, $O(n)$ storage for $I^n$ (against $O(1)$ in the new coding).

In other words, in real implementations the new coding has always a good advantage (although in some cases it is asymptotically equivalent).

# Chapter 6

# A literate IMP$(G)$ implementation

This chapter presents a series of tools for the manipulation of programs written in **IMP**$(G)$ which use the techniques developed in Chapter 5. The tools have been realized inside ASF+SDF, a system for developing specifications and manipulating rewriting systems created at the University of Amsterdam [Klint:1993]. The description, which alternates formal and natural language, is an example of *literate programming* [Knuth:1984], and has been realized using the tool ToLaTeX [Visser & Klint:1994]. The type-checker has been presented in [Vigna:1995].

The choice of ASF+SDF has the great advantage of making extending or changing the syntax of the language very easy, so that one can try, for instance, many different mathematical formalisms (e.g., $f \circ g$, $fg$ or $f; g$ for functional composition). The main disadvantage is the need of large resources, even for running relatively small programs. However, this is not a big issue, since our interest here is in implementing in detail the algorithmical and mathematical ideas described in Chapter 5. The description given in the ASF+SDF syntax, while being very near to the standard mathematical notation, has the advantage of being executable. A direct implementation in a more efficient language would be straightforward.

The reader interested in understanding every detail of the following pages should have some knowledge of the ASF+SDF system, but for a general understanding no particular background is necessary. We just remark that in the following sections the formal syntax is context-free (the rules are written "from right to left", i.e., they are seen as functions rather than productions), while the equational part specifies rewrite rules with equational conditions. "Execution" in the context of ASF+SDF means that a term is reduced using the given equations, thought of as reduction rules directed from left to right (no specified order should be assumed, except that the rules marked by **otherwise** will be the last to be executed). A rule is applied only if its conditions are all satisfied.

The key feature of ASF+SDF we shall use is the possibility of specifying *list sorts* and *list variables* (for examples of usage of lists, see for instance Section 6.7). When a list sort term is unified with one or more list sort variables, many splitting are usually possible. For instance, let $ABC$ be such a term; unification with $x^+y^*$ (where $x^+$ is a list sort variable which cannot be instantiated to an empty list, and $y^*$ is a list sort variable which can be instantiated to any list) can produce several different results, such as $x^+ = A$ and $y^* = CB$, or $x^+ = ABC$ and $y^*$ equal to the empty list. In this case, when ASF+SDF tries to satisfy the conditions of an equation all possible splittings are checked. On failure, the system *backtracks* and tries a new splitting, which could possibly lead to a positive result. This makes an implementation of the type-checking algorithm informally described in Chapter 5 strightforward, but with a drawback: it is not possible to detect ambiguous arrows. This is of course a limitation of the present implementation, and not of the algorithm itself.

Note that we omitted from the specification the modules providing layout tokens, boolean variables and integer support, since they are standard.

## 6.1 Identifiers

This module defines the identifiers allowed in the declaration and definition of objects and arrows. Instead of hard-coding the names of the structural arrows, we define one sort and one context free function per structural arrow, and assign a literal to each sort. Then, the literals are referred to through a variable of the corresponding sort. In this way we obtain three effects: the names can be easily redefined, we can name in several ways the same structural arrow, and the possibility of a wrong interpretation of a literal (i.e., as an basic arrow instead of a structural arrow) is avoided, due to the precedence rules of SDF.

**imports** Layout
**exports**
  **sorts** BASIC-ARROW BASIC-OBJ INITIAL TERMINAL IDENTITY
  FIRST-INJ SECOND-INJ FIRST-PROJ SECOND-PROJ
  INITIAL-MAP TERMINAL-MAP DIST

An arrow or object name must start with a letter (lower case for an arrow, upper case for an object). Note that we do not allow the usage of dashes, for they would create spurious syntax errors each time an object is not separated from a DART by a layout token (see Section 6.3). Circumflex accents and underscores are allowed in order to use (the TEX version of) mathematical notations such as $f_1$ or $k^3$.

**lexical syntax**
  [*a-z*][*A-Za-z0-9_^*]∗ → BASIC-ARROW
  [*A-Z*][*A-Za-z0-9_^*]∗ → BASIC-OBJ

It would be nice to be able to use "¡" for terminal maps, as it is available in ISO-8859.1 (i.e., Latin-1), but this would make the specification font-dependent; moreover, the current version of ASF+SDF does not support extended character codes. The alias used for *Terminal-Map*, however, is exactly "¡".

**context-free syntax**
  "*I*" → TERMINAL
  "*O*" → INITIAL
  "*id*" → IDENTITY
  "*inj_1*" → FIRST-INJ
  "*inj_2*" → SECOND-INJ
  "*proj_1*" → FIRST-PROJ
  "*proj_2*" → SECOND-PROJ
  "!" → INITIAL-MAP
  "*term*" → TERMINAL-MAP
  "*dist*" → DIST

The following variables are all aliased to a representation which matches the context free syntax. For instance, *First-Proj* becomes $\mathrm{proj}_1$. Note that *Dist* becomes $\delta^{-1}$.

**variables**
  *Initial* → INITIAL
  *Terminal* → TERMINAL
  *Identity* → IDENTITY
  *First-Inj* → FIRST-INJ
  *First-Proj* → FIRST-PROJ
  *Second-Inj* → SECOND-INJ
  *Second-Proj* → SECOND-PROJ
  *Initial-Map* → INITIAL-MAP

*Terminal-Map* → TERMINAL-MAP
*Dist* → DIST

## 6.2 Objects

An object is built by summing or multiplying together objects (starting from simple ones). We have standard assumptions about the relative priority of sum and product (note that we allow juxtaposition to denote product).

We note again that the ability to handle ISO-8859.1 codes would be very useful here, because the symbol "×" is part of its character set. Aliasing "*" to "×" is not possible, for also the occurrences of "*" in variable name definitions would be aliased.

**imports** Layout Identifiers[6.1]
**exports**
**sorts** STRUCTURAL-OBJ SIMPLE-OBJ OBJ
**context-free syntax**
TERMINAL → STRUCTURAL-OBJ
INITIAL → STRUCTURAL-OBJ

STRUCTURAL-OBJ → SIMPLE-OBJ
BASIC-OBJ → SIMPLE-OBJ

SIMPLE-OBJ → OBJ

OBJ "+" OBJ → OBJ {**left**}
OBJ "∗" OBJ → OBJ {**left**}
OBJ OBJ → OBJ {**left**}
"(" OBJ ")" → OBJ {**bracket**}
**priorities**
{**left**: OBJ OBJ → OBJ, OBJ "∗"OBJ → OBJ} > OBJ "+"OBJ → OBJ

Object variables are named with a single letter, in order to follow the mathematical custom. We use different group of letters in order to discriminate between basic, simple and compound entities. The underscore will be removed by aliasing, and has the sole purpose of avoiding clashes of object names (we suppose that names like $A$ or $X$ will be rather common in user programs). Note that since variables representing structural objects are seldom used, we prefer to name them in a different manner.

**variables**
"_"[*ABC*][*0-9*]∗[′]∗ → BASIC-OBJ
"_"[*ABC*]"∗"[*0-9*]∗[′]∗ → {BASIC-OBJ ","}∗
*Structural-Obj* [*0-9*]∗[′]∗ → STRUCTURAL-OBJ
"_"[*ST*][*0-9*]∗[′]∗ → SIMPLE-OBJ
"_"[*XYZ*][*0-9*]∗[′]∗ → OBJ
"_"[*XYZ*]"∗"[*0-9*]∗[′]∗ → {OBJ ","}∗

The only equation described here has the purpose of making a product denoted by juxtaposition explicit.

**equations**

[Obj1] $X\ Y\ =\ X * Y$

## 6.3 Arrows

An arrow is built by summing, multiplying, pairing (on sum or products), composing or iterating other arrows (starting from simple ones). DART-STARTER and DART-ENDER are visual markers which will be used in order to describe composition of arrows in Section 6.13.

Note that each structural arrow can be described in two ways: a *completely specified* structural arrow contains full information about its domain and codomain, while a *polymorphic* structural arrow is declared just by assigning a type (identity, first projection, and so on). The unification mechanism of the type-checker will find out a reasonable meaning for the arrow. Analogously, it is possible to use the **call**[ − ] operator in the short form **call**[ $f$ ].

Composition can be specified in two form: by means of the ";" operator, or using the "∘" operator. The meaning of $f; g$ is "do first $f$, then $g$," while the meaning of $f \circ g$ is "do first $g$, then $f$".

```
imports Layout Identifiers[6.1] Objects[6.2]
exports
  sorts STRUCTURAL-ARROW SIMPLE-ARROW ARROW
        DART-STARTER DART-ENDER DART
  lexical syntax
    [\−]+         → DART-STARTER
    [\−]+">"      → DART-ENDER
  context-free syntax
    IDENTITY                              → STRUCTURAL-ARROW
    FIRST-INJ                             → STRUCTURAL-ARROW
    FIRST-PROJ                            → STRUCTURAL-ARROW
    SECOND-INJ                            → STRUCTURAL-ARROW
    SECOND-PROJ                           → STRUCTURAL-ARROW
    INITIAL-MAP                           → STRUCTURAL-ARROW
    TERMINAL-MAP                          → STRUCTURAL-ARROW
    DIST                                  → STRUCTURAL-ARROW

    IDENTITY "(" OBJ ")"                  → STRUCTURAL-ARROW
    FIRST-INJ "(" OBJ "," OBJ ")"         → STRUCTURAL-ARROW
    FIRST-PROJ "(" OBJ "," OBJ ")"        → STRUCTURAL-ARROW
    SECOND-INJ "(" OBJ "," OBJ ")"        → STRUCTURAL-ARROW
    SECOND-PROJ "(" OBJ "," OBJ ")"       → STRUCTURAL-ARROW
    INITIAL-MAP "(" OBJ ")"               → STRUCTURAL-ARROW
    TERMINAL-MAP "(" OBJ ")"              → STRUCTURAL-ARROW
    DIST "(" OBJ "," OBJ "," OBJ ")"      → STRUCTURAL-ARROW

    BASIC-ARROW                           → SIMPLE-ARROW
    STRUCTURAL-ARROW                      → SIMPLE-ARROW

    SIMPLE-ARROW                          → ARROW

    ARROW "+" ARROW                       → ARROW        {right}
    ARROW "∗" ARROW                       → ARROW        {right}
    ARROW "|" ARROW                       → ARROW        {right}
    ARROW "," ARROW                       → ARROW        {right}
    ARROW ";" ARROW                       → ARROW        {right}
```

| | | |
|---|---|---|
| ARROW "∘" ARROW | → ARROW | {**right**} |
| "(" ARROW ")" | → ARROW | {**bracket**} |
| **call** "[" OBJ "," OBJ "," OBJ "," ARROW "]" | → ARROW | |
| **call** "[" ARROW "]" | → ARROW | |
| DART-STARTER ARROW DART-ENDER | → DART | |

The reader could have noticed that we allow to pair two arrows with "|" or "," *without external brackets*. Thus, $f \mid g$ and $f, g$ are valid arrows, while the standard notation would require $(f \mid g)$ and $(f, g)$. However, dropping the brackets allows to pair an arbitrary number of arrows instead of just two. Moreover, since we declare that the two kinds of pairing are reciprocally non-associative, and that they are at the bottom of the priority order, the user will be always forced to surround a pairing with brackets, except at the outermost level.

**priorities**
ARROW ";"ARROW → ARROW > ARROW "∗"ARROW → ARROW >
ARROW "+"ARROW → ARROW > {**non-assoc**: ARROW "|"ARROW → ARROW,
ARROW ","ARROW → ARROW}, ARROW "∘"ARROW → ARROW >
ARROW "∗"ARROW → ARROW

The same considerations of Section 6.2 about variable naming apply here. Note that *Dart-Starter* and *Dart-Ender* are aliased to − and →.

**variables**

| | |
|---|---|
| "_"[*abc*][*0-9*]∗[′]∗ | → BASIC-ARROW |
| "_"[*abc*]"∗"[*0-9*]∗[′]∗ | → {BASIC-ARROW ","}∗ |
| *Structural-Arrow* [*0-9*]∗[′]∗ | → STRUCTURAL-ARROW |
| "_"[*st*][*0-9*]∗[′]∗ | → SIMPLE-ARROW |
| "_"[*fgh*][*0-9*]∗[′]∗ | → ARROW |
| *Dart-Starter* | → DART-STARTER |
| *Dart-Ender* | → DART-ENDER |

The only equation described here has the purpose of reducing a composition denoted by $f \circ g$ to the other form.

**equations**

[Arr1] $g \circ f \;=\; f;\, g$

## 6.4 Support

This module contains syntax and equations which are needed by all high-level modules. In particular, it takes care of the conversion to flat normal form.

**imports** Identifiers[6.1] Objects[6.2] Arrows[6.3] Booleans Ints

The sort FOBJ (short for “flat object”) describes objects in a form which is much less readable, but also much less ambiguous, than the one described in Section 6.2. In practice, sums and products will be lists with explicit bracketing, and we shall arrange things in such a way that flat objects are always in *normal form*, i.e., they do not contain lists with just one element or sublists which could be merged by associativity. The operator **flat** turns an object into a flat object. The rôle of the additional symbols $\oplus$ and $\otimes$ and of the last four operators will be described in detail in the equational part.

**exports**
**sorts** FOBJ
**context-free syntax**

| | |
|---|---|
| BASIC-OBJ | $\rightarrow$ FOBJ |
| $\perp$ | $\rightarrow$ FOBJ |
| “$*$” “(” FOBJ$*$ “)” | $\rightarrow$ FOBJ |
| “$+$” “(” FOBJ$*$ “)” | $\rightarrow$ FOBJ |
| “$\oplus$” “(” FOBJ$*$ “)” | $\rightarrow$ FOBJ |
| “$\otimes$” “(” FOBJ$*$ “)” | $\rightarrow$ FOBJ |
| **flat** “[” OBJ “]” | $\rightarrow$ FOBJ |
| **len** “[” FOBJ$*$ “]” | $\rightarrow$ INT |
| **slen** “[” FOBJ “]” | $\rightarrow$ INT |
| **plen** “[” FOBJ “]” | $\rightarrow$ INT |
| *elem* INT *from* FOBJ$+$ | $\rightarrow$ FOBJ |

An arrow specification is given by a basic arrow name, a domain and a codomain. Instead of just writing down a triple, we prefer a syntax which mimicks the standard mathematical notation $f : X \rightarrow Y$. The sorts SPEC-LIST and BASIC-OBJ-LIST are used during the type-check in order to store the currently known entities (in other words, they are the symbol tables). Note the different notation ($X \rightarrow Y$ and $X \leadsto Y$) for basic arrows and arrows which have been compiled.

**sorts** SPEC SPEC-LIST BASIC-OBJ-LIST
**context-free syntax**

| | |
|---|---|
| BASIC-ARROW “:” FOBJ “$\rightarrow$” FOBJ | $\rightarrow$ SPEC |
| BASIC-ARROW “:” FOBJ “$\leadsto$” FOBJ | $\rightarrow$ SPEC |
| “[” {SPEC “,”}$*$ “]” | $\rightarrow$ SPEC-LIST |
| “[” {BASIC-OBJ “,”}$*$ “]” | $\rightarrow$ BASIC-OBJ-LIST |

The following function returns the domain and codomain of a basic arrow, taking them from a specification list; in case of failure, “$\perp$” is returned. Again, the domain and codomain are returned in a slightly different form if the arrow has been previously compiled.

**sorts** DOM-COD
**context-free syntax**

| | |
|---|---|
| “[” FOBJ “$\rightarrow$” FOBJ “]” | $\rightarrow$ DOM-COD |
| “[” FOBJ “$\leadsto$” FOBJ “]” | $\rightarrow$ DOM-COD |
| $\perp$ | $\rightarrow$ DOM-COD |
| *spec of* BASIC-ARROW *in* SPEC-LIST | $\rightarrow$ DOM-COD |

These three functions are *predicates*, i.e., they are guaranteed to be reducible to *true* or *false*. They are used by the type-checker in order to issue meaningful error messages when the user has typed the wrong name for an object or an arrow.

**context-free syntax**

BASIC-OBJ *is in* BASIC-OBJ-LIST → BOOL
OBJ *is built from* BASIC-OBJ-LIST → BOOL
ARROW *is built from* SPEC-LIST → BOOL

The variables denoting flat objects have names which are implemented directly as TEX commands. Thus, the choice of the actual symbols is left to the definition of the commands `\flX`, `\flY` and `\flZ`. The present choice ($\Xi$, $\Theta$ and $\Omega$) is motivated by the need of choosing symbols easily distinguishable from the ones assigned to (simple) objects (moreover, the author always loved the fact that $\Xi$ and $\Theta$ are disconnected).

**variables**

*Spec* [*0-9*]∗[′]∗ → SPEC
*Spec* "∗"[*0-9*]∗[′]∗ → {SPEC ","}∗
"Ξ"[*0-9*]∗[′]∗ → FOBJ
"Θ"[*0-9*]∗[′]∗ → FOBJ
"Ω"[*0-9*]∗[′]∗ → FOBJ
"Ξ""+"[*0-9*]∗[′]∗ → FOBJ+
"Θ""+"[*0-9*]∗[′]∗ → FOBJ+
"Ω""+"[*0-9*]∗[′]∗ → FOBJ+
"Ξ""∗"[*0-9*]∗[′]∗ → FOBJ∗
"Θ""∗"[*0-9*]∗[′]∗ → FOBJ∗
"Ω""∗"[*0-9*]∗[′]∗ → FOBJ∗

**equations**

**(De)normalization of flat objects.** Our first set of equations defines the rules for normalizing flat objects. First of all, we discard useless operators; then, we merge associatively sublists. A term in normal form with respect to these rules is guaranteed to be a flat object in normal form.

[Snfs1] $+(\Xi) \;=\; \Xi$
[Snfp1] $*(\Xi) \;=\; \Xi$

[Snfs2] $+(\Xi^* \; +(\Theta^*) \; \Omega^*) \;=\; +(\Xi^* \; \Theta^* \; \Omega^*)$
[Snfp2] $*(\Xi^* \; *(\Theta^*) \; \Omega^*) \;=\; *(\Xi^* \; \Theta^* \; \Omega^*)$

These two rules, on the contrary, aims at *denormalizing* a flat object in normal form. When we shall be interested in seeing a given flat object as a sum (possibly with just one summand) we shall apply the $\oplus$ operator to the object. Since $\oplus$ and $+$ merge, but $\oplus$ is not cancelled when applied to a single object, the result will always be a $\oplus$-list (the same, of course, is true for $\otimes$).

[Snfos3] $\oplus(+(\Xi^*)) \;=\; \oplus(\Xi^*)$
[Snfop3] $\otimes(*(\Xi^*)) \;=\; \otimes(\Xi^*)$

Finally, this function inductively traverses an object tree, mapping binary operators to lists of two elements. The result is a flat object which will be normalized by the previous equations. Note that the initial and terminal objects are represented by the empty $+$-list and by the empty $\times$-list, respectively.

[Sfl1] $\mathbf{flat}[A] = A$

[Sfl2] $\mathbf{flat}[\varnothing] = +()$

[Sfl3] $\mathbf{flat}[I] = *()$

[Sfl4] $\mathbf{flat}[X * Y] = *(\mathbf{flat}[X]\ \mathbf{flat}[Y])$

[Sfl5] $\mathbf{flat}[X + Y] = +(\mathbf{flat}[X]\ \mathbf{flat}[Y])$

**Calculating the length of a flat object.** The following rules specify how to compute the *length* of a list of flat objects and the *length as a sum* (or a product) of a flat object (a sum is seen as a one-factor product, and a product is seen as a one-summand sum). Note the application of the $\oplus/\otimes$ operators.

[Slen1] $\mathbf{len}[] = 0$

[Slen2] $\mathbf{len}[\Xi\ \Xi^*] = 1 + \mathbf{len}[\Xi^*]$

[Sslen] $\oplus(\Xi) = \oplus(\Xi^*) \Rightarrow \mathbf{slen}[\Xi] = \mathbf{len}[\Xi^*]$

[Splen] $\otimes(\Xi) = \otimes(\Xi^*) \Rightarrow \mathbf{plen}[\Xi] = \mathbf{len}[\Xi^*]$

This function returns the $n$-th element of a list of flat objects (the first element having index 0). It is useful when type-checking the data corresponding to a sum.

[Sel1] $\mathit{elem}\ 0\ \mathit{from}\ \Xi\ \Xi^* = \Xi$

[Sel2] $n \neq 0 \Rightarrow \mathit{elem}\ n\ \mathit{from}\ \Xi\ \Xi^+ = \mathit{elem}\ n-1\ \mathit{from}\ \Xi^+$

**Extracting the specification of a simple arrow.** Here we have to discover the domain and codomain of a given simple arrow by scanning the given specification list. If the arrow is not known we return "$\perp$".

[Ssp1] $\mathit{spec\ of}\ a\ \mathit{in}\ [] = \perp$

[Ssp2] $\mathit{spec\ of}\ a\ \mathit{in}\ [a : \Xi \to \Theta,\ \mathit{Spec}^*] = [\Xi \to \Theta]$

[Ssp3] $\mathit{spec\ of}\ a\ \mathit{in}\ [a : \Xi \rightsquigarrow \Theta,\ \mathit{Spec}^*] = [\Xi \rightsquigarrow \Theta]$

[Ssp4] $a \neq b \Rightarrow \mathit{spec\ of}\ a\ \mathit{in}\ [b : \Xi \to \Theta,\ \mathit{Spec}^*] = \mathit{spec\ of}\ a\ \mathit{in}\ [\mathit{Spec}^*]$

[Ssp5] $a \neq b \Rightarrow \mathit{spec\ of}\ a\ \mathit{in}\ [b : \Xi \rightsquigarrow \Theta,\ \mathit{Spec}^*] = \mathit{spec\ of}\ a\ \mathit{in}\ [\mathit{Spec}^*]$

**Predicates.** The following predicate returns *true* if the given basic object is in the given basic object list.

[Sii1] $A\ \mathit{is\ in}\ [] = \mathit{false}$

[Sii2] $A\ \mathit{is\ in}\ [A,\ A^*] = \mathit{true}$

[Sii3] $A \neq B \Rightarrow A\ \mathit{is\ in}\ [B,\ A^*] = A\ \mathit{is\ in}\ [A^*]$

The following two predicates go one step further, and scan inductively an object (arrow) in order to establish if it has been built using valid simple objects (arrows, respectively). Note the difference between variables denoting simple and compound entities.

[Sobf1] $A$ *is built from* $[A^*]$ $=$ $A$ *is in* $[A^*]$

[Sobf2] *Structural-Obj is built from* $[A^*]$ $=$ *true*

[Sobf3] $X + Y$ *is built from* $[A^*]$ $=$ $X$ *is built from* $[A^*] \wedge Y$ *is built from* $[A^*]$

[Sobf4] $X * Y$ *is built from* $[A^*]$ $=$ $X$ *is built from* $[A^*] \wedge Y$ *is built from* $[A^*]$

[Sabf1] $$\frac{\textit{spec of } a \textit{ in } [\textit{Spec}^*] = \bot}{a \textit{ is built from } [\textit{Spec}^*] = \textit{false}}$$

[Sabf2] $$\frac{\textit{spec of } a \textit{ in } [\textit{Spec}^*] \neq \bot}{a \textit{ is built from } [\textit{Spec}^*] = \textit{true}}$$

[Sabf3] $s$ *is built from* $[\textit{Spec}^*]$ $=$ *true*

[Sabf4] $f + g$ *is built from* $[\textit{Spec}^*]$ $=$ $f$ *is built from* $[\textit{Spec}^*] \wedge g$ *is built from* $[\textit{Spec}^*]$

[Sabf5] $f * g$ *is built from* $[\textit{Spec}^*]$ $=$ $f$ *is built from* $[\textit{Spec}^*] \wedge g$ *is built from* $[\textit{Spec}^*]$

[Sabf6] $f \mid g$ *is built from* $[\textit{Spec}^*]$ $=$ $f$ *is built from* $[\textit{Spec}^*] \wedge g$ *is built from* $[\textit{Spec}^*]$

[Sabf7] $f, g$ *is built from* $[\textit{Spec}^*]$ $=$ $f$ *is built from* $[\textit{Spec}^*] \wedge g$ *is built from* $[\textit{Spec}^*]$

[Sabf8] $f; g$ *is built from* $[\textit{Spec}^*]$ $=$ $f$ *is built from* $[\textit{Spec}^*] \wedge g$ *is built from* $[\textit{Spec}^*]$

[Sabf9] **call**$[X, Y, Z, f]$ *is built from* $[\textit{Spec}^*]$ $=$ $f$ *is built from* $[\textit{Spec}^*]$

[Sabf10] **call**$[f]$ *is built from* $[\textit{Spec}^*]$ $=$ $f$ *is built from* $[\textit{Spec}^*]$

## 6.5 DomCod

This module is devoted to the machinery which is necessary in order to compute the domain and codomain of an arrow.

**imports** Objects[6.2] Arrows[6.3] Booleans Support[6.4]

It is most important to note that these functions work assuming that their argument is well-typed. If this is not the case, the result should be checked using the type-checker (see Section 6.9).

**exports**
  **context-free syntax**
    **dom** “[” ARROW “]” using SPEC-LIST → FOBJ
    **cod** “[” ARROW “]” using SPEC-LIST → FOBJ

The general idea of the algorithm we shall use is as follows: we assume that the given arrow is well-typed, and we apply the standard equations which give the the domain and codomain of an arrow of a free distributive category. When we get to a structural arrow, there is a definite possibility that we cannot compute its domain or codomain. In this case, we return $\bot$. Any flat object contaning $\bot$ is reduced to $\bot$ by the first two equations, so we can safely apply structural induction.

**equations**

[DCnf1] $+(\Xi^* \perp \Theta^*) \;=\; \perp$

[DCnf2] $*(\Xi^* \perp \Theta^*) \;=\; \perp$

**Building the (co)domain of a basic or compiled arrow.** We can always build the (co)domain of a basic arrow, or of a compiled arrow, by looking at the given specification list.

[DCda1] $$\frac{\textit{spec of } a \textit{ in } [\textit{Spec}^*] = [\Xi \to \Theta]}{\mathbf{dom}[a] \text{ using } [\textit{Spec}^*] = \Xi}$$

[DCda2] $$\frac{\textit{spec of } a \textit{ in } [\textit{Spec}^*] = [\Xi \leadsto \Theta]}{\mathbf{dom}[a] \text{ using } [\textit{Spec}^*] = \Xi}$$

[DCda3] $$\frac{\textit{spec of } a \textit{ in } [\textit{Spec}^*] = \perp}{\mathbf{dom}[a] \text{ using } [\textit{Spec}^*] = \perp}$$

[DCca1] $$\frac{\textit{spec of } a \textit{ in } [\textit{Spec}^*] = [\Xi \leadsto \Theta]}{\mathbf{cod}[a] \text{ using } [\textit{Spec}^*] = \Theta}$$

[DCca2] $$\frac{\textit{spec of } a \textit{ in } [\textit{Spec}^*] = [\Xi \to \Theta]}{\mathbf{cod}[a] \text{ using } [\textit{Spec}^*] = \Theta}$$

[DCca3] $$\frac{\textit{spec of } a \textit{ in } [\textit{Spec}^*] = \perp}{\mathbf{cod}[a] \text{ using } [\textit{Spec}^*] = \perp}$$

**Building the (co)domain of a structural arrow.** The following list of equations is rather trivial. We just mention that initial and terminal maps are the only ones for which we can compute something in the polymorphic case.

[DCdid] $\mathbf{dom}[\mathbf{1}] \text{ using } [\textit{Spec}^*] \;=\; \perp$

[DCcid] $\mathbf{cod}[\mathbf{1}] \text{ using } [\textit{Spec}^*] \;=\; \perp$

[DCdID] $\mathbf{dom}[\mathbf{1}(X)] \text{ using } [\textit{Spec}^*] \;=\; \mathbf{flat}[X]$

[DCcID] $\mathbf{cod}[\mathbf{1}(X)] \text{ using } [\textit{Spec}^*] \;=\; \mathbf{flat}[X]$

[DCdi1] $\mathbf{dom}[\mathrm{inj}_1] \text{ using } [\textit{Spec}^*] \;=\; \perp$

[DCci1] $\mathbf{cod}[\mathrm{inj}_1] \text{ using } [\textit{Spec}^*] \;=\; \perp$

[DCdI1] $\mathbf{dom}[\mathrm{inj}_1(X, Y)] \text{ using } [\textit{Spec}^*] \;=\; \mathbf{flat}[X]$

[DCcI1] $\mathbf{cod}[\mathrm{inj}_1(X, Y)] \text{ using } [\textit{Spec}^*] \;=\; \mathbf{flat}[X + Y]$

[DCdi2] $\mathbf{dom}[\mathrm{inj}_2]$ using $[Spec^*] = \bot$
[DCci2] $\mathbf{cod}[\mathrm{inj}_2]$ using $[Spec^*] = \bot$
[DCdI2] $\mathbf{dom}[\mathrm{inj}_2(X, Y)]$ using $[Spec^*] = \mathbf{flat}[Y]$
[DCcI2] $\mathbf{cod}[\mathrm{inj}_2(X, Y)]$ using $[Spec^*] = \mathbf{flat}[X + Y]$

[DCdp1] $\mathbf{dom}[\mathrm{proj}_1]$ using $[Spec^*] = \bot$
[DCcp1] $\mathbf{cod}[\mathrm{proj}_1]$ using $[Spec^*] = \bot$
[DCdP1] $\mathbf{dom}[\mathrm{proj}_1(X, Y)]$ using $[Spec^*] = \mathbf{flat}[X * Y]$
[DCcP1] $\mathbf{cod}[\mathrm{proj}_1(X, Y)]$ using $[Spec^*] = \mathbf{flat}[X]$

[DCdp2] $\mathbf{dom}[\mathrm{proj}_2]$ using $[Spec^*] = \bot$
[DCcp2] $\mathbf{cod}[\mathrm{proj}_2]$ using $[Spec^*] = \bot$
[DCdP2] $\mathbf{dom}[\mathrm{proj}_2(X, Y)]$ using $[Spec^*] = \mathbf{flat}[X * Y]$
[DCcP2] $\mathbf{cod}[\mathrm{proj}_2(X, Y)]$ using $[Spec^*] = \mathbf{flat}[Y]$

[DCdini] $\mathbf{dom}[!]$ using $[Spec^*] = +()$
[DCcini] $\mathbf{cod}[!]$ using $[Spec^*] = \bot$
[DCdINI] $\mathbf{dom}[!(X)]$ using $[Spec^*] = +()$
[DCcINI] $\mathbf{cod}[!(X)]$ using $[Spec^*] = \mathbf{flat}[X]$

[DCdter] $\mathbf{dom}[\mathrm{i}]$ using $[Spec^*] = \bot$
[DCcter] $\mathbf{cod}[\mathrm{i}]$ using $[Spec^*] = *()$
[DCdTER] $\mathbf{dom}[\mathrm{i}(X)]$ using $[Spec^*] = \mathbf{flat}[X]$
[DCcTER] $\mathbf{cod}[\mathrm{i}(X)]$ using $[Spec^*] = *()$

[DCdd1] $\mathbf{dom}[\delta^{-1}]$ using $[Spec^*] = \bot$
[DCcd2] $\mathbf{cod}[\delta^{-1}]$ using $[Spec^*] = \bot$
[DCdD1] $\mathbf{dom}[\delta^{-1}(X, Y_1, Y_2)]$ using $[Spec^*] = \mathbf{flat}[X * (Y_1 + Y_2)]$
[DCcD2] $\mathbf{cod}[\delta^{-1}(X, Y_1, Y_2)]$ using $[Spec^*] = \mathbf{flat}[X * Y_1 + X * Y_2]$

**Building the (co)domain of a derived arrow.** Composition, sum and product of functions have very simple rewrite rules (remember that any flat object containing $\bot$ reduces to $\bot$).

[DCds] $\mathbf{dom}[f + g]$ using $[Spec^*] = +(\mathbf{dom}[f]$ using $[Spec^*]$ $\mathbf{dom}[g]$ using $[Spec^*])$
[DCcs] $\mathbf{cod}[f + g]$ using $[Spec^*] = +(\mathbf{cod}[f]$ using $[Spec^*]$ $\mathbf{cod}[g]$ using $[Spec^*])$

[DCdp] $\mathbf{dom}[f * g]$ using $[Spec^*] = *(\mathbf{dom}[f]$ using $[Spec^*]$ $\mathbf{dom}[g]$ using $[Spec^*])$
[DCcp] $\mathbf{cod}[f * g]$ using $[Spec^*] = *(\mathbf{cod}[f]$ using $[Spec^*]$ $\mathbf{cod}[g]$ using $[Spec^*])$

[DCdco] $\mathbf{dom}[f; g]$ using $[Spec^*] = \mathbf{dom}[f]$ using $[Spec^*]$
[DCcco] $\mathbf{cod}[f; g]$ using $[Spec^*] = \mathbf{cod}[g]$ using $[Spec^*]$

The case of the $(- \mid -)$ and $(-, -)$ operators is more interesting. When computing the codomain of $(f \mid g)$, we can take the information either from $f$ or from $g$. If $(f \mid g)$ is well-typed and both $f$ and $g$ have a computable codomain, then the information will be the same, but it is possible that the codomain of $f$ or $g$ is not computable because of polymorphism. This is why we analyze all possible cases. Analogous (but dual) considerations can be done about the $(-, -)$ operator.

[DCdb] $\mathbf{dom}[f \mid g] \text{ using } [\mathit{Spec}^*] \; = \; +(\mathbf{dom}[f] \text{ using } [\mathit{Spec}^*] \; \mathbf{dom}[g] \text{ using } [\mathit{Spec}^*])$

[DCcb1] $$\frac{\mathbf{cod}[f] \text{ using } [\mathit{Spec}^*] \neq \bot}{\mathbf{cod}[f \mid g] \text{ using } [\mathit{Spec}^*] = \mathbf{cod}[f] \text{ using } [\mathit{Spec}^*]}$$

[DCcb2] $$\frac{\mathbf{cod}[g] \text{ using } [\mathit{Spec}^*] \neq \bot}{\mathbf{cod}[f \mid g] \text{ using } [\mathit{Spec}^*] = \mathbf{cod}[g] \text{ using } [\mathit{Spec}^*]}$$

[DCcb3] $$\frac{\mathbf{cod}[f] \text{ using } [\mathit{Spec}^*] = \bot, \;\; \mathbf{cod}[g] \text{ using } [\mathit{Spec}^*] = \bot}{\mathbf{cod}[f \mid g] \text{ using } [\mathit{Spec}^*] = \bot}$$

[DCdc1] $$\frac{\mathbf{dom}[f] \text{ using } [\mathit{Spec}^*] \neq \bot}{\mathbf{dom}[f, g] \text{ using } [\mathit{Spec}^*] = \mathbf{dom}[f] \text{ using } [\mathit{Spec}^*]}$$

[DCdc2] $$\frac{\mathbf{dom}[g] \text{ using } [\mathit{Spec}^*] \neq \bot}{\mathbf{dom}[f, g] \text{ using } [\mathit{Spec}^*] = \mathbf{dom}[g] \text{ using } [\mathit{Spec}^*]}$$

[DCdc3] $$\frac{\mathbf{dom}[f] \text{ using } [\mathit{Spec}^*] = \bot, \;\; \mathbf{dom}[g] \text{ using } [\mathit{Spec}^*] = \bot}{\mathbf{dom}[f, g] \text{ using } [\mathit{Spec}^*] = \bot}$$

[DCcc] $\mathbf{cod}[f, g] \text{ using } [\mathit{Spec}^*] \; = \; *(\mathbf{cod}[f] \text{ using } [\mathit{Spec}^*] \; \mathbf{cod}[g] \text{ using } [\mathit{Spec}^*])$

**Building the (co)domain of call$[f]$.** Unless an application of **call**$[-]$ has been completely specified, there is nothing we can say about the domain and codomain of an iterated arrow. We could try to break the domain of $f$ in two summands in all possible ways, but nothing guarantees that the computation of the codomain would be coherent with the chosen splitting (so for $f : X{+}U{+}U \to U{+}U{+}Y$ we could obtain $\mathbf{dom}(\mathbf{call}[f]) = X$ but $\mathbf{cod}(\mathbf{call}[f]) = U + Y$).

[DCdCALL] $\mathbf{dom}[\mathbf{call}[X, Y, Z, f]] \text{ using } [\mathit{Spec}^*] \; = \; \mathbf{flat}[X]$

[DCcCALL] $\mathbf{cod}[\mathbf{call}[X, Y, Z, f]] \text{ using } [\mathit{Spec}^*] \; = \; \mathbf{flat}[Z]$

[DCdcall] $\mathbf{dom}[\mathbf{call}[f]] \text{ using } [\mathit{Spec}^*] \; = \; \bot$

[DCccall] $\mathbf{cod}[\mathbf{call}[f]] \text{ using } [\mathit{Spec}^*] \; = \; \bot$

## 6.6 Data

This module specifies the representation of the elements of a flat object in normal form. It is essentially a description of the operators of the free $\boldsymbol{N}$-motor [Vigna:1991] on the set of terms of sort BASIC-DATA. Note that the empty forest, which represents the only element of $I$, is the empty term, and that the terms of sort SINGLE-DATA are exactly the trees.

Should syntactical ambiguities arise (we do not know which are the subsorts of BASIC-DATA, because they are user-specified) the generators (i.e., the terms of sort BASIC-DATA) can be enclosed in **{** and **}**.

**imports** Layout Ints
**exports**
  **sorts** DATA BASIC-DATA SINGLE-DATA DATA-DATA
  **context-free syntax**
    "**{**" BASIC-DATA "**}**" $\rightarrow$ BASIC-DATA {**bracket**}
    BASIC-DATA $\rightarrow$ SINGLE-DATA

    "$\triangleleft$" INT "," DATA "$\triangleright$" $\rightarrow$ SINGLE-DATA

    SINGLE-DATA$*$ $\rightarrow$ DATA

There will be situations in which we shall want to concatenate two forests. The dot operator serves this purpose.

  **context-free syntax**
    DATA "$\cdot$" DATA $\rightarrow$ DATA
  **variables**
    "_"[*e*][*0-9*]$*$[$'$]$*$ $\rightarrow$ BASIC-DATA
    "_"[*u*][*0-9*]$*$[$'$]$*$ $\rightarrow$ SINGLE-DATA
    "_"[*u*]"$+$"[*0-9*]$*$[$'$]$*$ $\rightarrow$ SINGLE-DATA$+$
    "_"[*u*]"$*$"[*0-9*]$*$[$'$]$*$ $\rightarrow$ SINGLE-DATA$*$
    "_"[*d*][*0-9*]$*$[$'$]$*$ $\rightarrow$ DATA

The first equation of this module is fundamental; it normalizes the representation of the elements of a sum when they are injected into another sum. The other equation implements trivially the concatenation of forests.

**equations**

[Dnf1] $\triangleleft\, m,\ \triangleleft\, n,\ d\ \triangleright\ \triangleright\ =\ \triangleleft\, m+n,\ d\ \triangleright$

[Dnf2] $u^*\cdot u^{*\prime}\ =\ u^*\ u^{*\prime}$

## 6.7 Errors

This module provides error messages and error lists for the compiler and the type-checker. Beside duplicate or missing identifiers, we could discover that an arrow does not start or end where it should. We also define a dot operator which concatenates lists of errors.

**imports** Objects[6.2] Arrows[6.3]
**exports**
**sorts** ERROR ERROR-LIST
**context-free syntax**

| | |
|---|---|
| "*object␣name*" SIMPLE-OBJ "*already␣used*" | → ERROR |
| "*arrow␣name*" SIMPLE-ARROW "*already␣used*" | → ERROR |
| "*object*" OBJ "*contains␣undeclared␣basic␣objects*" | → ERROR |
| "*arrow*" ARROW "*contains␣undeclared␣basic␣arrows*" | → ERROR |
| "*arrow*" ARROW "*is␣not␣from*" OBJ "*to*" OBJ | → ERROR |
| "*arrow*" ARROW "*is␣ambiguous*" | → ERROR |
| "[" {ERROR ","}∗ "]" | → ERROR-LIST |
| ERROR-LIST "·" ERROR-LIST | → ERROR-LIST {**left**} |

**variables**
*Error* [*0-9*]∗[ ′]∗ → ERROR
*Error* "∗"[*0-9*]∗[ ′]∗ → {ERROR ","}∗

**equations**

[Err1] $[Error^*_1] \cdot [Error^*_2] \;=\; [Error^*_1, \, Error^*_2]$

## 6.8 IMP-syntax

The syntax of an **IMP**($G$) program is very simple: we start with a comma-separated list of basic objects, followed by a comma-separated list of *references* to basic arrows (a reference being made of a name, a domain and a codomain). Finally, a colon-separated list of arrow definitions is given. We use **lib** for introducing list of references because, in essence, the basic arrows (i.e., the basic graph $G$) have the same rôle of a library in a standard programming language.

In this implementation, an arrow is defined by juxtaposing a number of *steps*; each step is formed by a *dart*, i.e., an ARROW surrounded by some graphical syntactic sugar, and by an object (representing the codomain of the current step, and the domain of the following one). The sugar aims at mimicking the standard notation

$$f : X \xrightarrow{g} Y \xrightarrow{h} Z$$

for $f = h \circ g$, because the semantics of juxtaposition of darts is composition.

**imports** Arrows[6.3] Objects[6.2] Identifiers[6.1]
**exports**
**sorts** REF STEP DEF PROGRAM
**context-free syntax**

| | |
|---|---|
| BASIC-ARROW ":" OBJ DART-ENDER OBJ | → REF |
| DART OBJ | → STEP |
| BASIC-ARROW ":" OBJ STEP+ | → DEF |

**obj** {BASIC-OBJ ","}∗ ";"
**lib** {REF ","}∗ ";"

**def** {DEF ";"}∗ "." → PROGRAM
**variables**
*Def* [*0-9*]∗[′]∗ → DEF
*Def* "∗"[*0-9*]∗[′]∗ → {DEF ";"}∗
*Step* "∗"[*0-9*]∗[′]∗ → STEP∗
*Step* "+"[*0-9*]∗[′]∗ → STEP+
*Ref* [*0-9*]∗[′]∗ → REF
*Ref* "∗"[*0-9*]∗[′]∗ → {REF ","}∗
*Program* [*0-9*]∗[′]∗ → PROGRAM

## 6.9 IMP-compile

The compilation function ◦— we are going to specify will transform an arrow into a form which can be used by the execution module (see Section 6.11). We shall construct inductively a CODE which mimicks the structure of an arrow $f$, but uses as building blocks simple operations on forests called *instructions*. At the same time, we shall infer the type of the polymorphic arrows, if it is possible.

**imports** IMP-syntax[6.8] Support[6.4] DomCod[6.5]

The *basic codes* are those instructions which can be executed independently of the basic arrows chosen and of the arrows which have been previously defined. Distinguishing them greatly simplifies the rules for compilation and execution. In order to reduce ambiguity, we surround with braces all instructions and constructors (but in the latter case the braces are aliased away for better readability).

**exports**
**sorts** BASIC-CODE CODE

Some of the basic codes and of the code constructors contain a great deal of information, and are used internally by the compiler. A peephole optimization will completely eliminate the redundant instructionss from the final result, so we do not export their syntax. The names follow closely the categorical notation, and the parameters will be explained in the equational part.

**hiddens**
**context-free syntax**
"`INJ1`" "[" INT "," INT "]" → BASIC-CODE
"`INJ2`" "[" INT "," INT "]" → BASIC-CODE
"`PROJ1`" "[" INT "," INT "]" → BASIC-CODE
"`PROJ2`" "[" INT "," INT "]" → BASIC-CODE
"`DIST`" "[" INT "," INT "," INT "," INT "]" → BASIC-CODE

CODE "," "[" INT "," INT "]" CODE → CODE

For each instruction described here, instead, the execution module (Section 6.11) contain rules for execution. The only exeption is the `INIT` instruction, which can never be executed, since its domain contains no elements. The `TREE` instruction corresponds to the $\lhd n, - \rhd$ operator, while `RIGHTDEL` and `LEFTDEL` delete the rightmost or leftmost $n$ trees of a forest.

**exports**
**context-free syntax**
BASIC-ARROW → BASIC-CODE
"`NOP`" → BASIC-CODE
"`INIT`" → BASIC-CODE
"`TERM`" → BASIC-CODE

“`DIST`” “[” INT “,” INT “,” INT “]” → BASIC-CODE
“`TREE`” “[” INT “]” → BASIC-CODE
“`RIGHTDEL`” “[” INT “]” → BASIC-CODE
“`LEFTDEL`” “[” INT “]” → BASIC-CODE

“{” BASIC-CODE “}” → CODE

“{” “`COMP`” “,” BASIC-ARROW “}” → CODE
“{” “`ITER`” “,” CODE “,” INT “,” INT “,” INT “}” → CODE

CODE “ | ” “[” INT “,” INT “]” CODE → CODE
CODE “,” CODE → CODE

CODE “;” CODE → CODE

**priorities**
CODE “;”CODE → CODE > {**non-assoc**: CODE “,”“[”INT “,”INT “]”CODE → CODE,
CODE “ | ”“[”INT “,”INT “]”CODE → CODE}

The result of the compilation of a program is a list of pairs whose first element is an arrow identifier, and the second element is the corresponding code.

**sorts** COMP-ARROW COMP-ARROW-LIST
**context-free syntax**
“[” BASIC-ARROW “,” CODE “]” → COMP-ARROW
“[” {COMP-ARROW “,”}∗ “]” → COMP-ARROW-LIST

Finally, we specify the sorts to which the compilation function can be applied (note that it is possible to compile structural arrows without knowing the arrow specification list). The result of a compilation is a pair formed by a boolean term and a code; however, the boolean term will always be *true*, because no rules are provided which return *false*. This is intentional: if no compilation is possible, the term will not reduce, and will not unify with a pair.

**sorts** BOOL-CODE
**context-free syntax**
“[” BOOL “,” CODE “]” → BOOL-CODE

“Φ” “⟦” ARROW “:” FOBJ “→” FOBJ “⟧” using SPEC-LIST → BOOL-CODE
“Φ” “⟦” STRUCTURAL-ARROW “:” FOBJ “→” FOBJ “⟧” → BOOL-CODE
“Φ” “⟦” OBJ STEP+ “⟧” using SPEC-LIST → BOOL-CODE
“Φ” “⟦” **lib** {REF “,”}∗ “;” “⟧” → SPEC-LIST
“Φ” “⟦” **def** {DEF “;”}∗ “.” “⟧” using SPEC-LIST → COMP-ARROW-LIST
“Φ” “⟦” PROGRAM “⟧” → COMP-ARROW-LIST

**variables**
“$\gamma$”[*0-9*]∗[′]∗ → CODE
*Basic-Code* [*0-9*]∗[′]∗ → BASIC-CODE
*Bool-Code* [*0-9*]∗[′]∗ → BOOL-CODE
*Comp* “∗”[*0-9*]∗[′]∗ → {COMP-ARROW “,”}∗

The equations described here take care of the compilation of an arrow, or of a whole program. All polymorphic arrows must be instantiated now. Note that, given an arrow, we *cannot* in general build its domain and codomain, due to the polymorphism of structural arrows. But we can, using the backtracking caused by list matching, check if a given arrow has given domain and codomain. This is a much weaker statement, which we can usually prove. If we cannot, we shall not be able to rewrite the term we have to compile.

The usefulness of flat objects becomes immediately apparent in the rules which inductively try to assign the correct domain and codomain to structural or derived arrows. Since we need to forget the associativity isomorphisms, all possible ways of breaking in two a sum of a product have to be tried; backtracking takes care of that. Note that when a list variable matches a single flat object, condition evaluation will use the normalization equations [Snfs1] and [Snfp1] of Section 6.4 in order to bring the flat object into normal form.

For what matters structural arrows, we have to declare the *forms* of domain and codomain which are acceptable. For instance, the identity must have the same domain and codomain, while a first projection must start from a product and end on the first part of it.

Moreover, for each completely specified structural arrow we add a rule which instead of *assigning* inductively a domain and a codomain just *checks* whether the actual domain and codomain of the arrow match with the specified ones.

The general strategy of the compiler is to create redundant information in the code, which however avoids special cases during compilation. Then, a simple sequence of peephole optimizations handles special cases and reduces redundancy.

**equations**

**Compiling a basic (or defined) arrow.** For basic arrows compilation is very simple, since it is assumed that somewhere in the system the application of a basic arrow to its possible arguments has already been specified equationally. If the arrow has been previously defined in the program, it appears in the specification list with the special symbol $\rightsquigarrow$, and in this case we use the `COMP` instruction in order to remember to search for the arrow in the compiled arrow list at execution time.

$$\text{[Ca1]}\quad \frac{\textit{spec of } a \textit{ in } [\textit{Spec}^*] = [\Xi \to \Theta]}{\Phi \; [\![ \; a : \Xi \to \Theta \; ]\!] \text{ using } [\textit{Spec}^*] = [\textit{true}, \{a\}]}$$

$$\text{[Ca2]}\quad \frac{\textit{spec of } a \textit{ in } [\textit{Spec}^*] = [\Xi \rightsquigarrow \Theta]}{\Phi \; [\![ \; a : \Xi \to \Theta \; ]\!] \text{ using } [\textit{Spec}^*] = [\textit{true}, \{\texttt{COMP}, a\}]}$$

**Compiling a structural arrow.** We do not need the specification list in order to compile a structural arrow, so we discard it; in this way the rules are much less cluttered.

$$\text{[Cstr]}\quad \Phi \; [\![ \; \textit{Structural-Arrow} : \Xi \to \Theta \; ]\!] \text{ using } [\textit{Spec}^*] \;=\; \Phi \; [\![ \; \textit{Structural-Arrow} : \Xi \to \Theta \; ]\!]$$

We now compile the structural arrows following the mathematical definition of $\circ -$. At the same time, we check for a possible instantiation of polymorphic arrows. Compilation succeeds only if such an instantiation exists.

Note that there is an instruction corresponding to the initial map $!_X : \varnothing \to X$, but it should never happen that it has to be executed, unless an arrow is applied to badly-typed data. Correspondingly, Section 6.11 contains no rule for the execution of `INIT`.

Let us begin with identities: they must start and end on the same object, but in order to avoid unwanted assignments, the object cannot be initial or terminal in the polymorphic case.

$$\text{[Cid]}\quad \frac{\Xi \neq +(), \;\; \Xi \neq *()}{\Phi \; [\![ \; \mathbf{1} : \Xi \to \Xi \; ]\!] = [\textit{true}, \{\texttt{NOP}\}]}$$

$$\text{[CID]}\quad \frac{\mathbf{flat}[X] = \Xi}{\Phi \; [\![ \; \mathbf{1}(X) : \Xi \to \Xi \; ]\!] = [\textit{true}, \{\texttt{NOP}\}]}$$

If a first (second) injection is completely specified, we just have to check that the actual domain and codomain correspond to the specified ones. Otherwise, using the $\oplus$ operator (see Section 6.4) we break the given codomain into two summands in such a way that the domain is first (second) summand. If we succeed, we output an `INJ`$n$ instruction which records the length (as a sum) of each summand. With projections we keep a dual behaviour.

Note that in the polymorphic case we do not allow empty lists; in other words, we never instantiate a polymorphic arrow to a trivial arrow, such as $\mathrm{proj}_1 : A \times I \to A$. This is necessary in order to avoid unnatural assignments. For instance, $X + X \xrightarrow{(\mathrm{inj}_2, \mathrm{inj}_1)} X + X$ could be instantiated to an identity, with $\mathrm{inj}_2 : \varnothing \to X + X + \varnothing$ and $\mathrm{inj}_1 : X + X \to X + X + \varnothing$.

[Ci1] $$\frac{\oplus(\Xi) = \oplus(\Xi^+),\ \ \Omega = +(\Xi^+\ \Theta^+)}{\Phi\ \llbracket\ \mathrm{inj}_1 : \Xi \to \Omega\ \rrbracket = [\mathit{true},\ \{\mathtt{INJ1}[\mathbf{len}[\Xi^+],\ \mathbf{len}[\Theta^+]]\}]}$$

[CI1] $$\frac{\mathbf{flat}[X] = \Xi,\ \ \mathbf{flat}[Y] = \Theta,\ \ +(\Xi\ \Theta) = \Omega}{\Phi\ \llbracket\ \mathrm{inj}_1(X,\ Y) : \Xi \to \Omega\ \rrbracket = [\mathit{true},\ \{\mathtt{INJ1}[\mathbf{slen}[\Xi],\ \mathbf{slen}[\Theta]]\}]}$$

[Ci2] $$\frac{\oplus(\Theta) = \oplus(\Theta^+),\ \ \Omega = +(\Xi^+\ \Theta^+)}{\Phi\ \llbracket\ \mathrm{inj}_2 : \Theta \to \Omega\ \rrbracket = [\mathit{true},\ \{\mathtt{INJ2}[\mathbf{len}[\Xi^+],\ \mathbf{len}[\Theta^+]]\}]}$$

[CI2] $$\frac{\mathbf{flat}[X] = \Xi,\ \ \mathbf{flat}[Y] = \Theta,\ \ +(\Xi\ \Theta) = \Omega}{\Phi\ \llbracket\ \mathrm{inj}_2(X,\ Y) : \Theta \to \Omega\ \rrbracket = [\mathit{true},\ \{\mathtt{INJ2}[\mathbf{slen}[\Xi],\ \mathbf{slen}[\Theta]]\}]}$$

[Cp1] $$\frac{\otimes(\Xi) = \otimes(\Xi^+),\ \ \Omega = *(\Xi^+\ \Theta^+)}{\Phi\ \llbracket\ \mathrm{proj}_1 : \Omega \to \Xi\ \rrbracket = [\mathit{true},\ \{\mathtt{PROJ1}[\mathbf{len}[\Xi^+],\ \mathbf{len}[\Theta^+]]\}]}$$

[CP1] $$\frac{\mathbf{flat}[X] = \Xi,\ \ \mathbf{flat}[Y] = \Theta,\ \ *(\Xi\ \Theta) = \Omega}{\Phi\ \llbracket\ \mathrm{proj}_1(X,\ Y) : \Omega \to \Xi\ \rrbracket = [\mathit{true},\ \{\mathtt{PROJ1}[\mathbf{plen}[\Xi],\ \mathbf{plen}[\Theta]]\}]}$$

[Cp2] $$\frac{\otimes(\Theta) = \otimes(\Theta^+),\ \ \Omega = *(\Xi^+\ \Theta^+)}{\Phi\ \llbracket\ \mathrm{proj}_2 : \Omega \to \Theta\ \rrbracket = [\mathit{true},\ \{\mathtt{PROJ2}[\mathbf{len}[\Xi^+],\ \mathbf{len}[\Theta^+]]\}]}$$

[CP2] $$\frac{\mathbf{flat}[X] = \Xi,\ \ \mathbf{flat}[Y] = \Theta,\ \ *(\Xi\ \Theta) = \Omega}{\Phi\ \llbracket\ \mathrm{proj}_2(X,\ Y) : \Omega \to \Theta\ \rrbracket = [\mathit{true},\ \{\mathtt{PROJ2}[\mathbf{plen}[\Xi],\ \mathbf{plen}[\Theta]]\}]}$$

Initial and terminal maps have corresponding instructions. `TERM` just outputs the empty forest; `INIT` is never executed, but it is kept for optimization purposes. Again we avoid unwanted instantiations.

[Cini] $\Xi \neq +() \Rightarrow \Phi\ \llbracket\ ! : +() \to \Xi\ \rrbracket\ =\ [\mathit{true},\ \{\mathtt{INIT}\}]$

[CINI] $$\frac{\mathbf{flat}[X] = \Xi}{\Phi\ \llbracket\ !(X) : +() \to \Xi\ \rrbracket = [\mathit{true},\ \{\mathtt{INIT}\}]}$$

[Cter] $\Xi \neq *() \Rightarrow \Phi\ \llbracket\ \mathsf{i} : \Xi \to *()\ \rrbracket\ =\ [\mathit{true},\ \{\mathtt{TERM}\}]$

$$[\text{CTER}]\ \frac{\mathbf{flat}[X] = \Xi}{\Phi\ [\![\ \mathtt{i}(X) : \Xi \to *()\ ]\!] = [\mathit{true},\ \{\mathtt{TERM}\}]}$$

The four parameters recorded when compiling a distributivity isomorphism are the length as a product of the first factor of the domain, the length as a sum of the two summands which make the second factor of the domain, and the length as a sum of the first summand of the codomain.

Note that many distributivity isomorphisms trivialize when one of this numbers is zero. This is the reason why the rule for the instantiation of a polymorphic map requires nonempty lists everyhwhere.

$$[\text{Cd}]\ \frac{*(\Xi^+\ +(\Theta_1^+)) = +(\Omega_1^+),\ \ *(\Xi^+\ +(\Theta_2^+)) = +(\Omega_2^+)}{\begin{array}{l}\Phi\ [\![\ \delta^{-1} : *(\Xi^+\ +(\Theta_1^+\ \Theta_2^+)) \to +(\Omega_1^+\ \Omega_2^+)\ ]\!] = \\ [\mathit{true},\ \{\mathtt{DIST}[\mathbf{len}[\Xi^+],\ \mathbf{len}[\Theta_1^+],\ \mathbf{len}[\Theta_2^+],\ \mathbf{len}[\Omega_1^+]]\}]\end{array}}$$

$$[\text{CD}]\ \frac{\begin{array}{c}\mathbf{flat}[X] = \Xi,\ \ \mathbf{flat}[Y_1] = \Theta_1,\ \ \mathbf{flat}[Y_2] = \Theta_2, \\ *(\Xi\ +(\Theta_1\ \Theta_2)) = \Omega_1,\ \ +(*(\Xi\ \Theta_1)\ *(\Xi\ \Theta_2)) = \Omega_2\end{array}}{\begin{array}{l}\Phi\ [\![\ \delta^{-1}(X,\ Y_1,\ Y_2) : \Omega_1 \to \Omega_2\ ]\!] = \\ [\mathit{true},\ \{\mathtt{DIST}[\mathbf{plen}[\Xi],\ \mathbf{slen}[\Theta_1],\ \mathbf{slen}[\Theta_2],\ \mathbf{slen}[*(\Xi\ \Theta_1)]]\}]\end{array}}$$

**Compiling a derived arrow.** Now we scan inductively an arrow. Note that product and sums of functions are compiled in terms of the more primitive $(-, -)$ and $(- \mid -)$ operators, following the categorical definition, and that there is a double backtracking: both source and destination must be broken into sublists.

$$[\text{Cs}]\ \frac{\begin{array}{c}\oplus(\Omega_1) = \oplus(\Xi_1^*\ \Theta_1^*),\ \ \oplus(\Omega_2) = \oplus(\Xi_2^*\ \Theta_2^*), \\ \Phi\ [\![\ f : +(\Xi_1^*) \to +(\Xi_2^*)\ ]\!] \text{ using } [\mathit{Spec}^*] = [\mathit{true},\ \gamma], \\ \Phi\ [\![\ g : +(\Theta_1^*) \to +(\Theta_2^*)\ ]\!] \text{ using } [\mathit{Spec}^*] = [\mathit{true},\ \gamma']\end{array}}{\begin{array}{l}\Phi\ [\![\ f + g : \Omega_1 \to \Omega_2\ ]\!] \text{ using } [\mathit{Spec}^*] = \\ [\mathit{true},\ \gamma\ ;\ \{\mathtt{INJ1}[\mathbf{len}[\Xi_2^*],\ \mathbf{len}[\Theta_2^*]]\}\ \mid [\mathbf{len}[\Xi_1^*],\ \mathbf{len}[\Theta_1^*]]\ \gamma'\ ;\ \{\mathtt{INJ2}[\mathbf{len}[\Xi_2^*],\ \mathbf{len}[\Theta_2^*]]\}]\end{array}}$$

$$[\text{Cp}]\ \frac{\begin{array}{c}\otimes(\Omega_1) = \otimes(\Xi_1^*\ \Theta_1^*),\ \ \otimes(\Omega_2) = \otimes(\Xi_2^*\ \Theta_2^*), \\ \Phi\ [\![\ f : *(\Xi_1^*) \to *(\Xi_2^*)\ ]\!] \text{ using } [\mathit{Spec}^*] = [\mathit{true},\ \gamma], \\ \Phi\ [\![\ g : *(\Theta_1^*) \to *(\Theta_2^*)\ ]\!] \text{ using } [\mathit{Spec}^*] = [\mathit{true},\ \gamma']\end{array}}{\begin{array}{l}\Phi\ [\![\ f * g : \Omega_1 \to \Omega_2\ ]\!] \text{ using } [\mathit{Spec}^*] = \\ [\mathit{true},\ \{\mathtt{PROJ1}[\mathbf{len}[\Xi_1^*],\ \mathbf{len}[\Theta_1^*]]\}\ ;\ \gamma\ ,[\mathbf{len}[\Xi_2^*],\ \mathbf{len}[\Theta_2^*]]\ \{\mathtt{PROJ2}[\mathbf{len}[\Xi_1^*],\ \mathbf{len}[\Theta_1^*]]\}\ ;\ \gamma']\end{array}}$$

$$[\text{Cb}]\ \frac{\begin{array}{c}\oplus(\Omega_1) = \oplus(\Xi_1^*\ \Theta_1^*), \\ \Phi\ [\![\ f : +(\Xi_1^*) \to \Omega_2\ ]\!] \text{ using } [\mathit{Spec}^*] = [\mathit{true},\ \gamma], \\ \Phi\ [\![\ g : +(\Theta_1^*) \to \Omega_2\ ]\!] \text{ using } [\mathit{Spec}^*] = [\mathit{true},\ \gamma']\end{array}}{\Phi\ [\![\ f \mid g : \Omega_1 \to \Omega_2\ ]\!] \text{ using } [\mathit{Spec}^*] = [\mathit{true},\ \gamma\ \mid [\mathbf{len}[\Xi_1^*],\ \mathbf{len}[\Theta_1^*]]\ \gamma']}$$

$$[\text{Cc}]\ \frac{\begin{array}{c}\otimes(\Omega_2) = \otimes(\Xi_2^*\ \Theta_2^*), \\ \Phi\ [\![\ f : \Omega_1 \to *(\Xi_2^*)\ ]\!] \text{ using } [\mathit{Spec}^*] = [\mathit{true},\ \gamma], \\ \Phi\ [\![\ g : \Omega_1 \to *(\Theta_2^*)\ ]\!] \text{ using } [\mathit{Spec}^*] = [\mathit{true},\ \gamma']\end{array}}{\Phi\ [\![\ f, g : \Omega_1 \to \Omega_2\ ]\!] \text{ using } [\mathit{Spec}^*] = [\mathit{true},\ \gamma\ ,[\mathbf{len}[\Xi_2^*],\ \mathbf{len}[\Theta_2^*]]\ \gamma']}$$

Composition is a different beast. Our problem now is that given an arrow $X \xrightarrow{g\circ f} Z$, we do not know how to make structural induction because we do not have a candidate for an object $Y$ such that $X \xrightarrow{f} Y$ and $Y \xrightarrow{g} Z$. The fundamental idea here is that *as long as* $X \xrightarrow{f} Y$ *and* $Y \xrightarrow{g} Z$ *pass the type-check,* $Y$ *can be chosen arbitrarily*. This means that we can define a domain and codomain constructors which work with a certain looseness, for instance supposing that their arguments are well-typed; this is done in Section 6.5: all we need is just a *reasonable guess* for $Y$.

The rule for composition is more complicated exactly because we must establish whether we can compile $f; g$ using as "middle object" the domain of $g$ or the codomain of $f$ (in general they will not be both computable).

$$[\text{Cco1}]\ \frac{\mathbf{cod}[f] \text{ using } [\mathit{Spec}^*] = \Theta,\ \Theta \neq \bot,\quad \Phi \ [\![\ f : \Xi \to \Theta\ ]\!] \text{ using } [\mathit{Spec}^*] = [\mathit{true}, \gamma],\quad \Phi\ [\![\ g : \Theta \to \Omega\ ]\!] \text{ using } [\mathit{Spec}^*] = [\mathit{true}, \gamma']}{\Phi\ [\![\ f; g : \Xi \to \Omega\ ]\!] \text{ using } [\mathit{Spec}^*] = [\mathit{true}, \gamma\ ;\ \gamma']}$$

$$[\text{Cco2}]\ \frac{\mathbf{dom}[g] \text{ using } [\mathit{Spec}^*] = \Theta,\ \Theta \neq \bot,\quad \Phi \ [\![\ f : \Xi \to \Theta\ ]\!] \text{ using } [\mathit{Spec}^*] = [\mathit{true}, \gamma],\quad \Phi\ [\![\ g : \Theta \to \Omega\ ]\!] \text{ using } [\mathit{Spec}^*] = [\mathit{true}, \gamma']}{\Phi\ [\![\ f; g : \Xi \to \Omega\ ]\!] \text{ using } [\mathit{Spec}^*] = [\mathit{true}, \gamma\ ;\ \gamma']}$$

**Compiling call**[ − ]. First of all, we note that the three numerical arguments used by the `ITER` instruction are the length (as a sum) of the input, local and output state space. If the application of **call**[ − ] is completely specified, we just have to compute them. Otherwise, we try to compute the domain (codomain) of $f$ and to break it into two summands in such a way that the second (first) summand is a first (second) summand for the codomain (domain). Of course, there is a possibility of backtracking here. Note that we impose that the output space is non-empty.

$$[\text{Ccall1}]\ \frac{\Omega \neq +(),\ \oplus(\Xi) = \oplus(\Xi^+),\ \mathbf{dom}[f] \text{ using } [\mathit{Spec}^*] = +(\Xi^+\ \Theta^+),\quad \Phi\ [\![\ f : +(\Xi^+\ \Theta^+) \to +(\Theta^+\ \Omega)\ ]\!] \text{ using } [\mathit{Spec}^*] = [\mathit{true}, \gamma]}{\Phi\ [\![\ \mathbf{call}[f] : \Xi \to \Omega\ ]\!] \text{ using } [\mathit{Spec}^*] = [\mathit{true}, \{\mathtt{ITER}, \gamma, \mathbf{len}[\Xi^+], \mathbf{len}[\Theta^+], \mathbf{slen}[\Omega]\}]}$$

$$[\text{Ccall2}]\ \frac{\oplus(\Omega) = \oplus(\Omega^+),\ \mathbf{cod}[f] \text{ using } [\mathit{Spec}^*] = +(\Theta^+\ \Omega^+),\quad \Phi\ [\![\ f : +(\Xi\ \Theta^+) \to +(\Theta^+\ \Omega^+)\ ]\!] \text{ using } [\mathit{Spec}^*] = [\mathit{true}, \gamma]}{\Phi\ [\![\ \mathbf{call}[f] : \Xi \to \Omega\ ]\!] \text{ using } [\mathit{Spec}^*] = [\mathit{true}, \{\mathtt{ITER}, \gamma, \mathbf{slen}[\Xi], \mathbf{len}[\Theta^+], \mathbf{len}[\Omega^+]\}]}$$

$$[\text{CCALL}]\ \frac{\mathbf{flat}[X] = \Xi,\ \mathbf{flat}[Y] = \Theta,\ \mathbf{flat}[Z] = \Omega,\ \Omega \neq +(),\quad \Phi\ [\![\ f : +(\Xi\ \Theta) \to +(\Theta\ \Omega)\ ]\!] \text{ using } [\mathit{Spec}^*] = [\mathit{true}, \gamma]}{\Phi\ [\![\ \mathbf{call}[X, Y, Z, f] : \Xi \to \Omega\ ]\!] \text{ using } [\mathit{Spec}^*] = [\mathit{true}, \{\mathtt{ITER}, \gamma, \mathbf{slen}[\Xi], \mathbf{slen}[\Theta], \mathbf{slen}[\Omega]\}]}$$

**Compiling a program.** The following rules describe how ∘− can be applied to a program, or to part of it. Note that little or no error checking is performed, because we assume that the program has been already type-checked.

When applied to a DART, ∘− returns a boolean and a code, while when applied to a list of definitions, it returns a list of COMP-ARROW.

$$[\text{Cst1}]\ \frac{\Phi\ [\![\ f : \mathbf{flat}[X] \to \mathbf{flat}[Y]\ ]\!] \text{ using } [\mathit{Spec}^*] = [\mathit{true}, \gamma]}{\Phi\ [\![\ X - f \to Y\ ]\!] \text{ using } [\mathit{Spec}^*] = [\mathit{true}, \gamma]}$$

$$\text{[Cst2]}\ \frac{\Phi\ [\![\ f : \mathbf{flat}[X] \to \mathbf{flat}[Y]\ ]\!]\ \text{using}\ [Spec^*] = [true,\ \gamma], \quad \Phi\ [\![\ Y\ Step^+\ ]\!]\ \text{using}\ [Spec^*] = [true,\ \gamma']}{\Phi\ [\![\ X\ -\ f\ \to\ Y\ Step^+\ ]\!]\ \text{using}\ [Spec^*] = [true,\ \gamma\ ;\ \gamma']}$$

$$\text{[Cdef1]}\ \Phi\ [\![\ \mathbf{def}\ \ .\ ]\!]\ \text{using}\ [Spec^*]\ =\ []$$

$$\text{[Cdef2]}\ \frac{\Phi\ [\![\ X\ Step^*\ -\ f\ \to\ Y\ ]\!]\ \text{using}\ [Spec^*] = [true,\ \gamma], \quad \Phi\ [\![\ \mathbf{def}\ Def^*\ .\ ]\!]\ \text{using}\ [a : \mathbf{flat}[X] \rightsquigarrow \mathbf{flat}[Y],\ Spec^*] = [Comp^*]}{\Phi\ [\![\ \mathbf{def}\ a : X\ Step^*\ -\ f\ \to\ Y;\ Def^*\ .\ ]\!]\ \text{using}\ [Spec^*] = [[a,\ \gamma],\ Comp^*]}$$

Finally, the compilation of a whole program is obtained by discarding the object declaration, gathering (without any check) the basic arrow declaration into a specification list, and applying $\circ\!-$ to each arrow definition.

$$\text{[Clib]}\ \Phi\ [\![\ \mathbf{lib}\ ;\ ]\!]\ =\ []$$

$$\text{[Cref]}\ \frac{\Phi\ [\![\ \mathbf{lib}\ Ref^*;\ ]\!] = [Spec^*]}{\Phi\ [\![\ \mathbf{lib}\ a : X \to Y,\ Ref^*;\ ]\!] = [a : \mathbf{flat}[X] \to \mathbf{flat}[Y],\ Spec^*]}$$

$$\text{[Cprog]}\ \Phi\ [\![\ \mathbf{obj}\ A^*;\ \mathbf{lib}\ Ref^*;\ \mathbf{def}\ Def^*\ .\ ]\!]\ =\ \Phi\ [\![\ \mathbf{def}\ Def^*\ .\ ]\!]\ \text{using}\ \Phi\ [\![\ \mathbf{lib}\ Ref^*;\ ]\!]$$

**Peephole optimization.** Here we are going to examine small fragments of code, in order to make them more efficient. The rules we apply can in turn activate other rules, until no other rule is applicable. This technique, known as *peephole optimization* in compiler theory (see [McKeeman:1965]), is particularly useful in our case because it allows to eliminate a series of trivial maps which would be very cumbersome to manage during compilation on a case-by-case basis. For instance, composition with NOP (i.e., with the identity) can be eliminated:

$$\text{[CONOP1]}\ \{\texttt{NOP}\}\ ;\ \gamma\ =\ \gamma$$
$$\text{[CONOP2]}\ \gamma\ ;\ \{\texttt{NOP}\}\ =\ \gamma$$

If one of the summands of the codomain of an injection is the empty set, the injection is really an identity or an initial map. Dually for projections.

$$\text{[COINJ1]}\ p \neq 0 \Rightarrow \{\texttt{INJ1}[p,\ 0]\}\ =\ \{\texttt{NOP}\}$$
$$\text{[COINJ2]}\ q \neq 0 \Rightarrow \{\texttt{INJ2}[0,\ q]\}\ =\ \{\texttt{NOP}\}$$

$$\text{[COINJ3]}\ \{\texttt{INJ1}[0,\ q]\}\ =\ \{\texttt{INIT}\}$$
$$\text{[COINJ4]}\ \{\texttt{INJ2}[p,\ 0]\}\ =\ \{\texttt{INIT}\}$$

$$\text{[COPR1]}\ p \neq 0 \Rightarrow \{\texttt{PROJ1}[p,\ 0]\}\ =\ \{\texttt{NOP}\}$$
$$\text{[COPR2]}\ q \neq 0 \Rightarrow \{\texttt{PROJ2}[0,\ q]\}\ =\ \{\texttt{NOP}\}$$

$$\text{[COPR3]}\ \{\texttt{PROJ1}[0,\ q]\}\ =\ \{\texttt{TERM}\}$$
$$\text{[COPR4]}\ \{\texttt{PROJ2}[p,\ 0]\}\ =\ \{\texttt{TERM}\}$$

Iteration starting from the empty set can only produce the initial map, and if the local state space is empty we can apply the normal form theorem.

[COITER1] $\{\mathtt{ITER},\ \gamma,\ 0,\ n,\ p\} = \{\mathtt{INIT}\}$

[COITER2] $m \neq 0 \Rightarrow \{\mathtt{ITER},\ \gamma,\ m,\ 0,\ p\} = \gamma$

Initial and terminal maps enjoy the following properties:

[COINIT] $\{\mathtt{INIT}\}\ ;\ \gamma = \{\mathtt{INIT}\}$

[CODELALL] $\gamma\ ;\ \{\mathtt{TERM}\} = \{\mathtt{TERM}\}$

Now we have a series of rules which eliminate trivial cases of the $(-, -)$ and $(- \mid -)$ operators in which the arrow really acting is just one.

[COc1] $\gamma\ ,[0,\ q]\ \gamma' = \gamma'$

[COc2] $\gamma\ ,[p,\ 0]\ \gamma' = \gamma$

[COb1] $\gamma\ \mid [0,\ q]\ \gamma' = \gamma'$

[COb2] $\gamma\ \mid [p,\ 0]\ \gamma' = \gamma$

The most trivialized map is the distributivity isomorphism. If one of its parameters is zero, it reduces to an identity or to an injection. For instance, the first rule correspond to the case $I \times (X + Y) \xrightarrow{\delta^{-1}} I \times X + I \times Y$, where the flat normal form for both the domain and the codomain is $+(X\ Y)$, and $\delta^{-1}$ is the identity, while the second rule correspond to the case $X \times (\varnothing + Y) \xrightarrow{\delta^{-1}} X \times \varnothing + X \times Y$ (with $X \neq I$), where the normal form of the domain is $\times(X\ Y)$, the normal form of the codomain is $+(\times(X\ +())\ \times (X\ Y))$ and $\delta^{-1}$ is the second injection.

[CODIST1] $\{\mathtt{DIST}[0,\ q,\ q',\ n]\} = \{\mathtt{NOP}\}$

[CODIST2] $p \neq 0 \Rightarrow \{\mathtt{DIST}[p,\ 0,\ q',\ n]\} = \{\mathtt{TREE}[1]\}$

[CODIST3] $p \neq 0 \Rightarrow \{\mathtt{DIST}[p,\ q,\ 0,\ n]\} = \{\mathtt{TREE}[0]\}$

We complete our optimization with the following *default rules*, which will be applied only when no other equation is applicable. First of all, we note that an injection just forms a tree, modulo the rewrite rule $\lhd j, \lhd k, u \rhd \rhd = \lhd j + k, u \rhd$ (this is true because we can assume $p$ and $q$ not equal to zero). Analogously, projections can be implemented as cancellation of part of a forest.

[COINJ1] $\{\mathtt{INJ1}[n,\ p]\} = \{\mathtt{TREE}[0]\}$ **otherwise**

[COINJ2] $\{\mathtt{INJ2}[n,\ p]\} = \{\mathtt{TREE}[n]\}$ **otherwise**

[COPR1] $\{\mathtt{PROJ1}[p,\ q]\} = \{\mathtt{RIGHTDEL}[q]\}$ **otherwise**

[COPR2] $\{\mathtt{PROJ2}[p,\ q]\} = \{\mathtt{LEFTDEL}[p]\}$ **otherwise**

Once we assume that all parameters are nonnull, we can reduce the information carried by the distributivity isomorphism. Note that if $n$ is zero the isomorphism will never be applied, for both its domain and its codomain would be empty.

[CODIST] $\{\mathtt{DIST}[p,\ q,\ q',\ n]\} = \{\mathtt{DIST}[q,\ q',\ n]\}$ **otherwise**

Finally, we discard the arguments of the $(-, -)$ operator.

[COc] $\gamma\ ,[p',\ q']\ \gamma' = \gamma\ ,\ \gamma'$ **otherwise**

## 6.10 IMP-typecheck

The type-check function we are going to specify can be applied to several entities, giving different results (which, however, always contain an error list; the only case in which a boolean term is returned is when checking if certain data correspond to the representation of an element of a certain flat object). We export only the case in which type-checking is applied to a whole program, or to some data.

**imports** IMP-syntax[6.8] IMP-compile[6.9] Data[6.6] Support[6.4] DomCod[6.5] Errors[6.7]
**exports**
  **context-free syntax**
    **tc** "⟦" PROGRAM "⟧" → ERROR-LIST
    **tc** "⟦" DATA "," FOBJ "⟧" → BOOL
**hiddens**
  **sorts** ERROR-LIST-BASIC-OBJ-LIST ERROR-LIST-SPEC-LIST
  **context-free syntax**
    ERROR-LIST BASIC-OBJ-LIST → ERROR-LIST-BASIC-OBJ-LIST
    ERROR-LIST SPEC-LIST → ERROR-LIST-SPEC-LIST

When we type-check the basic object list (see Section 6.13 for the description of the structure of a program) we return a list of declared objects, and a list of errors (duplicate identifiers).

  **context-free syntax**
    **tc** "⟦" **obj** {BASIC-OBJ ","}∗ ";" "⟧" → ERROR-LIST-BASIC-OBJ-LIST

The following syntax is useful when one or more objects have to be checked; the relative errors are accumulated in the resulting error list.

  **context-free syntax**
    **tc** "⟦" {OBJ ","}∗ "⟧" using BASIC-OBJ-LIST → ERROR-LIST

A list of basic arrows is checked on the basis of a list of basic objects (which we need in order to check for validity of the domains and codomains). The result is a list of arrow specifications, and a list of errors.

  **context-free syntax**
    **tc** "⟦" **lib** {REF ","}∗ ";" "⟧"
    using BASIC-OBJ-LIST → ERROR-LIST-SPEC-LIST

Finally, knowing completely the base graph, we check for arrow steps and definitions. We also need a function to extract the boolean part of the result of a compilation.

  **context-free syntax**
    **tc** "⟦" BOOL-CODE "⟧" → BOOL

    **tc** "⟦" OBJ STEP∗ "⟧"
    using BASIC-OBJ-LIST SPEC-LIST → ERROR-LIST

    **tc** "⟦" **def** {DEF ";"}∗ "." "⟧"
    using BASIC-OBJ-LIST SPEC-LIST → ERROR-LIST

The following equations perform the type-check of an **IMP**$(G)$ program and of **IMP**$(G)$ data. The program is scanned, and a error list is build incrementally. The functions and the predicates of Section 6.4 are at the basis of the conditions.

**equations**

**Checking data.** The type-check of data is rather simple. Essentially, we check whether the *structure* of the data corresponds to the structure of a flat object. Products must corresponds to concatenations of forests, and sums to creation of a tree whose root label is smaller than the number of summands. Note that the rule for simple objects does not really guarantee that the basic data $e$ has the same type of $A$.

$$\text{[Td1]}\ \mathbf{tc}\ \llbracket \lhd\, n,\, d \rhd ,\ +(\Xi^{+})\ \rrbracket \;=\; n < \mathbf{len}[\Xi^{+}] \wedge \mathbf{tc}\ \llbracket\ d,\ \mathit{elem}\ n\ \mathit{from}\ \Xi^{+}\ \rrbracket$$

$$\text{[Td2]}\ \mathbf{tc}\ \llbracket\ u\ u^{*},\ *(\Xi\ \Xi^{*})\ \rrbracket \;=\; \mathbf{tc}\ \llbracket\ u,\ \Xi\ \rrbracket \wedge \mathbf{tc}\ \llbracket\ u^{*},\ *(\Xi^{*})\ \rrbracket$$

$$\text{[Td3]}\ \mathbf{tc}\ \llbracket\ ,\ *()\ \rrbracket \;=\; \mathit{true}$$

$$\text{[Td4]}\ \mathbf{tc}\ \llbracket\ e,\ A\ \rrbracket \;=\; \mathit{true}$$

$$\text{[Td5]}\ \mathbf{tc}\ \llbracket\ d,\ \Xi\ \rrbracket \;=\; \mathit{false}\quad \textbf{otherwise}$$

**Checking the basic object list.** First of all, we scan the list of basic objects; if we meet a duplicate definition, a suitable error is appended to the error list.

$$\text{[Tbo1]}\ \mathbf{tc}\ \llbracket\ \mathbf{obj}\ ;\ \rrbracket \;=\; [\,]\ [\,]$$

$$\text{[Tbo2]}\ \frac{\mathbf{tc}\ \llbracket\ \mathbf{obj}\ A^{*};\ \rrbracket = [\mathit{Error}^{*}]\ [B^{*}],\ \ A\ \mathit{is\ in}\ [B^{*}] = \mathit{false}}{\mathbf{tc}\ \llbracket\ \mathbf{obj}\ A^{*},\ A;\ \rrbracket = [\mathit{Error}^{*}]\ [A,\ B^{*}]}$$

$$\text{[Tbo2]}\ \frac{\mathbf{tc}\ \llbracket\ \mathbf{obj}\ A^{*};\ \rrbracket = [\mathit{Error}^{*}]\ [B^{*}],\ \ A\ \mathit{is\ in}\ [B^{*}] = \mathit{true}}{\mathbf{tc}\ \llbracket\ \mathbf{obj}\ A^{*},\ A;\ \rrbracket = [\mathit{Error}^{*},\ \mathit{object_name}\ A\ \mathit{already_used}]\ [B^{*}]}$$

Here we obtain from a list of objects a list of errors. This will be useful in the next equations.

$$\text{[Tobl1]}\ \mathbf{tc}\ \llbracket\quad \rrbracket\ \text{using}\ [A^{*}] \;=\; [\,]$$

$$\text{[Tobl2]}\ \frac{X\ \mathit{is\ built\ from}\ [A^{*}] = \mathit{true}}{\mathbf{tc}\ \llbracket\ X,\ X^{*}\ \rrbracket\ \text{using}\ [A^{*}] = \mathbf{tc}\ \llbracket\ X^{*}\ \rrbracket\ \text{using}\ [A^{*}]}$$

$$\text{[Tobl3]}\ \frac{X\ \mathit{is\ built\ from}\ [A^{*}] = \mathit{false}}{\begin{array}{l}\mathbf{tc}\ \llbracket\ X,\ X^{*}\ \rrbracket\ \text{using}\ [A^{*}] = \\ [\mathit{object}\ X\ \mathit{contains_undeclared_basic_objects}] \cdot \mathbf{tc}\ \llbracket\ X^{*}\ \rrbracket\ \text{using}\ [A^{*}]\end{array}}$$

**Checking the basic arrow list.** Now each basic arrow declaration is checked in two ways: its domain and codomain must have been built using known simple objects, and the arrow name must not have been already used.

$$\text{[Tba1]}\ \mathbf{tc}\ \llbracket\ \mathbf{lib}\ ;\ \rrbracket\ \text{using}\ [A^{*}] \;=\; [\,]\ [\,]$$

$$\text{[Tba2]}\ \frac{\begin{array}{c}\mathbf{tc}\ \llbracket\ \mathbf{lib}\ \mathit{Ref}^{*};\ \rrbracket\ \text{using}\ [A^{*}] = [\mathit{Error}_1^{*}]\ [\mathit{Spec}^{*}], \\ \mathbf{tc}\ \llbracket\ X,\ Y\ \rrbracket\ \text{using}\ [A^{*}] = [\mathit{Error}_2^{*}], \\ \mathit{spec\ of}\ a\ \mathit{in}\ [\mathit{Spec}^{*}] = \bot\end{array}}{\begin{array}{l}\mathbf{tc}\ \llbracket\ \mathbf{lib}\ \mathit{Ref}^{*},\ a : X \to Y;\ \rrbracket\ \text{using}\ [A^{*}] = \\ [\mathit{Error}_1^{*},\ \mathit{Error}_2^{*}]\ [a : \mathbf{flat}[X] \to \mathbf{flat}[Y],\ \mathit{Spec}^{*}]\end{array}}$$

[Tba3]
$$\frac{\begin{array}{c}\mathbf{tc}\ [\![\ \mathbf{lib}\ \mathit{Ref}^*;\ ]\!]\ \text{using}\ [A^*] = [\mathit{Error}_1^*]\ [\mathit{Spec}^*],\\ \mathbf{tc}\ [\![\ X,\ Y\ ]\!]\ \text{using}\ [A^*] = [\mathit{Error}_2^*],\\ \mathit{spec\ of}\ a\ \mathit{in}\ [\mathit{Spec}^*] \neq \bot\end{array}}{\begin{array}{l}\mathbf{tc}\ [\![\ \mathbf{lib}\ \mathit{Ref}^*,\ a : X \to Y;\ ]\!]\ \text{using}\ [A^*] =\\ [\mathit{Error}_1^*,\ \mathit{Error}_2^*,\ \mathit{arrow}\text{\textvisiblespace}\mathit{name}\ a\ \mathit{already}\text{\textvisiblespace}\mathit{used}]\ [\mathit{Spec}^*]\end{array}}$$

**Checking arrow steps and definitions.** An arrow step is correctly typed when the arrow it encloses starts and ends on the specified domain and codomain. A precondition for this to happen is that the enclosed arrow must have been built using known simple arrows. The compiling function $\circ\!-$ is called in order to discover whether an arrow has given domain and codomain.

[Tnf1] $\mathbf{tc}\ [\![\ [\mathit{true},\ \gamma]\ ]\!] \ = \ \mathit{true}$

[Tnf2] $\mathbf{tc}\ [\![\ \mathit{Bool\text{-}Code}\ ]\!] \ = \ \mathit{false}$ **otherwise**

[Tst1] $\mathbf{tc}\ [\![\ X\ ]\!]\ \text{using}\ [A^*]\ [\mathit{Spec}^*] \ = \ \mathbf{tc}\ [\![\ X\ ]\!]\ \text{using}\ [A^*]$

[Tst2]
$$\frac{f\ \mathit{is\ built\ from}\ [\mathit{Spec}^*] = \mathit{false}}{\begin{array}{l}\mathbf{tc}\ [\![\ X\ -\ f\ \to Y\ \mathit{Step}^*\ ]\!]\ \text{using}\ [A^*]\ [\mathit{Spec}^*] =\\ \mathbf{tc}\ [\![\ X\ ]\!]\ \text{using}\ [A^*] \cdot [\mathit{arrow}\ f\ \mathit{contains}\text{\textvisiblespace}\mathit{undeclared}\text{\textvisiblespace}\mathit{basic}\text{\textvisiblespace}\mathit{arrows}]\\ \cdot\ \mathbf{tc}\ [\![\ Y\ \mathit{Step}^*\ ]\!]\ \text{using}\ [A^*]\ [\mathit{Spec}^*]\end{array}}$$

[Tst3]
$$\frac{\begin{array}{c}f\ \mathit{is\ built\ from}\ [\mathit{Spec}^*] = \mathit{true},\\ \mathbf{tc}\ [\![\ \Phi\ [\![\ f : \mathbf{flat}[X] \to \mathbf{flat}[Y]\ ]\!]\ \text{using}\ [\mathit{Spec}^*]\ ]\!] = \mathit{true}\end{array}}{\begin{array}{l}\mathbf{tc}\ [\![\ X\ -\ f\ \to Y\ \mathit{Step}^*\ ]\!]\ \text{using}\ [A^*]\ [\mathit{Spec}^*] =\\ \mathbf{tc}\ [\![\ X\ ]\!]\ \text{using}\ [A^*] \cdot \mathbf{tc}\ [\![\ Y\ \mathit{Step}^*\ ]\!]\ \text{using}\ [A^*]\ [\mathit{Spec}^*]\end{array}}$$

[Tst4]
$$\frac{\begin{array}{c}f\ \mathit{is\ built\ from}\ [\mathit{Spec}^*] = \mathit{true},\\ \mathbf{tc}\ [\![\ \Phi\ [\![\ f : \mathbf{flat}[X] \to \mathbf{flat}[Y]\ ]\!]\ \text{using}\ [\mathit{Spec}^*]\ ]\!] = \mathit{false}\end{array}}{\begin{array}{l}\mathbf{tc}\ [\![\ X\ -\ f\ \to Y\ \mathit{Step}^*\ ]\!]\ \text{using}\ [A^*]\ [\mathit{Spec}^*] =\\ \mathbf{tc}\ [\![\ X\ ]\!]\ \text{using}\ [A^*] \cdot [\mathit{arrow}\ f\ \mathit{is}\text{\textvisiblespace}\mathit{not}\text{\textvisiblespace}\mathit{from}\ X\ \mathit{to}\ Y]\\ \cdot\ \mathbf{tc}\ [\![\ Y\ \mathit{Step}^*\ ]\!]\ \text{using}\ [A^*]\ [\mathit{Spec}^*]\end{array}}$$

Then, we scan the list of arrow definitions. For an arrow definition to pass the type-check, its name must be new, and its arrow steps must be correct. If the name is new, the specification of the arrow is appended to the specification list (because other definitions could use it).

[Tdef1] $\mathbf{tc}\ [\![\ \mathbf{def}\ .\ ]\!]\ \text{using}\ [A^*]\ [\mathit{Spec}^*] \ = \ []$

[Tdef2]
$$\frac{\mathit{spec\ of}\ a\ \mathit{in}\ [\mathit{Spec}^*] = \bot}{\begin{array}{l}\mathbf{tc}\ [\![\ \mathbf{def}\ a : X\ \mathit{Step}^*\ -\ f\ \to Y;\ \mathit{Def}^*\ .\ ]\!]\ \text{using}\ [A^*]\ [\mathit{Spec}^*] =\\ \mathbf{tc}\ [\![\ X\ \mathit{Step}^*\ -\ f\ \to Y\ ]\!]\ \text{using}\ [A^*]\ [\mathit{Spec}^*]\\ \cdot\ \mathbf{tc}\ [\![\ \mathbf{def}\ \mathit{Def}^*\ .\ ]\!]\ \text{using}\ [A^*]\ [a : \mathbf{flat}[X] \to \mathbf{flat}[Y],\ \mathit{Spec}^*]\end{array}}$$

[Tdef3]
$$\frac{\mathit{spec\ of}\ a\ \mathit{in}\ [\mathit{Spec}^*] \neq \bot}{\begin{array}{l}\mathbf{tc}\ [\![\ \mathbf{def}\ a : X\ \mathit{Step}^*\ -\ f\ \to Y;\ \mathit{Def}^*\ .\ ]\!]\ \text{using}\ [A^*]\ [\mathit{Spec}^*] =\\ [\mathit{arrow}\text{\textvisiblespace}\mathit{name}\ a\ \mathit{already}\text{\textvisiblespace}\mathit{used}]\\ \cdot\ \mathbf{tc}\ [\![\ X\ \mathit{Step}^*\ -\ f\ \to Y\ ]\!]\ \text{using}\ [A^*]\ [\mathit{Spec}^*] \cdot \mathbf{tc}\ [\![\ \mathbf{def}\ \mathit{Def}^*\ .\ ]\!]\ \text{using}\ [A^*]\ [\mathit{Spec}^*]\end{array}}$$

**Checking the whole program.** Finally, we type-check a whole program by accumulating the errors of the three sections.

$$\text{[Tprog]}\ \frac{\begin{array}{c}\textbf{tc}\ [\![\ \textbf{obj}\ A^*;\ ]\!] = [Error_1^*]\ [B^*],\\ \textbf{tc}\ [\![\ \textbf{lib}\ Ref^*;\ ]\!]\ \text{using}\ [B^*] = [Error_2^*]\ [Spec^*],\\ \textbf{tc}\ [\![\ \textbf{def}\ Def^*\ .\ ]\!]\ \text{using}\ [B^*]\ [Spec^*] = [Error_3^*]\end{array}}{\textbf{tc}\ [\![\ \textbf{obj}\ A^*;\ \textbf{lib}\ Ref^*;\ \textbf{def}\ Def^*\ .\ ]\!] = [Error_1^*,\ Error_2^*,\ Error_3^*]}$$

## 6.11 IMP-exec

This module specifies the mechanism for executing a program. The syntax declaration is very simple: it consists of a utility function which extracts the compiled code of an arrow from a list of compiled arrows, and of some ways of applying code to data.

**imports** IMP-compile[6.9] Data[6.6]
**exports**
**context-free syntax**

*code of* BASIC-ARROW *in* COMP-ARROW-LIST → CODE

"{" BASIC-CODE "}" "[" DATA "]" → DATA

CODE "[" DATA "]" using COMP-ARROW-LIST → DATA

Again for reasons of efficiency and clarity, in the equational description of the code execution we suppose that the data are correctly typed.

**equations**

**Executing basic codes.** This first two equations explain how to extract the code of an arrow from a compiled arrow list.

[Ecode1] $\textit{code of } a \textit{ in } [[a,\ \gamma],\ Comp^*]\ =\ \gamma$

[Ecode2] $a \neq b \Rightarrow \textit{code of } a \textit{ in } [[b,\ \gamma],\ Comp^*]\ =\ \textit{code of } a \textit{ in } [Comp^*]$

Now we discuss the basic instructions. First of all we discard the compiled arrow list.

[Ebasic] $\{Basic\text{-}Code\}[d]\ \text{using}\ [Comp^*]\ =\ \{Basic\text{-}Code\}[d]$

The equations described here are rather obvious. Recall that the empty term (i.e., the empty forest) represents the only element of $I$.

[ENOP] $\{\mathtt{NOP}\}[d]\ =\ d$

[EINJ] $\{\mathtt{TREE}[n]\}[d]\ =\ \triangleleft\, n,\ d\, \triangleright$

[ETERM] $\{\mathtt{TERM}\}[d]\ =$

[ERDL1] $\{\mathtt{RIGHTDEL}[0]\}[u^*]\ =\ u^*$

[ELDL1] $\{\mathtt{LEFTDEL}[0]\}[u^*]\ =\ u^*$

[ERDL2] $n \neq 0 \Rightarrow \{\mathtt{RIGHTDEL}[n]\}[u^*\ u] \ = \ \{\mathtt{RIGHTDEL}[n-1]\}[u^*]$

[ELDL2] $n \neq 0 \Rightarrow \{\mathtt{LEFTDEL}[n]\}[u\ u^*] \ = \ \{\mathtt{LEFTDEL}[n-1]\}[u^*]$

On the contrary, it is not obvious at all that the following four definitions implement correctly the distributivity isomorphism. We just remark that the form given here is allowed by the peephole optimization of Section 6.9, which got rid of a series of embarassing special cases, and that the four rules implement exactly the mathematical definition given in Section 5.2.1.

The values listed here for $q$, $q'$ and $n$ are all the possible ones, given that $n < p + q$ and that the cases for which $p$ or $q$ are equal to zero have been eliminated by the optimization phase.

[EDIST1] $$\frac{q \neq 1,\ \ m < q = \mathit{true}}{\{\mathtt{DIST}[q,\ q',\ n]\}[u^+ \lhd m,\ u^* \rhd] = \lhd\, 0,\ u^+ \lhd m,\ u^* \rhd \rhd}$$

[EDIST2] $$\frac{q' \neq 1,\ \ m \geq q = \mathit{true}}{\{\mathtt{DIST}[q,\ q',\ n]\}[u^+ \lhd m,\ u^* \rhd] = \lhd\, n,\ u^+ \lhd m - q,\ u^* \rhd \rhd}$$

[EDIST3] $\{\mathtt{DIST}[1,\ q',\ n]\}[u^+ \lhd 0,\ u^* \rhd] \ = \ \lhd\, 0,\ u^+\ u^* \rhd$

[EDIST4] $\{\mathtt{DIST}[q,\ 1,\ n]\}[u^+ \lhd q,\ u^* \rhd] \ = \ \lhd\, n,\ u^+\ u^* \rhd$

The execution of a compiled arrow consists in obtaining its code from the compiled arrow list and applying the code to the given data.

[ECOMP] $\{\mathtt{COMP},\ a\}[d]$ using $[\mathit{Comp}^*]$ $=$ *code of a in* $[\mathit{Comp}^*][d]$ using $[\mathit{Comp}^*]$

Structural induction on the code is a bit tricky, mainly because of the representation problems relative to sums. Again, the cases listed here for $p$, $q$ and $n$ are all the possible ones.

[Eb1] $$\frac{p \neq 1,\ \ n < p = \mathit{true}}{\gamma\ \ |\,[p,\ q]\ \gamma'[\lhd n,\ d \rhd] \text{ using } [\mathit{Comp}^*] = \gamma[\lhd n,\ d \rhd] \text{ using } [\mathit{Comp}^*]}$$

[Eb2] $$\frac{q \neq 1,\ \ n \geq p = \mathit{true}}{\gamma\ \ |\,[p,\ q]\ \gamma'[\lhd n,\ d \rhd] \text{ using } [\mathit{Comp}^*] = \gamma'[\lhd n - p,\ d \rhd] \text{ using } [\mathit{Comp}^*]}$$

[Eb3] $\gamma\ \ |\,[1,\ q]\ \gamma'[\lhd 0,\ d \rhd]$ using $[\mathit{Comp}^*]$ $=$ $\gamma[d]$ using $[\mathit{Comp}^*]$

[Eb4] $\gamma\ \ |\,[p,\ 1]\ \gamma'[\lhd p,\ d \rhd]$ using $[\mathit{Comp}^*]$ $=$ $\gamma'[d]$ using $[\mathit{Comp}^*]$

Composition and products are much easier to manage:

[Ec] $\gamma\ ,\ \gamma'[d]$ using $[\mathit{Comp}^*]$ $=$ $\gamma[d]$ using $[\mathit{Comp}^*] \cdot \gamma'[d]$ using $[\mathit{Comp}^*]$

[Eco] $\gamma\ ;\ \gamma'[d]$ using $[\mathit{Comp}^*]$ $=$ $\gamma'[\gamma[d]$ using $[\mathit{Comp}^*]]$ using $[\mathit{Comp}^*]$

It is however when executing iteration that the new coding for the data of a distributive program is useful. Due to the representation of a sum, the check for termination is just a check for the integer index of the data being greater than the length as a sum of the local state space. Note that $m$, $n$ and $p$ are all nonzero, and that we must handle as a special case $p = 1$, for then the representation of the result must be modified.

[EITER1] $$\frac{\gamma[\lhd 0,\ d \rhd] \text{ using } [\mathit{Comp}^*] = \lhd q,\ d' \rhd,\ \ q < n = \mathit{true}}{\begin{array}{l}\{\mathtt{ITER},\ \gamma,\ m,\ n,\ p\}[d] \text{ using } [\mathit{Comp}^*] = \\ \{\mathtt{ITER},\ \gamma,\ m,\ n,\ p\}[\lhd q + m,\ d' \rhd] \text{ using } [\mathit{Comp}^*]\end{array}}$$

[EITER2] $$\frac{p \neq 1,\ \gamma[\triangleleft\, 0,\ d \,\triangleright] \text{ using } [Comp^*] = \triangleleft\, q,\ d' \,\triangleright,\ q \geq n = true}{\{\texttt{ITER},\ \gamma,\ m,\ n,\ p\}[d] \text{ using } [Comp^*] = \triangleleft\, q - n,\ d' \,\triangleright}$$

[EITER3] $$\frac{\gamma[\triangleleft\, 0,\ d \,\triangleright] \text{ using } [Comp^*] = \triangleleft\, n,\ d' \,\triangleright}{\{\texttt{ITER},\ \gamma,\ m,\ n,\ 1\}[d] \text{ using } [Comp^*] = d'}$$

## 6.12 IMP-Nat

This module describes the usage of natural numbers in **IMP**$(G)$, and it can be used as a basis in order to include new data types into the language. The syntax part has just to declare as a subsort of BASIC-DATA all the desired data types (in this case, just INT).

**imports** IMP-exec$^{6.11}$ Ints
**exports**
  **context-free syntax**
    INT $\rightarrow$ BASIC-DATA

The equations, on the other hand, must specify the effect of each basic arrow, following the syntax described in Section 6.11. Here we define the predecessor and successor functions, which must be declared as $s : I + N \rightarrow N$ and $p : N \rightarrow I + N$.

**equations**

[IMP-Nat1] $\{s\}[\triangleleft\, 1,\ n \,\triangleright] \ = \ n + 1$

[IMP-Nat2] $\{s\}[\triangleleft\, 0,\ \,\triangleright] \ = \ 0$

[IMP-Nat3] $\{p\}[0] \ = \ \triangleleft\, 0,\ \,\triangleright$

[IMP-Nat4] $n \neq 0 \Rightarrow \{p\}[n] \ = \ \triangleleft\, 1,\ n - 1 \,\triangleright$

The following equations implement sum, truncated difference and multiplications functions from $N \times N$ to $N$ and "greater", "greater or equal" and "equal" predicates from $N \times N$ to $I + I$ (the first copy of $I$ representing false). Note that the integer module imported here uses "-" for truncated subtraction; this could not be true for other modules providing an integer data type, and the equations should be modified accordingly.

[IMP-Nat5] $\{plus\}[p\ q] \ = \ p + q$

[IMP-Nat6] $\{minus\}[p\ q] \ = \ p - q$

[IMP-Nat7] $\{times\}[p\ q] \ = \ p * q$

[IMP-Nat8] $p > q = false \Rightarrow \{gt\}[p\ q] \ = \ \triangleleft\, 0,\ \,\triangleright$

[IMP-Nat9] $p > q = true \Rightarrow \{gt\}[p\ q] \ = \ \triangleleft\, 1,\ \,\triangleright$

[IMP-Nat10] $p \geq q = false \Rightarrow \{ge\}[p\ q] \ = \ \triangleleft\, 0,\ \,\triangleright$

[IMP-Nat11] $p \geq q = true \Rightarrow \{ge\}[p\ q] \ = \ \triangleleft\, 1,\ \,\triangleright$

[IMP-Nat12] $p \neq q \Rightarrow \{eq\}[p\ q] \ = \ \triangleleft\, 0,\ \,\triangleright$

[IMP-Nat13] $p = q \Rightarrow \{eq\}[p\ q] \ = \ \triangleleft\, 1,\ \,\triangleright$

## 6.13 IMP

This module is the parent of all modules. It imports all which is necessary in order to type-check, compile and execute **IMP**$(G)$ programs, and an arbitrary number of data type modules. Moreover, it describes some useful syntax for executing arrows.

**imports** IMP-typecheck[6.10] IMP-compile[6.9] IMP-exec[6.11] IMP-Nat[6.12]

The standard way of executing a program is to give the name of one of the arrows which have been defined, followed by the data it has to be applied to enclosed in square brackets. The output can be either an error list produced by the type-checker, or the resulting data.

**exports**
**sorts** ERROR-LIST-OR-DATA
**context-free syntax**

ERROR-LIST → ERROR-LIST-OR-DATA
DATA → ERROR-LIST-OR-DATA

BASIC-ARROW "[" DATA "]" using PROGRAM → ERROR-LIST-OR-DATA

**equations**

**Executing (part of) a program.** When applying an arrow to some data, reasons of efficiency impose that the compiler tries to compile directly the arrow. If the compilation fails, the type-checker is invoked and an error list is returned. It should be noted, however, that in some cases (for instance, duplicate references or object declarations) the compiler could blindly compile a program with minor syntactical errors. However, if the resulting compiled arrow list is executable the code will perform correctly.

[Ef1] $$\frac{\Phi \;[\![\; Program \;]\!] = [Comp^*]}{a[d] \text{ using } Program = \{\texttt{COMP},\, a\}[d] \text{ using } [Comp^*]}$$

[Ef2] $a[d]$ using $Program \;=\; \textbf{tc}\;[\![\; Program \;]\!]$ **otherwise**